\RequirePackage{fix-cm}
\documentclass[natbib]{svjour3}                     % onecolumn (standard format)
\smartqed  % flush right qed marks, e.g. at end of proof
\usepackage{graphicx}
\usepackage{amsmath}
\usepackage[caption=false,font=footnotesize,labelfont=sf,textfont=sf]{subfig}
\usepackage{url}

\usepackage{enumitem}
\usepackage[super]{nth}
\usepackage{multirow}
\usepackage{algorithm}
\usepackage[noend]{algorithmic}
\usepackage{booktabs}
\usepackage{soul} % for \hl{} - can remove once we're done
\usepackage{comment}

\newcommand{\bb}{\textbf} % DCW for easy bolding in table cells
\newcommand{\hashtagsanshash}[1]{\texttt{#1}}
\newcommand{\hashtag}[1]{\texttt{\##1}}
\newcommand{\mention}[1]{\texttt{@#1}}
\newcommand{\redact}{{\small $\langle$REDACTED$\rangle$}}
\newcommand{\aurl}{{\small $\langle$URL$\rangle$}}

\usepackage{xcolor}

\usepackage{etoolbox}
\apptocmd{\thebibliography}{\setlength{\itemsep}{0pt}}{}{}

\journalname{Social Network Analysis and Mining}
\begin{document}

\title{A General Method to Find Highly Coordinating Communities in Social Media through Inferred Interaction Links}
% \title{Searching for teams engaging in coordinated inauthentic behaviour in political charged online discussions}
\subtitle{}

\titlerunning{Finding Highly Coordinating Communities in Social Media}        % if too long for running head

\author{Derek Weber   \and
        Frank Neumann
}


\institute{    
    D.C. Weber 
    \at School of Computer Science, \\University of Adelaide / \\
        Defence Science and Technology Group\\
        Adelaide, South Australia, Australia\\
        ORCID ID: 0000-0003-3830-9014\\
        \email{derek.weber@\{adelaide.edu.au,dst.defence.gov.au\}}
    \and
    F. Neumann 
    \at School of Computer Science, \\University of Adelaide\\
        Adelaide, South Australia, Australia\\
        ORCID ID: 0000-0002-2721-3618\\
        \email{frank.neumann@adelaide.edu.au}
}

\date{Received: date / Accepted: date}

\maketitle

\begin{abstract}
Political misinformation, astroturfing and organised trolling are %examples of 
online malicious behaviours with significant real-world effects. 
Many previous approaches examining these phenomena have focused on %individual behaviours 
broad campaigns rather than the small groups responsible for instigating or sustaining them.  
%such as mass posting of duplicated content or hashtag abuse. 
To reveal latent (i.e., hidden) networks of cooperating accounts, we propose a novel temporal window %general window-based 
approach that relies on account interactions and metadata alone. 
It detects groups of accounts engaging in various behaviours that, in concert, come to execute different goal-based strategies, a number of which we describe. 
The approach relies upon a pipeline that extracts relevant elements from social media posts, infers connections between accounts based on criteria matching the coordination strategies to build an undirected weighted network of accounts, which is then mined for communities exhibiting high levels of evidence of coordination using a novel community extraction method.
% To reveal latent networks of coordinating accounts, we propose an approach based only on interactions common among OSNs and time window-based streaming graphs, 
% %We propose an approach based on interaction network topology and streaming graphs 
% %to search for latent networks of accounts, %whose behaviour appears coordinated, 
% where the criteria for such behaviour are %described as partially ordered 
% combinations of the individual interactions studied by others. 
% where the criteria for coordination consists of partially ordered sequences of the individual behaviours studied by others.
%The interactions in our criteria are generalised from actions common to prominent OSNs.
% Behaviours common to multiple OSNs are used to generalise the criteria.
%%%\color{red}A decaying sliding window is used to further emphasise consistent coordinated behaviour over time. \color{black} 
We address the temporal aspect of the data by using a windowing mechanism, which may be suitable for near real-time application. We further highlight consistent coordination with a sliding frame across multiple windows and application of a decay factor.
Our approach is compared with other recent similar processing approaches and community detection methods and is validated against two relevant datasets with ground truth data, %.%, and find that we successfully reveal groups of accounts engaging in particular coordination strategies. 
using content, temporal, and network analyses, as well as with the design, training and application of three one-class classifiers built using the ground truth; its utility is furthermore demonstrated in two case studies of contentious online discussions.
%Investigating two relevant datasets and validating our results with one-class classifiers trained on seed examplars as ground truth, the approach successfully highlights accounts exhibiting a common coordination strategy. % 146 words

\keywords{Coordinated Behaviour \and Online Social Networks \and Influence Operations}
\end{abstract}

% SECTION 1.1
\section{Introduction}\label{sec:intro}

Online social networks (OSNs) 
have established themselves %in modern life as a very flexible and accessible system to create and maintain social ties, they also serve 
as flexible and accessible systems for activity coordination and information dissemination.
This benefit was illustrated during the Arab Spring~\citep{carvin2013} but inherent dangers are increasingly apparent in ongoing political interference and disinformation~\citep{BessiFerrara2016,howard2016brexit,ferrara2017frelec,KellerICWSM2017,Neudert2018ger,morstatter2018alt}. 
%This has brought significant hope and change to parts of the world, such as during the Arab Spring~\citep{carvin2013}, but also facilitated significant interference in political processes~\citep{BessiFerrara2016,ferrara2017frelec,KellerICWSM2017} %~\citep{ratkiewicz2011,metaxas2012,woolley2016autopower,StarbirdWilson2020,PachecoFM2020whitehelmets,KellerICWSM2017,Keller2019} 
%~\citep{BessiFerrara2016,ferrara2017frelec,mueller2018} 
% through sowing discord and disinformation. 
Modern information campaigns are participatory activities, which aim to use their audiences to amplify their desired narratives, not just receive it \citep{StarbirdAW2019cscw}. Through cyclical reporting (i.e., social media feeding stories and narratives to traditional news media, which then sparks more social media activity), social media users can unknowingly become ``unwitting agents'' as ``sincere activists'' of state-based operations~\citep{StarbirdWilson2020}. %, and struggle to identify automated bot accounts used to amplify the messaging~\citep{rise2016,cresci2017}. 
The use of \emph{political} bots to influence the framing and discussion of issues in the mainstream media (MSM) 
%amplify fringe voices, thereby influencing which issues are discussed and how they are framed in the mainstream media (MSM), 
remains prevalent~\citep{BessiFerrara2016,woolley2016autopower,woolley2017us,DebateNightICWSM2018}. % ~\citep{BessiFerrara2016,woolley2016autopower,ferrara2017frelec,woolley2017us,HegelichJ2016ukranianbotnet,PachecoFM2020whitehelmets}. 
%To effectively create 
The use of bots to amplify individual voices above the crowd, known as the \emph{megaphone effect}, requires coordinated action and %this behaviour requires 
a degree of regularity that may leave traces in the digital record.

Relevant research has focused on high level analyses of campaign detection and classification~\citep{LeeCCS2013campext,varol2017campaigndetection,Alizadeh2020}, the identification of botnets and other dissemination groups~\citep{vo2017,Gupta2019,woolley2017us}, and coordination at the community level~\citep{KumarHLJ2018conflict,HineOCKLSSB2017kekcucks,Cresci2020}. 
% Relevant research has investigated campaign detection and classification~\citep{LeeCCS2013campext,varol2017campaigndetection,Giglietto2020}, as well as identifying large groups of malicious retweeting accounts~\citep{vo2017,Gupta2019} and other botnets~\citep{woolley2017us}, %,HegelichJ2016ukranianbotnet}, 
% and the identification of conflicting communities~\citep{KumarHLJ2018conflict,HineOCKLSSB2017kekcucks}%~\citep{morstatter2018alt,KumarHLJ2018conflict,DattaA19conflictnetwork,HineOCKLSSB2017kekcucks}.
%Other research has 
Some have considered generalised approaches to social media analytics~\citep[e.g.,][]{rapid2018,weber2019coord,Pacheco2020arxiv,graham2020asnac,Nizzoli2020}, % and campaign detection, %~\citep{Pacheco2020arxiv,weber2019coord,AssenmacherATG2020}, 
but unanswered questions regarding the clarification of coordination strategies remain.% according to their goals and execution methods and a strategy-level general approaches for detecting such behaviour.

% To address these questions, we present a set of observed coordination strategies, including their intent and execution methods, and provide validated techniques for identifying groups of accounts using them.

Expanding on our work presented at ASONAM'20 \citep{weber2020coord}, we present a novel approach to detect groups engaging in potentially coordinated activities, revealed through anomalously high levels of coincidental behaviour. Links in the groups are inferred from behaviours that, when used intentionally, are used to execute a number of identifiable coordination strategies. We use a range of techniques to validate our new technique on two relevant datasets, as well as comparison with ground truth and a synthesized dataset, and show it successfully identifies coordinating communities.

%We present a new approach to identify coordination strategies. In particular, our approach is able to point out the intent and execution methods of the coordination strategies. We validate our new technique on various datasets and show that it successfully identifies groups of coordinated accounts within these datasets.

% Our contributions include:
% \begin{enumerate}%[itemsep=1pt]%[noitemsep]
%     \item categorising observed coordination strategies including their intent and methods of execution; and
%     \item providing validated techniques to identify groups of accounts using in them.
%     % \item describing coordination strategies observed online and mechanisms by which they are enacted;
%     % \item a technique to identify the sets of accounts using these strategies, generalisable to major OSNs; and % social media platforms; and
%     % \item validation of the technique against two relevant datasets collected with different strategies, one including a clearly defined ground truth subset.\marginpar{Four datasets?}
% \end{enumerate}

% We propose a general methodology %\hl{approach}\marginpar{approach or technique?} 
% that
Our approach infers ties between accounts to construct \emph{latent coordination networks} (LCNs) of accounts, using criteria specific to different coordination strategies. These are sometimes referred to as user similarity networks \citep[e.g.,][]{Nizzoli2020}. 
The accounts may not be directly connected, thus we use the term `latent' to mean `hidden' when describing these connections. The inference of connections is performed solely 
on the accounts' 
activity, i.e., not their content, only metadata and temporal information, though it could be expanded to make use of these, similar to \citet{Pacheco2020arxiv} and \citet{graham2020asnac}. 
% The ties are used to construct \emph{latent connection networks} (LCNs) of accounts, 
\emph{Highly coordinating communities} (HCCs) are then detected and extracted from the LCN. 
We propose a variant of \emph{focal structures analysis} \citep[FSA,][]{Sen2016fsa} to do this, in order to take advantage of FSA's focus on finding influential sets of nodes in a network, while also reducing the computational complexity of the algorithm. 
A window-based approach is used to enforce temporal constraints.%on the activities
%%%{\color{red}, and an additional decaying sliding window %historical\marginpar{historical?} view %is used to emphasise consistently coordinating accounts
%%%emphasises persistent coordination}%
% .

% % To evaluate this approach, we applied it to 
% Comparison of  
% two relevant %Twitter 
% datasets%(though its application to other OSNs will be clear) and validate it against a 
% , including 
% labeled ground truth, with a randomised dataset provides validation. % dataset. To guide the evaluation and validation, we used the following research questions:
The following research questions guided our evaluation:

\begin{description}%[itemsep=1pt]%[noitemsep]
    \item[{\small \textbf{RQ1}}] How can HCCs be found in an LCN?
    \item[{\small \textbf{RQ2}}] How do the discovered communities differ? %, depending on the method used?
    % \item[{\small RQ2}] How does the method used affect the HCCs discovered? %, depending on the method used?
    \item[{\small \textbf{RQ3}}] Are the %discovered communities 
    HCCs 
    internally or externally focused? %How does this validate the findings?
    \item[{\small \textbf{RQ4}}] How consistent is the HCC messaging?
    % \item[{\small RQ4}] How much variation is there in the content posted by the communities?
    \item[{\small \textbf{RQ5}}] What evidence is there of consistent coordination?
    \item[{\small \textbf{RQ6}}] How well can HCCs in one dataset inform the discovery of HCCs in another? %To what extent can known coordinating communities be used as exemplars in different datasets?

    %%%\item[{\small {\color{red} RQ5}}] {\color{red} How persistent is the HCC behaviour?}
    % \item[{\small {\color{red} RQ5}}] {\color{red} How consistent is the behaviour over time?}
    % To what extent does a sliding window mechanism emphasise HCCs? %Do the groups vary their membership across time or is it the same accounts working together? (perhaps this last is a future question)
\end{description}

This paper expands upon \citet{weber2020coord} by providing further methodological detail and experimental validation, and case studies in which the technique is applied to new real world datasets relating to contentious political issues, as well as consideration of algorithmic complexity and comparison with several similar techniques. Prominent among the extra validation provided is the use of machine learning classifiers to show that our datasets contain similar coordination to our ground truth, and the application of a sliding frame across the time windows as a way to search for consistent coordination.

This paper provides an overview of relevant literature, followed by a discussion of online coordination strategies and their execution. Our approach is then explained, and its experimental validation is presented\footnote{See \url{https://github.com/weberdc/find_hccs} for code and data.}. Following the validation, the algorithmic complexity and performance of the technique is presented, two case studies are explored, demonstrating the utility of the approach in two real-world politically-relevant datasets, and we compare our technique to those of \citet{Pacheco2020arxiv}, \citet{graham2020asnac}, \citet{Nizzoli2020} and \citet{Giglietto2020}.
% setup, presentation of results and detailed analyses attending to the research questions. Concluding remarks and options for future work are offered.

\subsection{A Motivating Example} \label{sec:motivation}

\begin{figure}[!ht]
    \centering
        \subfloat[One upset partner.\label{fig:one_upset_partner_eg}]{%
            \includegraphics[width=0.49\columnwidth]{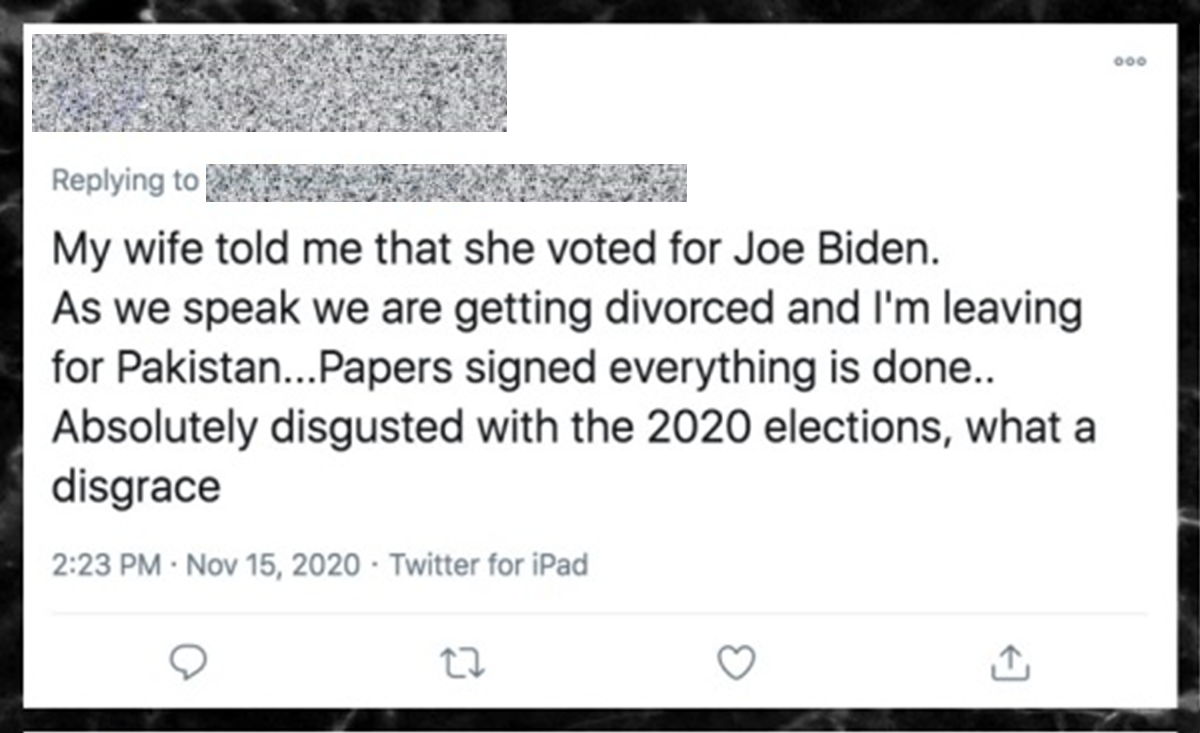}
        }
        \hfill
        \subfloat[Many upset partners.\label{fig:many_upset_partners_eg}]{%
            \includegraphics[width=0.49\columnwidth]{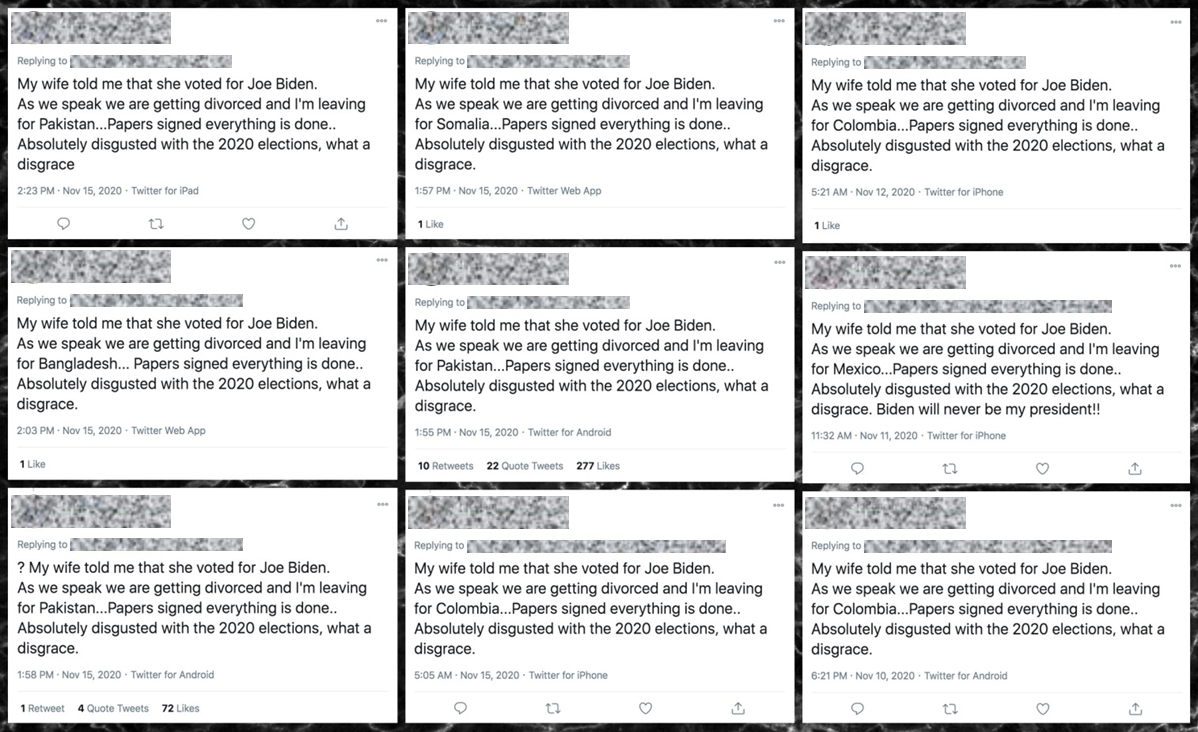}
        }
    \caption{Copypasta tweets noticed in the aftermath of the 2020 US Presidential election, which may bely a coordinated campaign to undermine confidence in American society's ability to accept electoral outcomes, or may just be a prank similar to a flashmob.}
    \label{fig:motivating_eg}
\end{figure}

In the aftermath of the 2020 US Presidential election, a data scientist noticed a pattern emerging on Twitter\footnote{Tweeted 2020-11-17: \url{https://twitter.com/conspirator0/status/1328479128908132358}}.
Figure~\ref{fig:one_upset_partner_eg} shows a tweet by someone who was so upset with their partner voting for Joe Biden in the election that they decided to divorce them immediately and move to Pakistan (in the midst of the COVID-19 pandemic). This might seem an extreme reaction, but the interesting thing was that the person was not alone. The researcher had identified dozens of similar, but not always identical, tweets by people leaving for other cities but for the same reason (Figure~\ref{fig:many_upset_partners_eg}). Analysis of these accounts also revealed they were not automated accounts. This kind of pattern of tweeting identical text is sometimes referred to as ``copypasta'' and can be used to give the appearance of a genuine grassroots movement on a particular issue. It had been previously used by ISIS terrorists as they approached the city of Mosul, Iraq, in 2014, which they took over and occupied for several years when the local forces believed a giant army was on its way based on the level of relevant online activity \citep{wargoesviral2016}.

It is unclear whether this ``copypasta'' campaign is a deliberate information operation, designed to damage trust in the electoral system and ability of Americans to accept the loss of a preferred political party in elections, or simply a group of like-minded jokers starting a viral gag or engaging in a kind of flashmob. At the very least, it is important to be able to identify which accounts are participating in the event, and how they are coordinating their actions.

\subsection{Online Information Campaigns and Related Work}

Social media has been increasingly used for communication in recent years (particularly political communication), and so the market has followed, with media organisations using it for cheap, wide dissemination and consumers increasingly looking to it for news \citep{ShearerGrieco2019pew}. Over that same time period, people have begun to explore how to exploit the features of the internet and social media that bring us benefits: the ability to target marketing to specific audiences that connects businesses with the most receptive customers also enables highly targeted non-transparent political advertising \citep{woolley2017us} and the ability to expose people to propaganda and recruit them to extremist organisations \citep{Badawy2018}; the anonymity that supports the voiceless in society to express themselves also enables trolls to attack others without repercussions \citep{HineOCKLSSB2017kekcucks,BurgessMF2016gamergate}; and the automation that enables news aggregators also facilitates social and political bots \citep{rise2016,woolley2016autopower,Cresci2020}. In summary, targeted marketing and automation coupled with anonymity provide the tools required for potentially significant influence in the online sphere, perhaps enough to swing an election.

Effective influence campaigns relying on these capabilities will somehow coordinate the actions of their participants. Early work on the concept of coordination by \citet{MaloneCrowston1994coordsurvey} described it as the dependencies between the tasks and resources required to achieve a goal. One task may require the output of another task to complete. Two tasks may share, and require exclusive access to, a resource or they may both need to use the resource simultaneously. 

At the other end of the spectrum, sociological studies of influence campaigns can reveal their intent and how they are conducted, but they consider coordination at a much higher level. \citet{StarbirdAW2019cscw} highlight three kinds of campaigns: \emph{orchestrated}, centrally controlled campaigns that are run from the top down \citep[e.g., paid teams,][]{theagency2015,king_pan_roberts_2017}; \emph{cultivated} campaigns that infiltrate existing issue-based movements to drive them to particular extreme positions; and \emph{emergent} campaigns arising from enthusiastic communities centred around a shared ideology (e.g., conspiracy groups and other fringe movements). Though their strategies differ, they use the same online interactions as normal users, but their patterns differ.

Disinformation campaigns effectively trigger human cognitive heuristics, such as individual and social biases to believe what we hear first (\emph{anchoring}) and what we hear frequently and can remember easily (\emph{availability} cascades) \citep{TverskyKahneman1973,KuranSunstein1999}; thus the damage is already done by the time lies are exposed. Recent experiences of false information moving beyond social media during Australia's 2019-2020 bushfires highlight that identifying these campaigns as they occur can aid OSN monitors and the media to better inform the public \citep{GrahamKeller2020conv}.%\footnote{\url{https://theconversation.com/bushfires-bots-and-arson-claims-australia-flung-in-the-global-disinformation-spotlight-129556}}.

In between task level coordination and entire campaigns, at the level of social media interactions, as demonstrated by \citet{GrahamKeller2020conv}, we can directly observe the online actions and effects of such campaigns, and infer links between accounts based on pre-determined criteria.
% A variety of research fields inform the study of online influence campaigns, including computer science, psychology, sociology, and political and military science. 
% From a sociological standpoint, Starbird \emph{et al.} highlight three types of coordination while studying \emph{strategic information operations} (SIOs) online~\citep{StarbirdAW2019cscw}: \emph{orchestrated} campaigns which are centrally managed and highly structured (e.g., Russia's Internet Research Agency~\citep{theagency2015}%,mueller2018,king_pan_roberts_2017}
% ); \emph{cultivated} campaigns, which rely on infiltrating and guiding existing movements; and \emph{emergent} campaigns which arise from communities surrounding ideologues and narratives (e.g., conspiracy groups). 
% % Such a distinction lends itself to offline phenomenon such as flash mobs (orchestrated) or love locks on bridges (emergent). 
% Observing campaigns at this level allows analysis of cross-platform campaigns~\citep{BurgessMF2016gamergate}%,BandeliAgarwal2018}
% , including ones that rely on the MSM~\citep{StarbirdWilson2020}, from an organisational level. 
Relevant efforts in computer science has focused on detecting information operations on social media via automation~\citep{rise2016,grimme2017,davis2016botornot,Cresci2020}, 
campaign detection via content~\citep{LeeCCS2013campext,AssenmacherATG2020,Alizadeh2020}, URLs~\citep{ratkiewicz2011,CaoCLGC2015urlsh,Giglietto2020} and hashtags~\citep{ratkiewicz2011,BurgessMF2016gamergate,varol2017campaigndetection,WeberNFM2020bushfiresspringer}, 
%campaign detection~\citep{LeeCCS2013campext,CaoCLGC2015urlsh,varol2017campaigndetection,Alizadeh2020},  %,WeberNFL2020bushfiresarxiv}, 
temporal patterns~\citep{chavoshi2017,HineOCKLSSB2017kekcucks,nasim2018,Mazza2019rtbust,PachecoFM2020whitehelmets},
and community detection~\citep{vo2017,morstatter2018alt,Gupta2019}. Other studies have explored how bots and humans interact in political settings~\citep{BessiFerrara2016,DebateNightICWSM2018,woolley2017us}, %woolley2016autopower,woolley2017us,BessiFerrara2016,ferrara2017frelec,HegelichJ2016ukranianbotnet}, 
including %using network analysis to explore 
exploring 
how deeply embedded bots appear in the network and their degree of organisation% when supporting different campaigns
~\citep{woolley2017us}. 
There is, however, a research gap: the computer science study of the ``orchestrated activities'' of accounts in general, regardless of their degree of automation%, as described by Grimme \emph{et al.}
~\citep{GrimmeAA2018perspectives,Alizadeh2020,Nizzoli2020,Vargas2020}. 
% Increasingly, however, researchers are calling for study of accounts' ``orchestrated activities''~\citep{GrimmeAA2018perspectives} rather that simply the detection of automated activity, which 
It must be noted that bot activity, even coordinated activity, may be entirely benign and even useful~\citep{rise2016,graham2017}. 

Though some studies have observed the existence of strategic behaviour in and between online groups \citep[e.g.,][]{KellerICWSM2017,KumarHLJ2018conflict,HineOCKLSSB2017kekcucks,Keller2019,Giglietto2020}, the challenge of identifying a broad range of strategies and their underpinning execution methods remains relatively unstudied. 

Inferring social networks from OSN data requires attendance to the temporal aspect to understand information (and influence) flow and degrees of activity~\citep{holme2012}. Real time processing of OSN posts can enable tracking narratives via text clusters~\citep{AssenmacherATG2020}, but to process networks requires graph streams~\citep{McGregor2014graphstreams} or window-based pipelines \citep[e.g.,][]{weber2019coord}, otherwise processing is limited to post-collection activities \citep{graham2020asnac,Pacheco2020arxiv}.

This work contributes to the identification of strategic coordination behaviours,  %provides a contribution to this challenge, 
along with a general technique to enable detection of groups using them.

% \section{Background}
\section{Coordination Strategies} \label{sec:coordination_strategies}

% \subsection{Dissemination and Engagement}

% Martin has suggested cutting this paragraph down and providing a citation for the dissemination/engagement idea (which I don't have because I made it up myself)
% Influencing others online with a specific narrative 
Online influence 
relies on two primary mechanisms: \emph{dissemination} %(of which the megaphone effect is an example) 
and \emph{engagement}. 
%For example, an investigation of social media activity following terrorist attacks in the UK in 2017\footnote{\url{https://crestresearch.ac.uk/resources/russian-influence-uk-terrorist-attacks/}} identified accounts promulgating false information regarding the attacks, partly to inflame racial tensions and partly to promote tolerance, in order to sow division. By engaging aggressively, the accounts drew in participants who spread the narrative further. %These have legitimate uses too, however, such as genuine activism and political campaigning.
For example, an investigation of social media activity following UK terrorist attacks in 2017\footnote{\url{https://crestresearch.ac.uk/resources/russian-influence-uk-terrorist-attacks/}} identified accounts promulgating contradictory narratives, inflaming racial tensions and simultaneously promoting tolerance to sow division. By engaging aggressively, the accounts drew in participants who then spread the message.

\textbf{Dissemination} aims to maximise audience, to convince through repeated exposure and, in the case of malicious use, to cause outrage, polarisation and confusion, or at least attract attention to distract from other content.% in the community (e.g., the anti- and pro-tolerance content spread after the 2017 UK terrorist attacks\footnote{\url{https://crestresearch.ac.uk/resources/russian-influence-uk-terrorist-attacks/}}). Dissemination can be enhanced if the OSN itself, rather than just other users, can be convinced to promulgate content (via gaming trend algorithms).

% \bb{Dissemination.} The aim of dissemination is to maximise the audience of the content, to convince others of its veracity through repetitive exposure and, especially in the case of malicious use, to cause outrage, polarisation and confusion in the community (e.g., the anti- and pro-tolerance content spread after the 2017 UK terrorist attacks\footnote{\url{https://crestresearch.ac.uk/resources/russian-influence-uk-terrorist-attacks/}}). Dissemination can be enhanced if the OSN itself, rather than just other users, can be convinced to promulgate content (via gaming trend algorithms).

\textbf{Engagement} is a subset of dissemination that solicits a response. It relies on targeting individuals or communities through mentions, replies and the use of hashtags as well as rhetorical approaches that invite responses (e.g., inflammatory comments or, as present in the UK terrorist example above, pleas to highly popular accounts).% The activity after the 2017 UK terrorist attacks included appeals to highly popular accounts to engage in the discussion to increase reach.

% \bb{Engagement.} Engagement can be thought of as a subset of dissemination that solicits a response. For this reason, it relies on targeting individuals or communities through mentions, replies and the use of hashtags as well as rhetorical approaches that invite responses (e.g., inflammatory comments or, as present in the UK terrorist example above, pleas to highly popular accounts).% The activity after the 2017 UK terrorist attacks included appeals to highly popular accounts to engage in the discussion to increase reach.

%To identify small sets of actively cooperating accounts, rather than community-level trends, it can be helpful to consider combinations of behaviours that contribute in concert as the execution of coordination strategies. 

\begin{figure}[t!]
    \centering
    \includegraphics[width=0.99\columnwidth]{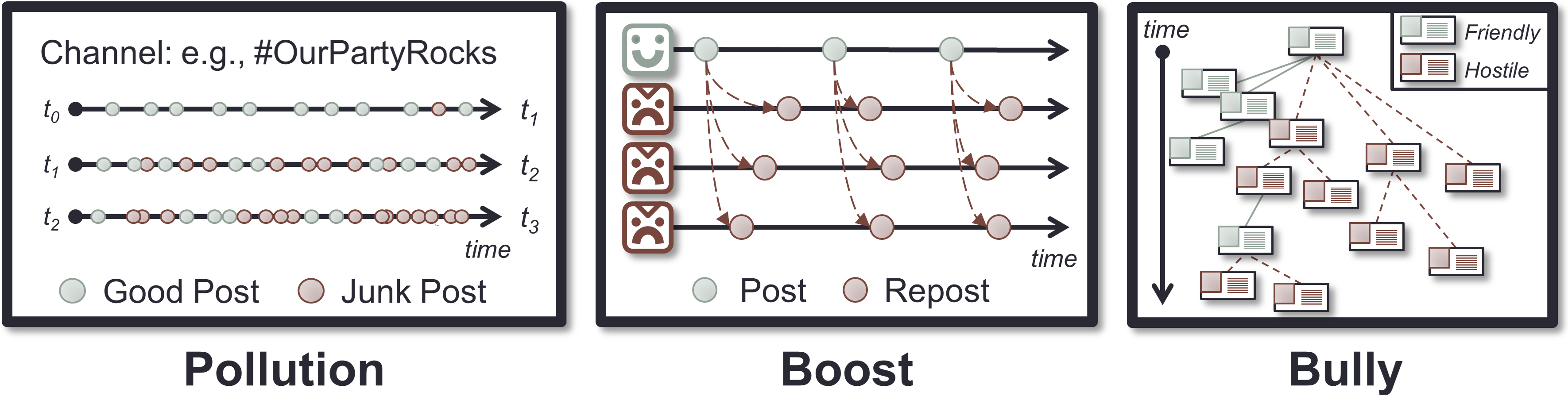}
    \caption{Patterns matching the mentioned coordination strategies. 
    %Blue posts and avatars are benign, red are malign.
    Green posts and avatars are benign, whereas red or maroon ones are malign.}
    \label{fig:coord_strategies}
\end{figure}

A number of online coordination strategies have been observed in the literature making use of both dissemination and engagement, including: %, executed through individual or combinations of actions. 
% These include:

\begin{enumerate}%[itemsep=1pt]%[noitemsep]
    \item %\textit{Channel pollution}:
    \textit{Pollution}: flooding a community with repeated or objectionable content, causing the OSN to shut it down~\citep{ratkiewicz2011,woolley2016autopower,HegelichJ2016ukranianbotnet,HineOCKLSSB2017kekcucks,nasim2018,fisher2018netwar,MaricontiSBCKLS2019cscw}
    \item %\textit{Amplify by repost}:
    \textit{Boost}: heavily reposting content to make it appear popular%, aiming to game trending algorithms
    ~\citep{ratkiewicz2011,CaoCLGC2015urlsh,varol2017campaigndetection,vo2017,Gupta2019,Keller2019,graham2020virus,AssenmacherATG2020};
    \item %\textit{Gang-up}: 
    \textit{Bully}: groups of individuals harassing another individual or community \citep{BurgessMF2016gamergate,HineOCKLSSB2017kekcucks,KumarHLJ2018conflict,DattaA19conflictnetwork,MaricontiSBCKLS2019cscw}; and
    \item \textit{Metadata Shuffling}: groups of accounts changing, and even swapping, %their names and other identifying 
    metadata to %confuse OSN auto-linking features, thus hiding account
    hide their identities~\citep{Mariconti2017,ferrara2017frelec}.
\end{enumerate}

\begin{table}
    \centering%\small
    \caption{Social media interaction equivalents}
    \label{tab:primitives}
    \resizebox{\columnwidth}{!}{%
        \begin{tabular}{@{}l|llllll@{}}
            \toprule
            OSN      & POST  & REPOST    & REPLY       & MENTION    & TAG        & LIKE \\
            \midrule
            Twitter  & tweet & retweet   & reply tweet & @mention   & \#hashtags & favourite \\
            Facebook & post  & share     & comment     & mention    & \#hashtag  & reactions \\
            Tumblr   & post  & repost    & comment     & @mention   & \#tag      & heart \\
            Reddit   & post  & crosspost & comment     & /u/mention & subreddit  & up/down vote \\
            \bottomrule
        \end{tabular}
    } % end resizebox/textwidth
\end{table}

Different behaviour primitives (e.g., Table~\ref{tab:primitives}) can be used to execute these strategies. Dissemination can be carried out by reposting, using hashtags, or mentioning highly connected individuals in the hope they spread a message further. Accounts doing this covertly will avoid direct connections, and thus inference is required for identification. 
\citet{Giglietto2020} propose detecting anomalous levels of coincidental URL use as a way to do this; we expand this approach to other interactions. 

Some strategies require more sophisticated detection: detecting bullying through \emph{dogpiling} 
(e.g., as happened during the \#GamerGate incident, studied by \citet{BurgessMF2016gamergate}, or to those posing questions to public figures at political campaign rallies\footnote{https://www.bbc.co.uk/bbcthree/article/72686b6d-abd2-471b-ae1d-8426522b1a97}) requires collection of (mostly) entire conversation trees, which, while trivial to obtain on forum-based sites (e.g., Facebook and Reddit), are difficult on stream-of-post sites (e.g., Twitter, Parler and Gab). 
Detecting metadata shuffling requires long term collection on broad issues to detect the same accounts being active in different contexts.
%If consistent co-commenting on conversations is the execution of \emph{Bully}, detecting it differs inasmuch as it requires the collection and analysis of entire conversation trees. On Reddit, Facebook or Tumblr, this requires simply collecting posts and their replies, but collecting such structures on Twitter requires more sophisticated strategies\footnote{E.g., ICWSM'19 tutorial on ``Generative models of online discussion threads''. \url{https://www.icwsm.org/2019/program/tutorial/#t2}}. Metadata shuffling requires long term collection on broad (often political) issues, in order to detect the same accounts being active in different contexts, such as bots supporting Donald Trump in 2016 and then denigrating Emmanuel Macron in 2017~\citep{ferrara2017frelec}.

Figure~\ref{fig:coord_strategies} shows representations of %the following 
several 
strategies, %illustrating 
giving clues about 
how they might be identified. 
To detect \emph{Pollution}, we match the authors of posts mentioning the same (hash)tag. This way we can reveal not just those who are using the same hashtags with significantly greater frequency than the average but also those who use more hashtags than is typical. 
To detect \emph{Boost}, we match authors reposting the same original post, and can explore which sets of users not only repost more often than the average, but those who repost content from a relatively small pool of accounts. Alternatively, we can match authors who post identical, or near identical text, as seen in our motivating example (Section~\ref{sec:motivation}); \citet{graham2020asnac} have recently developed open sourced methods for this kind of matching. 
Considering reposts like retweets, however, it is unclear whether platforms deprioritise them when responding to stream filter and search requests, so special consideration may be required when designing data collection plans.
Finally, to detect \emph{Bully}, we match authors whose replies are transitively rooted in the same original post, thus they are in the \emph{same conversation}. This requires collection strategies that result in complete conversation trees, and also stipulates a somewhat strict definition of `conversation'. On forum-based OSNs, the edges of a `conversation' may be relatively clear: by commenting on a post, one is `joining' the `conversation'. Delineating smaller sets of interactions within all the comments on a post to find smaller conversations may be achieved by regarding each top-level comment and its replies as a conversation, but this may not be sufficient. Similarly, on stream-based OSNs, a conversation may be engaged in by a set of users if they all mention each other in their posts, as it is not possible to \emph{reply} to more than one post at a time. 
%1) \emph{Pollution} involves searching for accounts that use the same (hash)tags; 2) \emph{Boost} relies on reposts of the same post; and 3) \emph{Bully} requires collection strategies that obtain (near) complete conversation trees, rooted at the same initial post. 
%Another potential strategy, observed by the authors, is replying with a hashtag, drawing the original post to a hashtag community.
% All of these strategies would become more effective through the exploitation of automated accounts.

\subsection{Problem Statement} \label{sec:prob_stmt}

A clarification of our challenge at this point is:
\begin{quote}
    \emph{To identify groups of accounts whose behaviour, though typical in nature, is anomalous in degree.}
\end{quote}
There are two elements to this. The first is \emph{discovery}. How can we identify not just behaviour that appears more than coincidental, but also the accounts responsible for it? That is the topic of the next section.
The second element is \emph{validation}. Once we identify a group of accounts via our method, what guarantee do we have that the group is a real, coordinating set of users? This is especially difficult given inauthentic behaviour is hard for humans to judge by eye \citep{cresci2017}.

\section{Methodology}%{Approach}

The major OSNs share a number of features, primarily in how they permit users to interact. 
By focusing on these commonalities, it is possible to develop approaches that generalise across the OSNs that offer them. % (e.g., Table~\ref{tab:primitives}). 

Traditional social network analysis relies on long-standing relationships between actors \citep{wasserman1994social,borgatti2009network}. On OSNs these are typically friend/follower relations. These are expensive to collect and quickly degrade in meaning if not followed with frequent activity. By focusing on active interactions, it is possible to understand not just who is interacting with whom, but to what degree. This provides a basis for constructing (or inferring) social networks, acknowledging they may be transitory.

LCNs are built from inferred links between accounts. Supporting criteria relying on interactions alone, as observed in the literature \citep{ratkiewicz2011,Keller2019}, include retweeting the same tweet (\emph{co-retweet}), using the same hashtags (\emph{co-hashtag}) or URLs (\emph{co-URL}), or mentioning the same accounts (\emph{co-mention}). To these we add joining the same `conversation' (a tree of \emph{reply} chains with a common root tweet) (\emph{co-conv}). As mentioned earlier, other ways to link accounts rely on similar or identical content \citep[e.g.,][]{LeeCCS2013campext,Keller2019,PachecoFM2020whitehelmets,graham2020asnac} or metadata \citep[][]{Mariconti2017,ferrara2017frelec}.

% % SECTION 3.1
\subsection{The LCN / HCC Pipeline} \label{sec:pipeline}
% Discovery Pipeline}

\begin{figure}[t!]
    \centering
    \includegraphics[width=0.95\columnwidth]{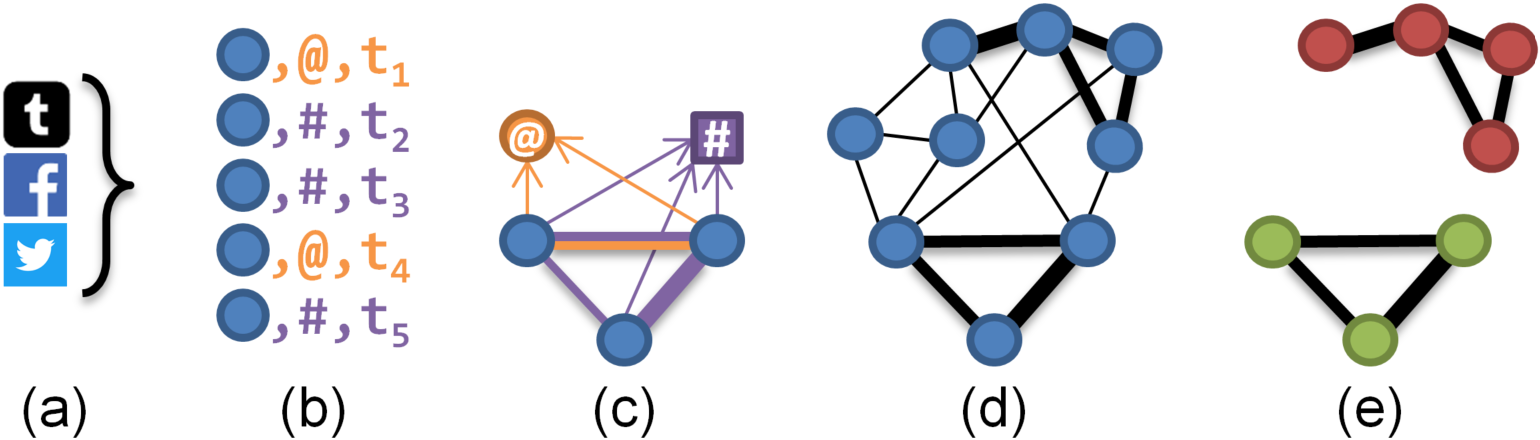}
    \caption{Conceptual LCN construction and HCC discovery process.}
    \label{fig:graph_construction}
\end{figure}

The key steps to %constructing LCNs and extracting 
extract HCCs from raw social media data are shown in Figure~\ref{fig:graph_construction} and documented in Algorithm~\ref{alg:pipeline}. The example in Figure~\ref{fig:graph_construction} is explained after the algorithm has been explained, in Section~\ref{sec:explaining_the_example}.

% \marginpar{Algo 1 removed}
% \marginpar{I could remove Algo 1, but then it's v. similar to Pacheco et al, 2020, arxiv}
\begin{algorithm}[t!]
\caption{FindHCCs}
\label{alg:pipeline}%find_hccs_in_sm}
\textbf{Input}: $P$: Social media posts, $C$: Coordination criteria, $\theta$: Extraction parameter \\ %$\theta$: HCC threshold \\
\textbf{Output}: $H$: A list of HCCs
\begin{algorithmic}[1] %[1] enables line numbers
    \STATE $I_{all} \leftarrow \text{ParseInteractionsFrom}(P)$  \label{alg:step1_convert}
    % \STATE $G^I \leftarrow \text{ConstructInteractionGraph}(interactions, C)$  \label{alg:step2_int_net}
    \STATE $I_C \leftarrow \text{FilterInteractions}(I_{all}, C)$  \label{alg:step2_int_net}
    \STATE $M \leftarrow \text{FindCoordination}(I_C, C)$  \label{alg:step3_coord_evidence}
    \STATE $L \leftarrow \text{ConstructLCN}(M)$ \label{alg:step4_lcn}
    \STATE $H \leftarrow \text{ExtractHCCs}(L, \theta)$ \label{alg:step5_ext_hccs}
    % \STATE \textbf{return} $hcc\_list$
\end{algorithmic}
\end{algorithm}

\paragraph{\textbf{Step 1.}} %\ref{alg:step1_convert}.} 
Convert social media posts $P$ %from their raw format 
to common interaction primitives, $I_{all}$. % (e.g., Table~\ref{tab:primitives}). %By using parsers specific to different social media data, t
This step removes extraneous data and provides an opportunity for the fusion of sources by standardising all interactions (thus including only the elements required for the coordination being sought). %, though the challenge of matching entities remains. 

\paragraph{\textbf{Step 2.}} %\ref{alg:step2_int_net}.} 
{\sloppy From $I_{all}$, filter the interactions, $I_C$, relevant to the set $C$=$\{c_1, c_2, ..., c_q\}$ of criteria (e.g., co-mentions and co-hashtags).} 
%%%% Illustrated in Figure~\ref{fig:graph_construction}b are the filtered mentions (in orange) and hashtag uses (in purple), ordered according to timestamp. %%%%{\color{blue}Each filtered record consists of the post author's account, the account mentioned or hashtag used, and the timestamp of the post.}
%For example, Figure~\ref{fig:graph_construction}b highlights timestamped mentions (in orange) and hashtag uses (in purple). The timestamps offer a natural ordering and can be used to constrain interactions to windows.

% \iffalse
% \color{red}
% An optional part of this step, which may apply when considering hashtags or mentioned accounts, is to remove interactions with any hashtags or account names used in the dataset's collection criteria (filter terms). These entities will be highly connected and may obscure the network around them. Removing these interactions will improve processing performance (there being fewer to process) but may impact the effectiveness of searching for particular coordination strategies (e.g., \emph{Pollution} or \emph{Boost}).
% \color{black}
% \fi

\paragraph{\textbf{Step 3.}} %\ref{alg:step3_coord_evidence}.} 
Infer links between accounts %based on the criteria in
given 
$C$, ensuring links are typed by criterion. % and weighted based on the relevant evidence found. %$C$ is a set of coordination criteria $C = \{c_1, c_2, \ldots, c_q\}$ of size $q$. 
The result, %ing matches, 
$M$, is a collection of inferred pairings. The count of inferred links between accounts $u$ and $v$ due to criterion $c \in C$ is $\beta^c_{\{u,v\}}$. 
%%%% Figure~\ref{fig:graph_construction}c shows inferred links between accounts 
% (blue circles) because they used the same hashtag (the purple node and edges) and mentioned the same account (the orange node and edges).
%%%% due to common interactions. %%%% {\color{blue}Three accounts co-use a hashtag while only two of them co-mention another account.}

\paragraph{\textbf{Step 4.}} \ref{alg:step4_lcn} 
Construct an LCN, $L$, from the pairings in $M$. 
This network $L$=$(V,E)$ is a set of vertices $V$ representing accounts connected by undirected weighted edges $E$ of inferred links. 
These edges represent evidence of different criteria linking the adjacent vertices. 
%We refer to 
The weight of each edge $e \in E$ %$e_{\{u,v\}} \in E$ 
between vertices representing accounts $u$ and $v$ 
for each criterion $c$ is $w^c(e)$, %$w^c_{\{u,v\}}$, 
and is equal to $\beta^c_{\{u,v\}}$.%\marginpar{Could be $w^c_e$ and $\beta^c_e$ instead?} %Thus,

% \begin{equation} \label{eq:lcn_uv_c_weight}
%     w^c_{\{u,v\}} = \beta^c_{\{u,v\}}.
% \end{equation}

% and the set of edge weights between vertices $u,v \in V$ is\marginpar{Eq 2 neeeded?}

% \begin{equation}
%     w_{\{u,v\}} = \{ w^1_{\{u,v\}}, w^2_{\{u,v\}}, \ldots w^q_{\{u,v\}} \}.
% \end{equation}

Most community detection algorithms will require the multi-edges be collapsed to single edges. 
The edge weights are incomparable (e.g., retweeting the same tweet is not equivalent to using the same hashtag), however, for practical purposes, the inferred links can be collapsed %to a single edge 
and the weights combined for cluster detection using a simple summation, e.g., Equation~(\ref{eq:lcn_uv_combined_weight}), or a more complex process like varied criteria weighting. %Thus, a final combined weight $w_{\{u,v\}}$ could be simply

\vspace{-0.5em}
\begin{equation} \label{eq:lcn_uv_combined_weight}
    % w_{\{u,v\}} = \sum_{c=1}^{q} w^c_{\{u,v\}}
    w(e) = \sum_{c=1}^{q} w^c(e)
\end{equation}
\vspace{-0.5em}

% Figure~\ref{fig:graph_construction}d shows an LCN with combined edges.

Some criteria may result in highly connected LCNs, even if its members never interact directly. % in $I_C$.
Not all types of coordination will be meaningful -- people will co-use the same hashtag repeatedly if that hashtag defines the topic of the discussion (e.g., \hashtag{auspol} for Australian politics), in which case it is those accounts who co-use it significantly more often than others which are of interest. 
If required, the final step filters out these coincidental connections. %Consider searching for coordinated dissemination in Facebook by associating all accounts that share the same post: the LCN will contain an edge for every pair of accounts sharing the post, which will result in a very dense LCN if the post is popular.% (Figure~\ref{fig:act_to_coord}).

% \begin{figure}[t!]
%     \centering
%     \includegraphics[scale=0.15]{resources/activity_to_coordination_v3.png}
%     \caption{Each pair of blue vertices that interact with the red vertex (a) are added to and connected in the LCN (b) resulting in $n(n+1)/2$ edges for $n$ blue vertices.}
%     \label{fig:act_to_coord}
% \end{figure}

\paragraph{\textbf{Step 5.}} %\ref{alg:step5_ext_hccs}.} 
Identify the highest coordinating communities, $H$, in $L$ (Figure~\ref{fig:graph_construction}e), using a suitable community detection algorithm, such as \citet{blondel2008}'s Louvain algorithm \citep[used by][]{morstatter2018alt,nasim2018}, \emph{k nearest neighbour} ($kNN$) \citep[used by][]{CaoCLGC2015urlsh}, edge weight thresholding \citep[used by][]{LeeCCS2013campext,Pacheco2020arxiv}, or FSA \citep{Sen2016fsa}, an algorithm from the Social Network Analysis community  that focuses on extracting sets of highly influential nodes from a network. Depending on the size of the dataset under consideration, algorithms suitable for very large networks may need to be considered \citet{Fang2019biggraphcd}. Some  algorithms may not require the LCN's multi-edges to be merged \cite[e.g.,][]{Bacco2017multigraphcd}. 
%\color{blue} A variety of community detection methods can be used at this point, including simply using connected components, the Louvain method~\citep{blondel2008}, edge weight filtering, \textit{k nearest neighbour} or Focal Structures Analysis (FSA)~\citep{Sen2016fsa}. 
We present a variant of FSA~\citep{Sen2016fsa}, FSA\_V (Algorithm~\ref{alg:extract_hccs}), because it is designed to take advantage of FSA's benefits, while addressing some of its costs. FSA does not just divide a network into communities (so that every node belongs to a community), but extracts only sets of nodes that form influential communities within the overall network. FSA\_V reduces the computational complexity introduced by FSA, which recursively applies Louvain to divide the network into smaller components and then, under certain circumstances, stitches them back together. The reason for this is to make it more suitable for application to a streaming scenario, in which execution speed is a priority.

Similar to FSA, FSA\_V initially divides $L$ into communities using the Louvain algorithm but then builds candidate HCCs within each, starting with the `heaviest' (i.e., highest weight) edge (representing the most evidence of coordination). It then attaches the next heaviest edge until the candidate's mean edge weight (MEW) is no less than $\theta$ ($0 < \theta \leq 1$) of the previous candidate's MEW, or is less than $L$'s overall MEW. In testing, edge weights appeared to follow a power law, so $\theta$ was introduced to identify the point at which the edge weight drops significantly; $\theta$ requires tuning. A final filter ensures no HCC with a MEW less than $L$'s is returned. Unlike in FSA~\citep{Sen2016fsa}, recursion is not used, nor stitching of candidates, resulting in a simpler algorithm.

\begin{algorithm}[tb]
\caption{ExtractHCCs (FSA\_V)}
\label{alg:extract_hccs}
\textbf{Input}: $L$=$(V,E)$: An LCN, $\theta$: HCC threshold \\
\textbf{Output}: $H$: Highly coordinating communities 
\begin{algorithmic}[1] %[1] enables line numbers
    \STATE $E' \leftarrow \text{MergeMultiEdges}(E)$
    \STATE $g\_mean \leftarrow \text{MeanWeight}(E')$
    \STATE $louvain\_communities \leftarrow \text{ApplyLouvain}(L)$
    \STATE Create new list, $H$
    \FOR {$l \in louvain\_communities$}
        \STATE Create new community candidate, $h=(V_h,E_h)$
        \STATE Add heaviest edge $e \in l$ to $h$
        \STATE $growing \leftarrow$ \TRUE
        \WHILE{$growing$}
            \STATE Find heaviest edge $\vec{e} \in l$ connected to $h$ not in $h$
            \STATE $old\_mean \leftarrow \text{MeanWeight}(E_h)$
            % \STATE $new\_edges \leftarrow \text{Concatenate}(E_h, \vec{e})$
            % \STATE $new\_mean \leftarrow \text{MeanWeight}(new\_edges)$
            \STATE $new\_mean \leftarrow \text{MeanWeight}(\text{Concatenate}(E_h, \vec{e}))$
            \IF{$new\_mean < g\_mean$ \OR \\ $new\_mean < (old\_mean \times \theta)$}
                \STATE $growing \leftarrow$ \FALSE
            \ELSE
                \STATE Add $\vec{e}$ to $h$
            \ENDIF
        \ENDWHILE
        \IF{MeanWeight$(E_h) > g\_mean$}
            \STATE Add $h$ to $H$
        \ENDIF
    \ENDFOR
    % \STATE \textbf{return} $hcc\_list$
\end{algorithmic}
\end{algorithm}

This algorithm prioritises edge weights while maintaining an awareness of the network topology by examining adjacent edges, something ignored by simple edge weight filtering. 
Our goal is to find sets of strongly coordinating users, so it is appropriate to prioritise strongly tied communities 
%%%as long as tight communities of weak ties are still considered (e.g., $1,000$ accounts paid to retweet one tweet).%~\citep{granovetter1973weakties}.
while still acknowledging coordination can also be achieved with weak ties (e.g., $100$ accounts paid to retweet a single tweet).

The complexity of the entire pipeline is low order polynomial %, O($kn^2$), 
due primarily to the pairwise comparison of accounts to infer links in Step~$3$, which can be constrained by window size when addressing the temporal aspect. For large networks (meaning networks with many accounts), that may be too costly to be of practical use; the solution for this relies on the application domain inasmuch as it either requires a tighter temporal constraint (i.e., a smaller time window) or tighter stream filter criteria, causing a reduction in the number of accounts, potentially along with a reduction in posts. The complexity of the pipeline and HCC extraction is discussed in more detail in Section~\ref{sec:complexity}.%, along with experimental runtime performance, in Section~\ref{sec:performance}.

% \iffalse
% \color{red}
% An optional part of constructing $G^I$ is to remove vertices representing filter terms used to build the dataset (e.g., \texttt{ACCOUNT}s or \texttt{TAG}s), as these may be highly connected and obscure the network around them. Reducing the edges will improve processing performance but may impact the effectiveness of searching for particular types of coordination (e.g., \emph{%channel p
% Pollution} of a \texttt{TAG}-based filter term).
% \color{black}
% \fi

% SECTION 3.1.1
\subsubsection{Addressing the Temporal Aspect} \label{sec:temporal}
% \subsection{Persistent Coordination}

% In searching for many types of coordinated behaviour, not only is the order of expected interactions important, but also the period within which they occur---e.g., to get a hashtag to trend on Twitter, frequent posts with the hashtag must occur in a short period of time~\citep{GrimmeAA2018perspectives,varol2017campaigndetection};
% %the likelihood accounts posting the same news article URL are coordinating their actions is greater if their actions occur minutes apart rather than days~\citep{GrimmeAA2018perspectives}; 
% spikes of online activity are often regarded as anomalous, if not necessarily malign. The timestamps of interactions can be used to address the question of ordering. We address time constraints by segmenting the incoming posts into discrete time windows and using a decaying sliding window approach. 
% %\color{blue}An alternative to this is to process each edge as it arrives in real-time, but this introduces significant computational overheads~\citep{McGregor2014graphstreams}.\color{black} 

Temporal information is a key element of coordination%~\citep{MaloneCrowston1994coordsurvey}
, and thus is critical for effective coordination detection. Frequent posts within a short period may represent genuine discussion or deliberate attempts to game trend algorithms~\citep{GrimmeAA2018perspectives,varol2017campaigndetection,AssenmacherATG2020}. 
We treat the post stream as a series of discrete windows to constrain detection periods%%% {\color{red} and apply a decaying sliding historical window over these to identify persistent coordination}%
. An LCN is constructed from each window (Step~4), and these are then aggregated and mined for HCCs (Step~5). 
As we assume posts arrive in order, their timestamp metadata can be used to sort and assign them to windows.
% As we assume the completeness of the post stream (i.e., posts arrive in order)%\footnote{Some feeds, e.g., forum dumps, may not arrive in stream form.}
% , timestamps can be used for ordering. %%%, and we use discrete time windows to constrain our detections%%% {\color{red} plus a decayed sliding window mechanism to emphasise consistent behaviour}%
%%%.

\iffalse
\color{red}
Given a stream of windows $W$=$\{w_1, w_2, ..., w_{t-1}, w_t\}$, each of size $\gamma$ (minutes, seconds, etc.), where $w_t \in W$ is the current window, a correlating set of LCNs is constructed, $\mathcal{L}$=$\{L_1, L_2, ..., L_{t-1}, L_t\}$.
To address persistent coordination while constructing $L_t$, it is necessary to consider $\{L_1, ..., L_{t-1}\}$ using decay (through weights) to reduce the influence of temporally remote windows.
The decay weights are applied to $e \in E_t$, where $L_t$=$(V_t,E_t)$ by

\begin{equation} \label{eq:decayed_lcn_weight}
    w'_{e \in E_t} = \sum_{n=0}^{t} w_{e \in E_n} \times \alpha^{t-n} 
\end{equation}

where $\alpha$ is the decay factor ($0 < \alpha <= 1$) and $w'_{e \in E_t}$ is the new weight. Rather than consider all $L_i$ back to $i$=$0$, we can consider only the past $T$ windows by start $n$ in (\ref{eq:decayed_lcn_weight}) at $t$-$T$ (or keep it $0$ if $t$-$T$ is negative). If $T$=$1$, only the current window's edge weights are considered, and (\ref{eq:decayed_lcn_weight}) resolves to (\ref{eq:lcn_uv_combined_weight}) (i.e., $w'_{e \in E_t}=w_{e \in E_t}$).
\fi

% SECTION 3.1.2
\subsubsection{A Brief Example}
\label{sec:explaining_the_example}

Figure~\ref{fig:graph_construction} gives an example of searching for co-hashtag coordination across Facebook, Twitter, and Tumblr posts. The posts are converted to interaction primitives in Step~1, shown in Figure~\ref{fig:graph_construction}a. The information required from each post is the identity of the post's author\footnote{Linking identities across social media platforms is beyond the scope of this work, but the interested reader is referred to \citet{Adjali2020} for a recent contribution to the subject.}, the timestamp of the post for addressing the temporal aspect, and the hashtag mentioned (there may be many, resulting in separate records for each). This is done in Figure~\ref{fig:graph_construction}b, which shows the filtered mentions (in orange) and hashtag uses (in purple), ordered according to timestamp. %Once the extraneous information is removed, fusion is straightforward in such a case.

% {\color{blue}Each filtered record consists of the post author's account, the account mentioned or hashtag used, and the timestamp of the post.}
Step 3 in Figure~\ref{fig:graph_construction}c involves searching for evidence of coordination through searching for our target coordination strategies through pairwise examination of accounts and their interactions. Here, three accounts co-use a hashtag while only two of them co-mention another account.

By Step~4 in Figure~\ref{fig:graph_construction}d, the entire LCN has been constructed, and then Figure~\ref{fig:graph_construction}e shows its most highly coordinating communities. 

As mentioned above, to account for the temporal aspect, the LCNs produced for each time window in Figure~\ref{fig:graph_construction}d can be aggregated and then mined for HCCs, or HCCs could be extracted from each window's LCN and then they can be aggregated, or analysed in near real-time, as dictated by the application domain.

\subsection{Validation Methods}

As mentioned in Section~\ref{sec:prob_stmt}, the second element of addressing our research challenge is that of validation. Once HCCs have been discovered, it is necessary to confirm that what has been found are examples of genuine coordinating groups. This step is required before the further question of whether the coordination is authentic (e.g., grassroots activism) or inauthentic (e.g., astroturfing).

\subsubsection{Datasets}

In addition to relevant datasets, we make use of a ground truth (GT), in which we expect to find coordination \citep[\emph{cf.},][]{KellerICWSM2017,Vargas2020}. By comparing the evidence of coordination (i.e., HCCs) we find within the ground truth with the coordination we find in the other datasets, we can develop confidence that: a) our method finds coordination where we expect to find it (in the ground truth); and b) our method also finds coordination of a similar type where it was not certain to exist. 
Furthermore, to represent the broader population (which is not expected to exhibit coordination), similar to \citet{CaoCLGC2015urlsh}, we create a randomised HCC network from the non-HCC accounts in a given dataset, and then compare its HCCs with the HCCs that had been discovered by our method.

\subsubsection{Membership Comparison Across Variables}
% examining networks: nodes, edges, HCC counts
% comparison between hcc results based on which nodes are captured (they're the important bits)
% jaccard and overlap calculations

Our primary factors include the HCC extraction method (using FSA\_V, $kNN$, or thresholds), the window size, $\gamma$, and the strategy being targeted (\emph{Boost}, \emph{Pollution} or \emph{Bully}). Given our interest prioritises the sets of accounts over how they are connected, for each pair of variations we compare the membership of the HCCs discovered, as a whole, and the number of HCCs discovered, as well as the edge count. 
%Given the high number of possible combinations of these factors, a demonstrative selection of variable combinations are presented.
Also included are the exact numbers of HCC members common to each pair. These figures provide further context for the degree of overlap between the HCC members identified under different conditions (i.e., factor values). 
We use Jaccard and overlap similarity measures~\citep{Verma2020jaccard} to compare the accounts appearing in each (ignoring their groupings) and render them as heatmaps.  
The Jaccard similarity coefficient, $J$, is a value between $0.0$ and $1.0$ which represents the similarity between two sets of items, $X$ and $Y$:

\begin{equation} \label{eq:jaccard}
    J(X, Y) = \frac{|X \cap Y|}{|X \cup Y|} = \frac{|X \cap Y|}{|X|+|Y|-|X \cap Y|}.
\end{equation}

If there is significant imbalance in the sizes of $X$ and $Y$, then their similarity may be low, even if one is a subset of the other. An alternative measure, the Overlap or Szymkiewicz-Simpson coefficient \citep{Verma2020jaccard}, takes this imbalance into account by using the size of the smaller of the two sets as the denominator:

\begin{equation} \label{eq:overlap}
    overlap(X, Y) = \frac{|X \cap Y|}{min(|X|,|Y|)}.
\end{equation}

In a circumstance such as ours, it is unclear whether a longer time window will garner more results after HCC extraction is applied. The Jaccard and overlap coefficients can be used to quickly understand two facts about the sets of accounts identified as HCC members with different values of $\gamma$: 
\begin{itemize}
    \item \emph{Is one set a subset of the other?} If so, the overlap coefficient will reach $1.0$, while the Jaccard coefficient may not, if the two sets differ in size. If they are disjoint, it will be $0.0$, along with the Jaccard coefficient.
    \item \emph{Do the sets differ in size?} If the sets are different sizes, but one is a subset of the other, the overlap coefficient will hide this fact, while the Jaccard coefficient will expose it. If both coefficients have values close to $0.0$, then the sets are clearly different in membership and potentially also in size. If the coefficient values are very close, then the sets are close in size, because the denominators are similar in size, meaning $|X \cup Y| =~ min(|X|,|Y|)$, but this will only occur if they share many members (i.e., $|X \cap Y|$ is high).
\end{itemize}
The exact numbers of common accounts complement the degree of overlap visible in the heat maps to provide the reader a basis to easily compare how different variations compare with each other beyond examining them only as pairs and therefore illuminates the influence each particular factor has overall. For example, by being able to compare the results using each value for $\gamma$, it is possible to see the progression of results as the window size increases (both in raw numbers and with a colour scale in the heatmaps).

\subsubsection{Network Visualisation}

\sloppy A second subjective method of analysis for networks is, of course, to visualise them. We use two visualisation tools, \textit{visone} (https://visone.info) and Gephi (https://gephi.org), both of which make use of force directed layouts, which help to clarify clusters in the network structure. 
In particular, Gephi provides access to the Fruchterman-Reingold (FR) algorithm~\citep{Fruchterman1991}, which is force directed, but constrains the layout within a circle. This enables us to see to what degree individual clusters dominate the overall graph, in a way that standard force-directed layouts do not. Node colour is used to represent cluster membership. Clusters are identified with the Louvain method~\citep{blondel2008}. Node size can be used to represent degree or, for nodes representing accounts, the number of posts an account contributes to a corpus. Edge thickness and colour (darkness) are used to represent the weight of the edges.
% visualisation of hccs
As HCCs are weighted undirected networks and each connected component is an HCC, node colour can be used to represent the number of posts, and edge weight can be represented by thickness and, depending on the density of the network, colour darkness. Node shape can be also employed to represent a variable; for analyses that involve multiple criteria (e.g., co-conv and co-mention), we use shape to represent which combination of criteria an HCC is bound by (e.g., just co-mention or a combination of co-mention and co-conv or just co-conv). 

% Account / reason two-level networks
By extending the HCC account networks with nodes to represent the `reasons' or instances of evidence that link each pair of nodes, e.g., the tweets they retweet in common, or accounts they both mention or reply to, creating a two-level \emph{account-reason} network in doing so, we can investigate how HCCs relate to one another. In this case, the two-level network has two types of nodes (\emph{accounts} and \emph{reasons}) and two types of edges (`coordinates with' links between accounts and `caused by'\footnote{Or other appropriate phrasing to indicate that the account is linked to the reason, by which it is indirectly linked to other accounts. \textbf{N.B.} The reason itself may be an account, e.g., in the context of co-mentions, but for the purpose of the two-level network, the mentioned account is treated as a reason.} links between `reasons' and accounts). Visualising the two-level network by colouring nodes by their HCC and using a force-directed layout highlights how closely associated HCCs are with each other, not only revealing what reasons draw many HCCs together (i.e., many HCCs may be bound by a single reason, or an HCC may be entirely isolated from others in the broader community), but also how many reasons may bind them (i.e., many reasons may bind an HCC together). 
Deeper insights can be revealed from this point using multi-layer network analyses.

\subsubsection{Consistency of Content} \label{sec:method_int_messaging}
% internal consistency / text comparison
% use of a random hcc dataset
To help answer RQ2, it is helpful to look beyond network structures and consider how consistent the content within an HCC\footnote{I.e., the content produced by the HCC's members.} is relative to other HCCs and the population in general. This will be most applicable when the type of strategy the HCC is suspected to have engaged in requires repetition, e.g., co-retweeting or copypasta.
If HCCs are boosting a message, it is reasonable to assume the content of HCCs members will be more similar
internally that when compared externally, to the content of non-members. 
To analyse this internal consistency of content, we treat each HCC member's tweets as a single document and create a doc-term matrix using $5$~character n-grams for terms. Comparing the members' document vectors using cosine similarity in a pairwise fashion creates a $n \cdot n$ matrix where $n$ is the number of accounts in the HCC network. This approach was chosen for its performance with non-English corpora~\citep{Damashek1995}, and because using individual tweets as documents produced too sparse a matrix in a number of tests we conducted. 
The pairwise account similarity matrix can be visualised, using a range of colours to represent similarity. By ordering the accounts on both the $x$ and $y$ axes to ensure they are grouped within their HCCs, if our hypothesis is right that similarity within HCCs is higher than outside, then we should observe clear bright squares extending from the diagonal of the matrix. The diagonal itself will be the brightest because it represents each account's similarity with itself. 

If HCCs contribute few posts, which are similar or identical to other HCCs, then bright squares may appear off the diagonal, and this would be evidence similar to clusters of account nodes around a small number of reason nodes in the two-level account-reason networks mentioned in the previous section.

This method offers no indication of how active each HCC or HCC member is, so displays of high similarity may imply low levels of activity as well as high content similarity, just because of the lower likelihood that highly active accounts are highly similar in content (by contributing more posts, there are simply more opportunities for accounts' content to diverge). The use of the 5-character n-gram approach is designed to offset this because each tweet in common between two accounts will yield a large number of points of similarity, as will the case when the same two tweets are posted in the same order (i.e., two accounts both post tweet $t_1$ and then $t_2$), because the overlap between the tweets will yield at least 4 points of similarity.

\subsubsection{Variation of Content}
%content variation: Cumulative frequency of features (hashtags, urls, mentions, RTed accounts, domains)
Converse to the consistency of content within HCC is the question of content variation, and how does the variation observed in detected HCCs differ from that of RANDOM groupings. 
Highly coordinated behaviour such as co-retweeting involves reusing the same content frequently, resulting in low feature variation (e.g., hashtags, URLs, mentioned accounts), which can be measured as entropy~\citep{CaoCLGC2015urlsh}. A frequency distribution of each HCC's use of each feature type is used to calculate each entropy score. Low feature variation corresponds to low entropy values. As per~\citet{CaoCLGC2015urlsh}, we compare the entropy of features used by detected HCCs to RANDOM ones and visualise their cumulative frequency. Entries for HCCs which did not use a particular feature are omitted, as their scores would inflate the number of groups with $0$ entropy. 

\subsubsection{Hashtag Analysis}
% hcc hashtag use
Hashtags can be used to define discussion groups or chat channels \citep{woolley2016autopower}, so hashtag analysis can be used to study those communities. It is another aspect to content analysis that relies upon social media users declaring the topic of their post through the hashtags they include. At the minimum, we can plot the frequency of the most frequently used hashtags as used by the most active HCCs. In doing so, we can quickly see which hashtags different HCCs make use of, and how they relate by how they overlap. Some hashtags will be unique to HCCs, while others will be used by many. This exposes the nature of HCC behaviour: they may focus their effort on a single hashtag, perhaps to get it trending, or they may use many hashtags together, perhaps to spread their message to different communities.

% hashtag co-occurrence
To further explore how hashtags are used together, we perform \emph{hashtag co-occurrence analysis}, creating networks of hashtags linked when they are mentioned in the same tweet. These hashtag co-occurrence networks are sometimes referred to as \emph{semantic networks} \citep{Radicioni2020semnet}. When visualised it is possible to see themes in the groupings of hashtags, and to gain insights from how the theme clusters are connected (including when they are isolated from one another). To extend this approach, Louvain clustering \citep{blondel2008} can be applied and hashtag nodes can be coloured by cluster, providing a statistical measure to how related the hashtags are.

\subsubsection{Temporal Patterns} \label{sec:method_temp_avg}

Campaign types can exhibit different temporal patterns~\citep{LeeCCS2013campext}, so we use the same temporal averaging technique as \citet{LeeCCS2013campext} (dynamic time warping barycenter averaging) to compare the daily activities of the HCCs in the GT and RANDOM datasets with those in the test datasets. The temporal averaging technique produces a single time series made from the time series of all activities of each member in a set of accounts. Using this technique avoids averaging out of time series that are off-phase from one another, by aligning them before averaging them.

Another aspect of temporal analysis is the comparison of HCCs detected in different time windows, including specifically observing whether any HCCs between time windows share members and what the implications are for the behaviour of those members. This is non-trivial for any moderately large dataset, but examination of the ground truth can provide some insight into the behaviours exhibited by accounts known to be coordinating.

\subsubsection{Focus of Connectivity}

Groups that retweet or mention themselves create direct connections between their members, meaning if one is discovered, it may be trivial to find its collaborators. To be covert, therefore, it would be sensible to have a low \emph{internal retweet} and \emph{mention ratios} (IRR and IMR, respectively). 
Formally, if $RT_{int}$ and $M_{int}$ are the the sets of retweets and mentions of accounts within an HCC, respectively, and $RT_{ext}$ and $M_{ext}$ are the corresponding sets of retweets and mentions of accounts outside the HCC, then, for a single HCC, 
\begin{equation}
  IRR = \frac{|RT_{int}|}{|RT_{int}|+|RT_{ext}|} 
\end{equation}
and 
\begin{equation}
  IMR = \frac{|M_{int}|}{|M_{int}|+|M_{ext}|}.
\end{equation}

\subsubsection{Consistency of Coordination} \label{sec:sliding_window}
% Consistent coordination (alpha)
The method presented Section~\ref{sec:pipeline} highlights HCCs that coordinate their activity at a high level over an entire collection period. Further steps can be taken to determine which HCCs are coordinating their behaviour repeatedly and consistently across adjacent time windows. In this case, for each time window, we consider not just the nodes and edges from the current LCN, but additionally from previous windows, applying a degrading factor the contribution of their edge weights.
To build an LCN from a sliding frame of $T$ time windows, the new LCN includes the union of the nodes and edges of the individual LCNs from the current and previous windows, but to calculate the edge weights, we apply a decay factor, $\alpha$, to the weights of edges appearing in windows before the current one. In this way, we apply a multiplier of $\alpha^x$ to the edge weights, where $x$ is the number of windows into the past: the current window is $0$ windows into the past, so its edges are multiplied by $\alpha^0 = 1$; the immediate previous window is $1$ window back, so its edge multiplier is $\alpha^1$; the one before that uses $\alpha^2$, and so on until the farthest window back uses $\alpha^{T-1}$. Generalising from Step~\ref{alg:step4_lcn}, the weight $w^{c,t}(e)$ %$w^{c,t}_{\{u,v\}}$ 
for an edge $e \in E$ between accounts $u$ and $v$ for criterion $c$ at window $t$ and a sliding window $T$ windows wide is given by

\begin{equation} \label{eq:decayed_lcn_weight}
%   w^{c,t}_{\{u,v\}} = \sum_{x=0}^{T-1} w^{c,(t-x)}_{\{u,v\}} \cdot \alpha^{x}.
  w^{c,t}(e) = \sum_{x=0}^{T-1} w^{c,(t-x)}(e) \cdot \alpha^{x}.
\end{equation}

In this way, to create a baseline in which the sliding frame is only one window wide, one only need choose $T$=$1$, regardless of the value of $\alpha$. As~$\alpha \to 1$, the contributions of previous windows' are given more consideration.

\subsubsection{Machine Learning with One-Class Classifiers} \label{sec:method_classification}
%Classification: using one-class classifiers, F1, accuracy, precision/recall of COORDINATING class (cf. Vargas et al., 2020)
An approach that aids in the management of data with many features is classification through machine learning. This is an approach that has been used extensively in campaign detection, in which tweets are classified, rather than accounts \citep[e.g.,][]{LeeCCS2013campext,Chu2012,Wu2018}.
As our intent is to determine whether HCCs detected in datasets are similar to those detecting in ground truth, it is acceptable to rely on one-class classification (i.e., an HCC detected in a dataset is recognised as COORDINATING/positive, or NON-COORDINATING/unknown). The more common binary classification approach was used by \citet{Vargas2020}, however our approach has two distinguishing features:
\begin{enumerate}
    \item We rely on one-class classification because we have positive examples of what we regard as COORDINATING from the ground truth, and everything else is regarded as unknown, rather than definitely `not coordinating'. 
    % Consider the example of a book recommendation system in a library. 
    A one-class classifier can, for example, suggest a new book from a wide range (such as a library) based on a person's reading history. In such a circumstance, the classifier designer has access to positive examples (books the reader likes or has previously borrowed) but all other instances (books here) are either positive or negative.
    When our one-class classifier recognises HCC accounts as positive instances, it provides confidence that the HCC members are coordinating their behaviour in the same manner as the accounts in the ground truth. 

    \item We rely on features from both the HCCs and the HCC members and use the HCC members as the instances for classification, given it is unclear how many members an HCC may have, and accounts that are members of HCCs may have traits in common that are distinct from `normal' accounts. For this reason the feature vectors that our classifier is trained and tested on will comprise features drawn from the individual accounts and their behaviour as well as the behaviour of the HCC of which they are a member. Feature vectors for members of the same HCC will naturally share the feature values drawn from their grouping.
\end{enumerate}

Regarding the construction of the feature vector, at a group level, we consider not just features from the HCC itself, which is a weighted undirected network of accounts, but of the activity network built from the interactions of the HCC members within the corpus. The activity network is a multi-network (i.e., supports multiple edges between nodes) with nodes and edges of different types. The three node types are account nodes, URL nodes, and hashtag nodes. Edges represent interactions and the following types are modelled: hashtag uses, URL uses, mentions, repost/retweets, quotes (\textit{cf.} comments on a Facebook or Tumblr post), reply, and `in conversation', meaning that one account replied to a post that was transitively connected via replies to an initial post by an account in the corpus. This activity network therefore represents not just the members of the HCC but also their degree of activity. 

\paragraph{\textbf{Classifier algorithms.}}

We use the GT to train three classifiers. Two were well-known classifiers drawn from the \texttt{scikit-learn} Python library \citep{Pedregosa2011scikitlearn}: Support Vector Machine (SVM) and Random Forest (RF). The SVM and RF classifiers were trained with two-fold validation and the RF was configured to use $1,000$ trees (estimators). The third classifier was a bagging PU classifier \citep[BPU,][]{MordeletVert2014baggingPU}\footnote{Thanks to Roy Wright for his implementation: \url{https://github.com/roywright/pu_learning/blob/master/baggingPU.py}} also configured to use a $1,000$ RF classifier internally. 
Contrasting ``unlabeled'' training instances were created from the RANDOM dataset. %sampled from DS1 accounts not appearing in HCCs, randomly grouped as non-coordinating groups of similar sizes to the HCCs. 
Feature vector values were standardised prior to classification and upsampling was applied to create balanced training sets. 

The classifiers predict whether instances provided to them are in the positive or unlabeled classes, which we refer to, to aid readability, as `COORDINATING' and `NON-COORDINATING', respectively.

\paragraph{\textbf{Performance metrics.}}

The performance metrics used include the classifier's accuracy, $F_1$ scores for each class, % (COORDINATING and NON-COORDINATING), 
and the Precision and Recall measures that the $F_1$ scores are based upon. High precision implies the classifier is good at recognising samples correctly, and high Recall implies that a classifier does not miss instances of the class they are trained on in any testing data. For example, a good apple classifier will successfully recognise an apple when it is presented with one, and when presented with a bowl of fruit, it will successfully find all the apples in it. The $F_1$ score combines these two measures:

\begin{equation}
    F_1 = 2 \cdot \frac{Precision \cdot Recall}{Precision + Recall}
\end{equation}

and provides insight into to the balance between the classifiers precision and recall. The accuracy of a classifier is the proportion of instances in a test data set that the classifier labeled correctly. In this way, the accuracy is the most coarse of these measures, because it offers little understanding of whether the classifier is missing instances it should find (false negatives) or labeling non-matching instances incorrectly (false positives). The $F_1$ score begins to address this failing, but direct examination of the Precision and Recall provides the most insight into each classifier's performance.

\subsubsection{Bot Analysis}

Although coordinated behaviour in online campaigns is often conducted without automation \citep{StarbirdAW2019cscw}, automation is still commonly present in campaigns, especially in the form of social bots, which aim to present themselves as typical human users \citep{rise2016,Cresci2020}. For this reason, the technique presented here is a valid tool for exposing teams of cooperating bot and social bot accounts. We use the Botometer \citep{davis2016botornot} service to evaluate selected accounts for bot-like behaviour, which provides a detailed assessment of an account's \emph{bot}ness based on over $1,000$ features in six categories. Each category is assigned a value from $0$, meaning human, to $1$, meaning automated. Although other studies have relied on $0.5$ as a threshold for labelling an account as a bot, there is a a significant overlapping area between humans that act in a very bot-like manner, and bots that are quite human-like, so we adopt the approach of \citet{DebateNightICWSM2018} and regard scores below $0.2$ to be human and scores above $0.6$ to be bots.

\subsection{Complexity Analysis} \label{sec:complexity}

The steps in processing timeline presented in Section~\ref{sec:pipeline} are reliant on two primary factors: the size of the corpus of posts, $P$, being processed, and the size of the set of accounts, $A$, that posted them. Therefore $|A| \leq |P|$ Then the complexity of Step 1 is linear, $O(|P|)$, because it requires processing each post, one-by-one. The set of interactions, $I_{all}$, it produces may be larger than $|P|$, because a post may include many hashtags, mentions, or URLs, but given posts are not infinitely long (even long Facebook or Tumblr posts only include several thousand words), the number of interactions will also be linear, even if it $|I| = k|P|$, for some constant $k$. Step 2 filters these interactions down to only the ones of interest, $I_C$, based on the type of coordinated activity sought, $C$, so $|I_{all}| \geq |I_C|$, and again the complexity of this step is also linear, $O(|I_{all}|)$, as it requires each interaction to be considered. Step 3 seeks to find evidence of coordination between the accounts in the dataset, and so requires examining each filtered interaction and building up data structures to associated each account with their interactions ($O(|I_{all}|)$), then emitting pairs of accounts matching the coordination criteria, producing the set $M$, which requires the pairwise processing of all accounts, and so is $|A|^2$ with a complexity of $O(|A|^2)$. This, however, also depends on the pairwise comparison of each account's interactions, which is likely to be small, practically, but theoretically could be as large as $|I_C|$ if one user is responsible for every interaction in the corpus (but then $|A|$ would be $1$). On balance, as a result, we will regard the processing of each pair of users' interactions as linear with a constant factor $k$ (i.e., $O(k|A|^2)$ = $O(|A|^2)$). In Step 4, producing the LCN, $L$, from the criteria is a matter of considering each match one-by-one, so is again linear (though potentially large, depending on $|M|$). The final step (5) is to extract the HCCs from the LCN, and its performance and complexity very much depend upon the algorithm employed, but significant research has been applied in this field \cite[as considered in, e.g.,][]{Bedru2020}. For FSA\_V, which relies on the Louvain algorithm with complexity $O(|A|\log{}_2 |A|)$ \citep{blondel2008}, it considers edges within each community to build its HCC candidates, so has a complexity of less than $O(|E|)$, where $|E|$ is the number of edges in the LCN, meaning its complexity is linear. FSA\_V's complexity is therefore $O(|A|\log{}_2 |A| + |E|)$.

We regard the computation complexity of the entire pipeline as the highest complexity of its steps, which are:
\begin{enumerate}
    \item Extract interactions from posts: $O(|P|)$
    \item Filter interactions: $O(|I_{all}|)$
    \item Find evidence of coordination: $O(|A|^2)$
    \item Build LCN from the evidence: $O(|M|)$
    \item Extract HCCs from LCN using, e.g., FSA\_V: $O(|A|\log{}_2 |A| + |E|)$
\end{enumerate}
The maximum of these is Step 3, the search for evidence of coordination, $O(|A|^2)$. Though in theoretical terms the method is potentially highly costly, in practical terms, we are bound by the number of accounts in the collection (which is highly dependent on the manner in which the data was collected and the nature of online discussion it to which it pertains) and may be managed by constraining the time window, further reducing the number of posts (and therefore accounts) considered, as long as that suits the type of coordination being sought.

\section{Evaluation}% and Validation} 
\label{sec:4exp}

Our approach was evaluated in two phases: 
\begin{itemize}
    \item The first was conducted as an experiment using the validation methods mentioned above and two datasets known to include coordinated behaviour, as well as a ground truth dataset.
    \item The second phase involved two case studies applying the approach against datasets relating to politically contentious topics expected to include polarised groups.
\end{itemize}

The first stage of the evaluation involved searching for \emph{Boost} by co-retweet and other strategies %in two datasets, 
while varying window sizes ($\gamma$). %%% {\color{red} and the decay factor ($\alpha$)}%
FSA\_V was compared against two other community detection algorithms when applied to the LCNs built in Step 4 (aggregated). %{\color{red} We used a baseline in which only the current window was considered to build the LCN (i.e., no decay was required).} 
%Our approach was evaluated by applying Algorithm~\ref{alg:pipeline} to two datasets using three community detection algorithms at Step~\ref{alg:step5_ext_hccs} (including Algorithm~\ref{alg:extract_hccs}), and varying window sizes ($\gamma$) and decaying sliding window conditions, the configurations of which are detailed in Section~\ref{sec:setup}. 
We then validated the resulting HCCs through a variety of network, content, and temporal analyses and machine learning classification, guided by the research questions posed in Section~\ref{sec:intro}. Discussion of further applications and performance metrics is also presented. %, including how results varied across the community detection algorithms. 
% High level results are presented initially, followed by deeper analysis of a particular combination of parameters.

%The \emph{Boost} strategy was the focus for many of the analyses, but generality to other strategies is demonstrated.

\begin{table}[ht]
    \centering%\scriptsize %\small
    \caption{Experiment dataset statistics}
    \label{tab:dataset_stats}
    \resizebox{\columnwidth}{!}{%
    \begin{tabular}{@{}lrrrrrr@{}}
        \toprule
                    %   & Tweets (T) & \multicolumn{2}{c}{Retweets (RT)} & Accounts & T / Account / Day & RT / Account / Day  \\
                      & Tweets    & \multicolumn{2}{c}{Retweets} & Accounts & Tweets per    & Retweets per  \\
                      &           &          &                   &          & Account / Day & Account / Day \\
                       \cmidrule{2-7}
        DS1            & 115,913   & 63,164   & (54.5\%)          & 20,563   & 0.31          & 0.17          \\
        % (Ground Truth) & 4,193     & 2,505    & (59.7\%)          & 134      & 1.74          & 1.04          \\ 
        (GT)           & 4,193     & 2,505    & (59.7\%)          & 134      & 1.74          & 1.04          \\ 
        DS2            & 1,571,245 & 729,937  & (56.5\%)          & 1,381    & 3.12          & 1.45          \\ 
        \bottomrule
    \end{tabular}
    } % end resizebox/textwidth
\end{table}

% SECTION 4.1
\subsection{The Experiment Datasets}

The two real-world datasets selected (%the statistics of which are shown in 
Table~\ref{tab:dataset_stats}) represent two primary collection techniques: filtering a stream of posts using keywords direct from the OSN (DS1) and collecting the posts of specific accounts (DS2):

\begin{description}%[itemsep=1pt]%[noitemsep]
    \item [{\small \textbf{DS1}}] Tweets relating to a regional Australian election in March 2018, including a ground truth subset (GT); and %, including by known political candidate and party accounts; and
    % \item [{\small DS3}] A corpus of tweets collected during a much larger regional election in March 2019.
    \item [{\small \textbf{DS2}}] A large subset of the Internet Research Agency (IRA) dataset published by Twitter in October
    %2018\footnote{\url{https://about.twitter.com/en_us/values/elections-integrity.html\#data}}; and
    2018\footnote{\url{https://about.twitter.com/en_us/values/elections-integrity.html}}.
\end{description}

The data were collected, held and analysed in accordance with an approved ethics protocol\footnote{Protocol H-2018-045 was approved by the University of Adelaide's human research ethics committee.}.
% ANONYMISED:
% The data were collected, held and analysed in accordance with an approved ethics protocol\footnote{Protocol REDACTED was approved by REDACTED's human research ethics committee.}.

DS1 was collected using %the Real-time Analytics Platform for Interactive Data Mining (
RAPID~\citep{rapid2018} over an 18 day period (the election was on day 15) in March 2018. The filter terms included nine hashtags and $134$ political handles (candidate and party accounts)\footnote{Not included, but available on request, as per the ethics protocol.}. The dataset was expanded by retrieving all replied to, quoted and political account tweets posted during the collection period. The political account tweets formed our %validation 
% \emph{ground truth} (GT). 
ground truth.
% The GT accounts were responsible for 3.6\% of the tweets, but 4\% of the retweets (Table~\ref{tab:dataset_stats}). This high retweet to account ratio is expected, as election candidates and party accounts spruik their policies and campaign slogans.

% \iffalse
% \marginpar{Review this paragraph later}
% \color{red}
% As DS1 covers the general discussion surrounding an election, it was expected to contain peaks of highly popular tweets relevant to the campaign, which would result in highly connected LCN structures. % (\emph{cf.} Figure~\ref{fig:act_to_coord}). 
% It would also contain evidence of coincidental coordination forming LCNs with complex topologies. These incidental structures are unlikely to remain once the HCCs are extracted.
% \fi
% \color{blue}

The IRA tweets cover 2009 to 2018, but DS2 consists of all posted in 2016, the year of the US Presidential election. %, the US Presidential election year. 
Because DS2 consists entirely of IRA accounts~\citep{theagency2015}, %{mueller2018} %``believe[d] are connected to state-based information operations'' 
it was expected to include evidence of cooperation.%~\citep{mueller2018}.

% {\color{red}Real-world datasets were favoured over seeded synthetic ones, as it was felt the latter would not adequately demonstrate our approach.}

% SECTION 4.2
\subsection{Experimental Set Up} \label{sec:setup}

% Three primary variables were considered: community detection method, window size $\gamma$, decay factor $\alpha$. Our baseline used no decaying sliding window, and just considered the current window $t$ to build the $L_t$. When using the sliding window, we considered $5$ windows into the past, i.e., $T$=$5$. Window size $\gamma$ was varied at $\{15,60,360,1440\}$ minutes, while $\alpha$ took values $\{0.5,0.7,0.9\}$. The community detection methods used were FSA\_V with a $\theta$ value of $0.3$, \textit{k Nearest Neighbour} (\textit{kNN}) with $k$ approximately $ln(V)$ (\emph{cf.}~\citep{CaoCLGC2015urlsh}), and a simple threshold applied to the edge weights of $L$, retaining the heaviest $10\%$ edges. 

%%% Our independent variables were \color{blue}window size \color{black} $\gamma \in \{15,60,360,1440\}$ in minutes, {\color{red}$\alpha \in \{0.5,0.7,0.9\}$ 
% and $T \in \{1,5\}$}. 

Window size $\gamma$ was set at $\{15,60,360,1440\}$ (in minutes) and 
the three community detection methods used on the aggregated LCNs were:
\begin{itemize}
    \item FSA\_V ($\theta$=$0.3$);
    \item $kNN$ with $k$=$ln(|V|)$ \citep[\emph{cf.},][]{CaoCLGC2015urlsh}; %and
    \item a simple threshold retaining the heaviest edges by removing those with a normalised value below $0.1$. %$90\%$ of edges.
\end{itemize}
%FSA\_V ($\theta$=$0.3$), \textit{k nearest neighbour} (\textit{kNN}) with $k$=$ln(|V|)$ (\textit{cf.}~\citep{CaoCLGC2015urlsh}), and an edge weight threshold retaining the heaviest $10\%$ of edges. 
Values %of $0.3$ and $0.1$ For FSA\_V's $\theta$ and the threshold minimum, respectively, 
for $\theta$ and the threshold were 
based on experimenting with values in $[0.1,0.9]$, maximising the MEW to HCC size ratio, using the $\gamma$=$\{15,1440\}$ DS1 and DS2 aggregated LCNs. %A value of approximately $ln(|V|)$ was chosen for \textit{kNN} to match~\citep{CaoCLGC2015urlsh}.
% The threshold parameter $\theta$ was fixed at $0.5$. 
%%% The experiment conditions consisted of:
%%% \begin{description}[itemsep=1pt]%[noitemsep]
%%%     \item%[Single time window] 
%%%     [Baseline] $T$=$1$ {\color{red} and no $\alpha$} for all $\gamma$ values and community detection methods; {\color{red}and
%%%     \item[Decaying sliding window] $T$=$5$ for all $\alpha$ and $\gamma$ values and community detection methods}.
%%% \end{description}
%As mentioned, our independent variables were $\gamma$ and $\alpha$, while $T$ was kept constant at $5$, plus a control condition where $T$=$1$. 
Values for $\gamma$ were based on \citet{DBLP:conf/kdd/ZhaoEHRL15} observation that 75\% of retweets occur within six hours of posting. This implies that if attempts were made to boost a tweet, retweeting it in much shorter times would be required for it to stand out from typical traffic.
%Zhao \emph{et al.} observed that 75\% of retweets occur within six hours~\citep{DBLP:conf/kdd/ZhaoEHRL15}, which implies that if retweeting is going to be deliberately coordinated, it is likely to be in a much shorter time frame. 
\citet{varol2017campaigndetection} checked Twitter's trending hashtags every $10$ minutes, so values chosen for $\gamma$ ranged from 15 minutes to a day, growing by a factor of approximately four at each increment. Coordinated retweeting was %therefore likely 
expected to occur in the smaller windows, but then replaced by coincidental co-retweeting as the window size increases. % and the normalisation factor in Equation~\ref{eq:lcn_uv_c_weight} drops (as users have more opportunities to interact with others). 
%%%{\color{red}$\alpha$ values ranged from $0.5$ to $0.9$, %Due to the exponential drop off,  
%%%with the expectation that higher values would better expose consistent coordination behaviour. }
%$T$=$5$ was chosen so one decayed sliding window would equal approximately a single window at the next step up for $\gamma$.
%%%{\color{red}Parameter tuning is specific to datasets, so the values chosen for one dataset may not perform well with another.} %Methods for optimising parameter values for these methods is an element of future work.

% OPTIONAL
%\hl{The experiments were conducted on Windows 10 on a 7th generation Intel Core i7-7820HQ 2.9GHz processor with 32Gb of memory and the Anaconda Python~3.6.8 environment, relying primarily on the NetworkX~2.2 library.}

% SECTION 4.3
\subsection{Experimental Results}

% Key results
% - ground truth, SA elec, IRA, by window size
% - in common
% - ground truth, SA elec, IRA, by alpha
% - compare using different HCC methods
% - compare using text similarity
% ground truth deep dive - show groups and their party affiliations, decorated

% temporal stuff
% cf non-HCC stuff

% - 10s experiment with SA elec & IRA - found TAFE supporters in SA elec & team structures in IRA

% Research questions
% \item How can HCCs be found in an LCN?
% \item How do the discovered communities differ, depending on the method used?
% \item How internally focused are the discovered communities? How does this validate the findings?
% \item How much variation is there in the content posted by the communities?
% \item How consistent is the behaviour? To what extent does a sliding window mechanism emphasise HCCs? %Do the groups vary their membership 

% \item How can HCCs be found in an LCN?

% Our results consist of two distinct aspects: 1) what communities (if any) can be found within our test and ground truth datasets? and 2) how we can be confident we have found meaningful results? 
The research questions introduced in Section~\ref{sec:intro} guide our discussion, but we also present follow-up analyses. %throughout we need to consider whether the results are ``meaningful''.
% We explore these aspects through addressing the research questions introduced in Section~\ref{sec:intro}.

\subsubsection{HCC Detection (RQ1)} \label{sec:hcc_detection}

\paragraph{\textbf{Detecting different strategies.}}
The three detection methods all detected HCCs when searching for \emph{Boost} (co-retweets), \emph{Pollute} (co-hashtags), and \emph{Bully} (co-mentions), details of which are shown in %with $\gamma$ at $15$, $15$, and $60$, respectively 
Table~\ref{tab:strategies_info_1}. 
%Table~\ref{tab:strategies_info_1} shows the application of different HCC detection methods applied to LCNs of GT, DS1, and DS2 generated by searching for \emph{Boost} (co-retweets), \emph{Pollute} (co-hashtags), and \emph{Bully} (co-mentions) with $\gamma$ at $15$, $15$, and $60$, respectively. HCCs are identified by all strategies. 
Notably, \textit{kNN} consistently builds a single large HCC, highlighting the need to filter the network prior to applying it % Cao \emph{et al.}
~\citep[\textit{cf.},][]{CaoCLGC2015urlsh}. The \textit{kNN} HCC is also consistently nearly as large as the original LCN for DS2, %reduces the size of the LCN only minimally in DS2, 
perhaps due to the low number of accounts and the fact that every edge of the retained vertices is retained, regardless of weight. It is not clear, then, that \textit{kNN} is producing meaningful results, even if it can extract a community.

% \usepackage{graphicx}
\begin{table}[t]
    \caption{HCCs by coordination strategy}
    \label{tab:strategies_info_1}
    \resizebox{\columnwidth}{!}{%
    \begin{tabular}{@{}llr|rrr|rrr|rrr@{}}
        \toprule
        \multicolumn{3}{c}{}  & \multicolumn{3}{c}{GT} & \multicolumn{3}{c}{DS1} & \multicolumn{3}{c}{DS2} \\
        & Strategy & $\gamma$ & Nodes & Edges & Comp. & Nodes  & Edges     & Comp. & Nodes & Edges  & Comp. \\
        \midrule
        \multirow{3}{*}{\rotatebox[origin=c]{90}{LCN}} % puts LCN across 3 rows on the left
        & Boost    & 15       & 44    & 112   & 5    & 8,855  & 80,702    & 419  & 855   & 23,022 & 14 \\
        & Pollute  & 15       & 51    & 154   & 2    & 13,831 & 1,281,134 & 73   & 1,203 & 65,949 & 5 \\
        & Bully    & 60       & 70    & 482   & 1    & 16,519 & 1,925,487 & 222  & 1,103 & 37,368 & 5 \\
        \midrule
        \midrule
        \multirow{3}{*}{\rotatebox[origin=c]{90}{FSA\_V}} % puts FSA_V across 3 rows on the left
        & Boost    & 15       & 9     & 6     & 3    & 633   & 753     & 167  & 113   & 758   & 19 \\
        & Pollute  & 15       & 9     & 5     & 4    & 135   & 93      & 50   & 24    & 15    & 9 \\
        & Bully    & 60       & 11    & 7     & 4    & 338   & 280     & 119  & 109   & 1,123 & 16 \\
        \midrule
        \multirow{3}{*}{\rotatebox[origin=c]{90}{\textit{kNN}}} % puts knn across 3 rows on the left
        & Boost    & 15       & 9     & 21    & 1    & 1,041 & 33,621  & 1    & 675   & 22,494 & 1 \\
        & Pollute  & 15       & 11    & 37    & 1    & 724   & 153,424 & 1    & 1,040 & 65,280 & 1 \\
        & Bully    & 60       & 18    & 135   & 1    & 1,713 & 663,413 & 1    & 692   & 35,136 & 1 \\
        \midrule
        \multirow{3}{*}{\rotatebox[origin=c]{90}{{\tiny Threshold}}} % puts t_0.1 across 3 rows on the left
        & Boost    & 15       & 11    & 16    & 3    & 85    & 68     & 31   & 8     & 10    & 2 \\
        & Pollute  & 15       & 24    & 26    & 3    & 44    & 37     & 10   & 6     & 13    & 1 \\
        & Bully    & 60       & 15    & 19    & 3    & 25    & 23     & 8    & 10    & 10    & 3 \\
        \bottomrule
    \end{tabular}%
    }% end resizebox
\end{table}

\paragraph{\textbf{Varying window size $\gamma$.}} 
Different strategies may be executed over different time periods, based on their aims. \emph{Boost}ing a message to game trending algorithms requires the messages to appear close in time, whereas some forms of \emph{Bully}ing exhibit only consistency and low variation (mentioning the same account repeatedly). Polluting a user's timeline on Twitter can also be achieved by frequently joining their conversations over a sustained period. 
%It may be necessary to join the same conversations within a small time frame however (thereby suddenly \emph{polluting} an individual's feed). 

\begin{table}[t]
    \caption{HCCs by window size $\gamma$ (Boost, FSA\_V)}
    \label{tab:hccs_info_1}
    \resizebox{\columnwidth}{!}{%
    % \begin{tabular}{@{}lr|rrr|rrr|rrrr@{}}
    %     \toprule
    %     \multicolumn{2}{c}{} & \multicolumn{3}{c}{Graph Attributes} & \multicolumn{3}{c}{HCC Sizes} & \multicolumn{4}{c}{Nodes in common} \\
    %     & $\gamma$        & Nodes  & Edges  & HCCs & Min. & Median & Max. & $\gamma$=15 & $\gamma$=60 & $\gamma$=360 & $\gamma$=1440 \\
    %     \midrule
    %     \multirow{4}{*}{DS1} % puts DS1 across 4 rows on the left
    %     & 15   & 633 & 753 & 167 & 2 & 3 & 18 & 633 & 218 & 93 & 100           \\
    %     & 60   & 619 & 1,293 & 151 & 2 & 3 & 13 & - & 619 & 208 & 193           \\
    %     & 360  & 503 & 1,119 & 127 & 2 & 3 & 19 & - & - & 503 & 350           \\
    %     & 1440 & 815 & 2,019 & 141 & 2 & 3 & 110 & - & - & - & 815         \\
    %     \midrule
    %     \multirow{4}{*}{DS2} % puts DS2 across 4 rows on the left
    %     & 15   & 113 & 758 & 19 & 2 & 3 & 65 & 113 & 34 & 29 & 25           \\
    %     & 60   & 77 & 394 & 18 & 2 & 3 & 27 & - & 77 & 62 & 54           \\
    %     & 360  & 98 & 775 & 15 & 2 & 3 & 32 & - & - & 98 & 56            \\
    %     & 1440 & 69 & 380 & 15 & 2 & 3 & 27 & - & - & - & 69           \\
    %     \bottomrule
    % \end{tabular}%
    \begin{tabular}{@{}lr|rrr|rrrr|rrrr@{}}
        \toprule
        \multicolumn{2}{c}{} & \multicolumn{3}{c}{Graph Attributes} & \multicolumn{4}{c}{HCC Sizes} & \multicolumn{4}{c}{Nodes in common} \\
        & $\gamma$ & Nodes & Edges & HCCs & Min. & Max. & Mean & Std. Dev. & $\gamma$=15 & $\gamma$=60 & $\gamma$=360 & $\gamma$=1440 \\
        \midrule
        \multirow{4}{*}{\rotatebox[origin=c]{90}{GT}} % puts DS1 across 4 rows on the left
        & 15   &   9 &     6 &   3 & 3 &   3 & 3.00 &  0.00 &   9 &   9 &   8 &   8 \\
        & 60   &  14 &     9 &   5 & 2 &   3 & 2.80 &  0.40 &  -  &  14 &  10 &  12 \\
        & 360  &  13 &     9 &   5 & 2 &   3 & 2.60 &  0.49 &  -  &  -  &  13 &  12 \\
        & 1440 &  17 &    12 &   6 & 2 &   3 & 2.80 &  0.37 &  -  &  -  &  -  &  17 \\
        \midrule
        \multirow{4}{*}{\rotatebox[origin=c]{90}{DS1}} % puts DS1 across 4 rows on the left
        & 15   & 633 &   753 & 167 & 2 &  18 & 3.79 &  2.21 & 633 & 218 &  93 & 100 \\
        & 60   & 619 & 1,293 & 151 & 2 &  13 & 4.10 &  2.30 &  -  & 619 & 208 & 193 \\
        & 360  & 503 & 1,119 & 127 & 2 &  19 & 3.96 &  2.58 &  -  &  -  & 503 & 350 \\
        & 1440 & 815 & 2,019 & 141 & 2 & 110 & 5.78 & 12.60 &  -  &  -  &  -  & 815 \\
        \midrule
        \multirow{4}{*}{\rotatebox[origin=c]{90}{DS2}} % puts DS2 across 4 rows on the left
        & 15   & 113 &   758 &  19 & 2 &  65 & 5.95 & 13.94 & 113 &  34 &  29 &  25 \\
        & 60   &  77 &   394 &  18 & 2 &  27 & 4.28 &  5.64 &  -  &  77 &  62 &  54 \\
        & 360  &  98 &   775 &  15 & 2 &  32 & 6.53 &  9.13 &  -  &  -  &  98 &  56 \\
        & 1440 &  69 &   380 &  15 & 2 &  27 & 4.60 &  6.15 &  -  &  -  &  -  &  69 \\
        \bottomrule
    \end{tabular}%
    }
\end{table}

\begin{figure*}[t!]
    \centering
    \subfloat[GT Jaccard similarity.]{
        \includegraphics[width=0.48\columnwidth]{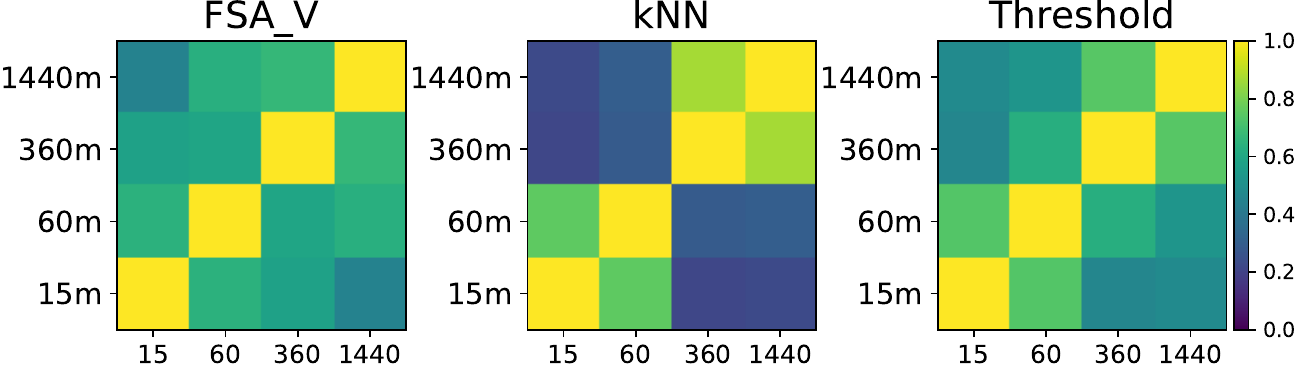}%
        \label{fig:sapol_sim_mtx_gamma_jaccard}
    } \hfill 
    \subfloat[GT overlap similarity.]{
        \includegraphics[width=0.48\columnwidth]{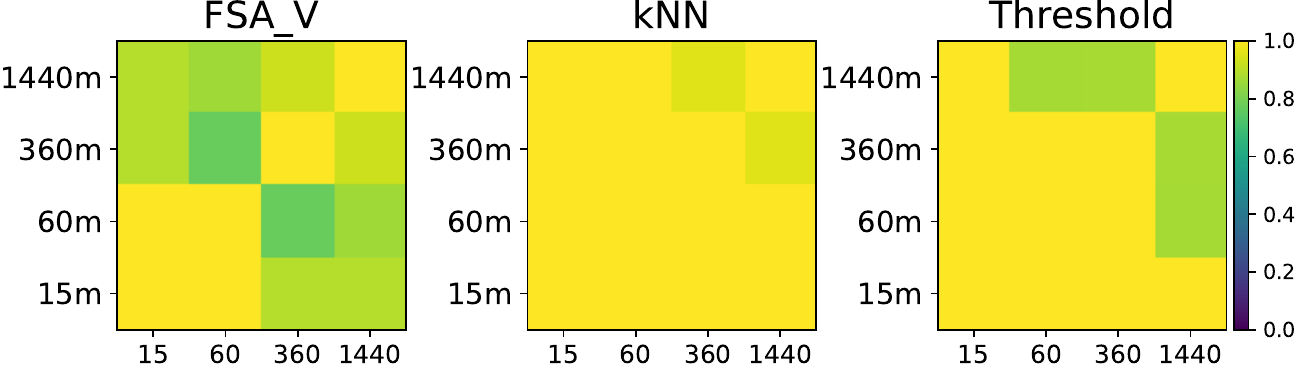}%
        \label{fig:sapol_sim_mtx_gamma_overlap}
    } \\ 
    \subfloat[DS1 Jaccard similarity.]{
        \includegraphics[width=0.48\columnwidth]{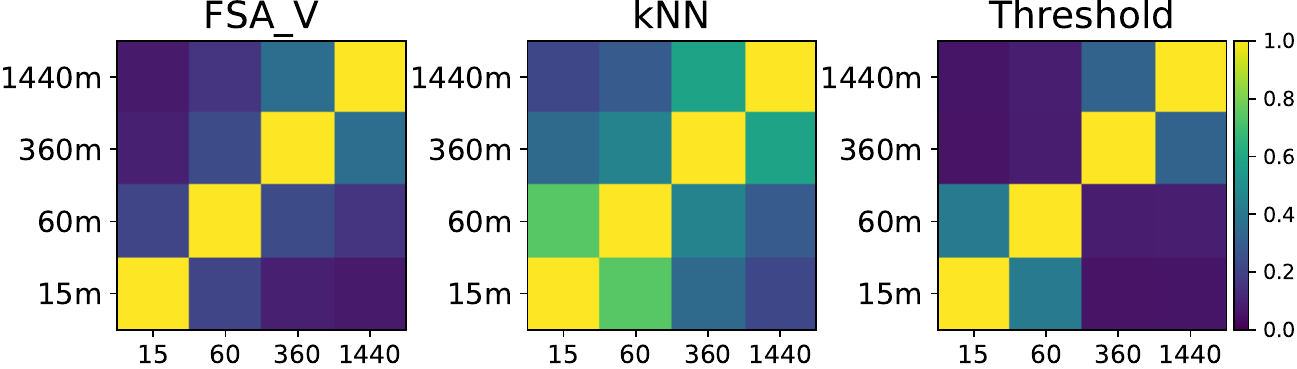}%
        \label{fig:saelec_sim_mtx_gamma_jaccard}
    } \hfill
    \subfloat[DS1 overlap similarity.]{
        \includegraphics[width=0.48\columnwidth]{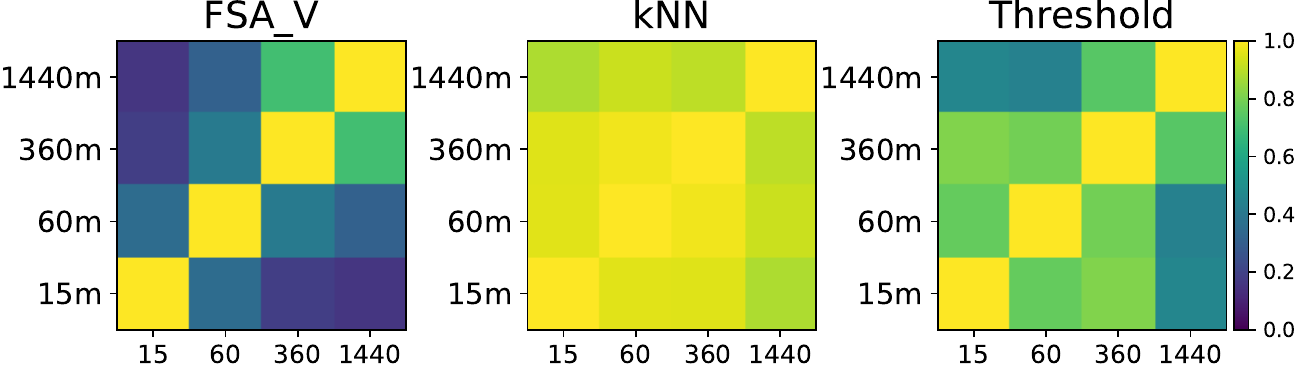}%
        \label{fig:saelec_sim_mtx_gamma_overlap}
    } \\ 
    \subfloat[DS2 Jaccard similarity.]{
        \includegraphics[width=0.48\columnwidth]{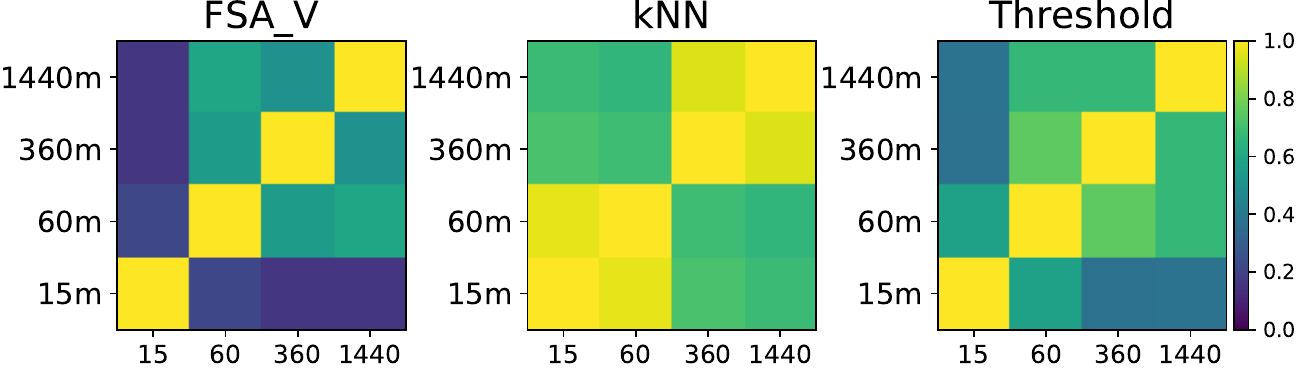}%
        \label{fig:ira_sim_mtx_gamma_jaccard}
    } \hfill
    \subfloat[DS2 overlap similarity.]{
        \includegraphics[width=0.48\columnwidth]{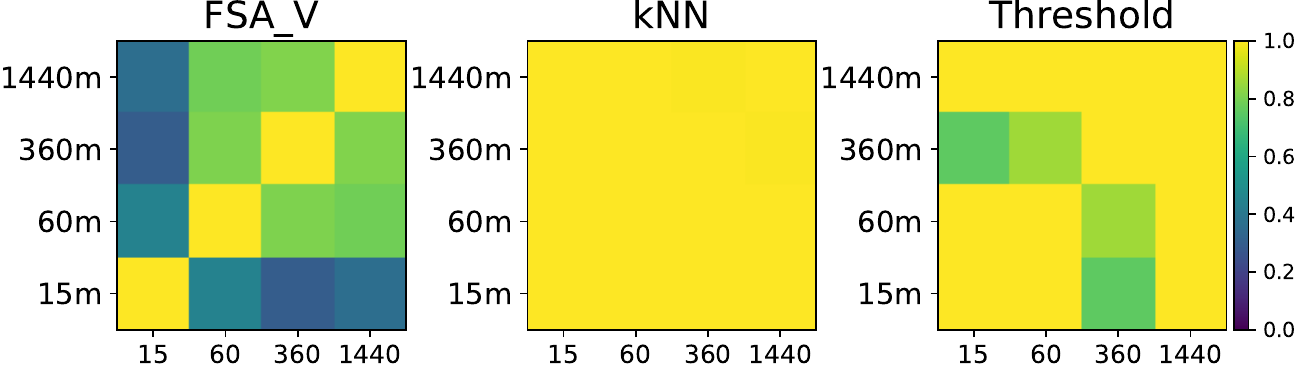}%
        \label{fig:ira_sim_mtx_gamma_overlap}
    }
    \caption{Similarity matrices of HCC account sets found using different window sizes (FSA\_V). The similarity measured here relates to the accounts found not to the similarity in groupings of accounts into HCCs. Yellow implies a high similarity (Jaccard: account sets are identical, Overlap: one set is a subset of the other), while blue implies low similarity (i.e., account sets are disjoint).}
    \label{fig:hcc_sim_mtxs_gamma}
\end{figure*}

Varying $\gamma$ searching for \emph{Boost}, we found different accounts were prominent over different time frames (Table~\ref{tab:hccs_info_1}); the overlap in the accounts detected in each time frame differed considerably even though the number of HCCs stayed relatively similar. 
Figure~\ref{fig:hcc_sim_mtxs_gamma} shows the Jaccard and overlap similarity between the sets of accounts appearing in each window size (agnostic of HCC membership).
Examining these %the Jaccard and overlap heatmaps in Figure~\ref{fig:hcc_sim_mtxs_gamma} 
reveals a number of facts. The overlap results for $kNN$ shows very high levels of similarity, but lower levels of Jaccard similarity. For all datasets, as $\gamma$ grows $kNN$ finds more and more HCC members, always finding the ones it found with smaller window sizes (overlap similarity values appear close to $1.0$, shown as yellow). The highest Jaccard similarities for $kNN$ seem to group the shorter periods ($\gamma = {15,60}$) and the longer periods ($\gamma = {360,1440}$). FSA\_V finds different sets of members in each time window, without significant overlap, though for DS2 it appears that the windows longer than $15$ minutes have some common members amongst themselves, but very few with the ones from $\gamma=15$. As might be expected, thresholding by LCN edge weight results in the identification of additional accounts as $\gamma$ increases, and the Jaccard similarity of GT and DS1 (Figure~\ref{fig:saelec_sim_mtx_gamma_jaccard}) reveals that accounts identified in the shorter time windows ($\gamma = {15,60}$) are very different to those from the longer time windows, but they still overlap somewhat (Figure~\ref{fig:saelec_sim_mtx_gamma_overlap}). This suggests that although there are some accounts that coordinate in short periods, other accounts coordinate \emph{more} over the longer time periods. These include media accounts that are consistently highly active over longer periods than the bursty kind of activity expected of active discussion participants who might log on to Twitter in the evening for a few hours.

%Most HCCs were very small (2-3 members), but typically included one or a few large components.
Other than in GT, which revealed very few HCCs, the sizes of the HCCs found seemed to follow a rough power law; most were very small but one or a few were very large (see the HCC Sizes section in Table~\ref{tab:hccs_info_1}). The number of HCCs did not vary significantly nor consistently as $\gamma$ increased. The number of edges retrieved tells us in DS1, as the window increased, more edges had high enough weights to be retained, whereas DS2 edge counts diminished correspondingly, implying that the LCNs were progressively dominated by a smaller number of very \emph{heavy} edges, while other remained relatively \emph{light}.

%  & Nodes & Edges & Components & Min. & Med. & Max. & G1 & G2 & G3 & G4
% G1 & 633 & 753 & 167 & 2 & 3 & 18 & 633 & 218 & 93 & 100
% G2 & 619 & 1293 & 151 & 2 & 3 & 13 & - & 619 & 208 & 193
% G3 & 503 & 1119 & 127 & 2 & 3 & 19 & - & - & 503 & 350
% G4 & 815 & 2019 & 141 & 2 & 3 & 110 & - & - & - & 815

% G1,saelec-retweets-15m-hccs-fsa_v_0.3
% G2,saelec-retweets-60m-hccs-fsa_v_0.3
% G3,saelec-retweets-360m-hccs-fsa_v_0.3
% G4,saelec-retweets-1440m-hccs-fsa_v_0.3

%  & Nodes & Edges & Components & Min. & Med. & Max. & G1 & G2 & G3 & G4
% G1 & 113 & 758 & 19 & 2 & 3 & 65 & 113 & 34 & 29 & 25
% G2 & 77 & 394 & 18 & 2 & 3 & 27 & - & 77 & 62 & 54
% G3 & 98 & 775 & 15 & 2 & 3 & 32 & - & - & 98 & 56
% G4 & 69 & 380 & 15 & 2 & 3 & 27 & - & - & - & 69

% G1,ira-retweets-15m-hccs-fsa_v_0.3
% G2,ira-retweets-60m-hccs-fsa_v_0.3
% G3,ira-retweets-360m-hccs-fsa_v_0.3
% G4,ira-retweets-1440m-hccs-fsa_v_0.3

\paragraph{\textbf{HCC detection methods.}}
Similarly, HCCs discovered by the three community extraction methods (Table~\ref{tab:hccs_info_2}) exhibit large discrepancies, suggesting that whichever method is used, tuning is required to produce interpretable results. %\color{red}
This is evident in the literature: \citeauthor{CaoCLGC2015urlsh} conducted significant pre-processing when identifying URL sharing campaigns% across two years of Twitter activity
~\citeyear{CaoCLGC2015urlsh}, and 
%Before manually categorising URL domains used in a corpus of 1.6b shared URLs, %Cao \emph{et al.}~
%\citep{CaoCLGC2015urlsh} used \textit{kNN} ($k = \ln(|V|)$) only after removing domains mentioned fewer than $50$ times and users who posted fewer than $50$ times. After applying \textit{kNN} they removed small clusters and broke down large ones (maximising modularity) before inspecting the resulting $2,775$ clusters. Similarly, 
Pacheco \emph{et al.} showed how specific strategies could identify groups in the online narrative surrounding the Syrian White Helmet organisation~\citep{PachecoFM2020whitehelmets}.
% demonstrated their approach by selecting specific strategies and their configuration for their datasets~\citep{Pacheco2020arxiv}. 
Here we present the variation in results while controlling methods and other variables and 
%HCCs discovered while varying the time window and 
keeping the coordination strategy constant, as our focus is the effectiveness of the method%%%{\color{red} and the addition of the decaying sliding window}
. 

% \usepackage{graphicx}
\begin{table}[t]
    \caption{HCCs by detection method (Boost, $\gamma$=$15$) %HCC attributes and nodes in common, by detection method, applied to the co-retweet LCNs ($\gamma$=$15$).
    }
    \label{tab:hccs_info_2}
    \resizebox{\columnwidth}{!}{%
    % \begin{tabular}{@{}ll|rrr|rrr|rrr@{}}
    %     \toprule
    %     \multicolumn{2}{c}{} & \multicolumn{3}{c}{Graph Attributes} & \multicolumn{3}{c}{HCC Sizes} & \multicolumn{3}{c}{Nodes in common} \\
    %     &               & Nodes  & Edges  & HCCs & Min. & Median & Max. & FSA\_V & $kNN$ & Threshold \\
    %     \midrule
    %     \multirow{3}{*}{DS1} % puts DS1 across 3 rows on the left
    %     & FSA\_V    & 633 & 753 & 167 & 2 & 3 & 18 & 633 & 56 & 36        \\
    %     & kNN       & 1,041 & 33,621 & 1 & 1,041 & 1,041 & 1,041 & - & 1,041 & 44         \\
    %     & Threshold & 85 & 68 & 31 & 2 & 2 & 14 & - & - & 85        \\
    %     \midrule
    %     \multirow{3}{*}{DS2} % puts DS2 across 3 rows on the left
    %     & FSA\_V    & 113 & 758 & 19 & 2 & 3 & 65 & 113 & 88 & 4         \\
    %     & kNN       & 675 & 22,494 & 1 & 675 & 675 & 675 & - & 675 & 8         \\
    %     & Threshold & 8 & 10 & 2 & 2 & 4 & 6 & - & - & 8         \\
    %     \bottomrule
    % \end{tabular}%
    \begin{tabular}{@{}ll|rrr|rr|rrr@{}}
        \toprule
        \multicolumn{2}{c}{} & \multicolumn{3}{c}{Graph Attributes} & \multicolumn{2}{c}{HCC Sizes} & \multicolumn{3}{c}{Nodes in common} \\
        &               & Nodes  & Edges  & HCCs & Min. & Max. & FSA\_V & $kNN$ & Threshold \\
        \midrule
        \multirow{3}{*}{\rotatebox[origin=c]{90}{DS1}} % puts DS1 across 3 rows on the left
        & FSA\_V    & 633 & 753 & 167 & 2 & 18 & 633 & 56 & 36        \\
        & kNN       & 1,041 & 33,621 & 1 & 1,041 & 1,041 & - & 1,041 & 44         \\
        & Threshold & 85 & 68 & 31 & 2 & 14 & - & - & 85        \\
        \midrule
        \multirow{3}{*}{\rotatebox[origin=c]{90}{DS2}} % puts DS2 across 3 rows on the left
        & FSA\_V    & 113 & 758 & 19 & 2 & 65 & 113 & 88 & 4         \\
        & kNN       & 675 & 22,494 & 1 & 675 & 675 & - & 675 & 8         \\
        & Threshold & 8 & 10 & 2 & 2 & 6 & - & - & 8         \\
        \bottomrule
    \end{tabular}%
    }
\end{table}

%  & Nodes & Edges & Components & Min. & Med. & Max. & G1 & G2 & G3
% G1 & 633 & 753 & 167 & 2 & 3 & 18 & 633 & 56 & 36
% G2 & 1041 & 33621 & 1 & 1041 & 1041 & 1041 & - & 1041 & 44
% G3 & 85 & 68 & 31 & 2 & 2 & 14 & - & - & 85

% G1,saelec-retweets-15m-hccs-fsa_v_0.3
% G2,saelec-retweets-15m-hccs-knn
% G3,saelec-retweets-15m-hccs-t_0.1

%  & Nodes & Edges & Components & Min. & Med. & Max. & G1 & G2 & G3
% G1 & 113 & 758 & 19 & 2 & 3 & 65 & 113 & 88 & 4
% G2 & 675 & 22494 & 1 & 675 & 675 & 675 & - & 675 & 8
% G3 & 8 & 10 & 2 & 2 & 4 & 6 & - & - & 8

% G1,ira-retweets-15m-hccs-fsa_v_0.3
% G2,ira-retweets-15m-hccs-knn
% G3,ira-retweets-15m-hccs-t_0.1

\begin{figure*}[t!]
    \centering
    \subfloat[DS1 HCCs (FSA\_V).]{
        \includegraphics[width=0.31\columnwidth]{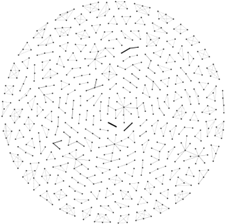}%
        \label{fig:ds1_boost_15m_fsa_v}
    } \hfill 
    \subfloat[DS1 HCCs ($kNN$).]{
        \includegraphics[width=0.31\columnwidth]{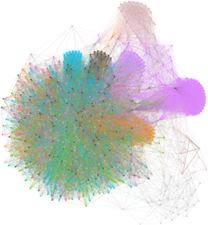}%
        \label{fig:ds1_boost_15m_knn}
    } \hfill 
    \subfloat[DS1 HCCs (Threshold).]{
        \includegraphics[width=0.31\columnwidth]{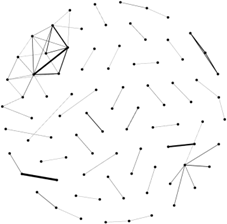}%
        \label{fig:ds1_boost_15m_threshold}
    } \\
    \subfloat[DS2 HCCs (FSA\_V).]{
        \includegraphics[width=0.31\columnwidth]{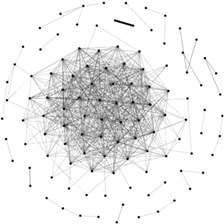}%
        \label{fig:ds2_boost_15m_fsa_v}
    } \hfill 
    \subfloat[DS2 HCCs ($kNN$).]{
        \includegraphics[width=0.31\columnwidth]{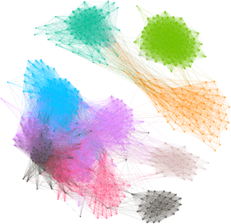}%
        \label{fig:ds2_boost_15m_knn}
    } \hfill
    \subfloat[DS2 HCCs (Threshold).]{
        \includegraphics[width=0.31\columnwidth]{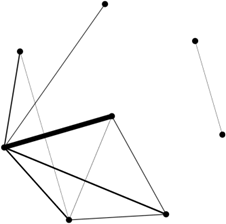}%
        \label{fig:ds2_boost_15m_threshold}
    }
    \caption{HCCs discovered using different methods in DS1 and DS2 (Boost, $\gamma$=$15$). The networks have been laid out with the Fruchterman–Reingold algorithm~\citep{Fruchterman1991} in Gephi (https://gephi.org), and Louvain~\citep{blondel2008} method has been used to colour detected clusters in the $kNN$ network, which each consist of a single connected component.}
    \label{fig:boost_15m_extracted_hccs}
\end{figure*}

The networks were visualised using the FR layout %and laid out with the Fruchterman-Reingold algorithm~\citep{Fruchterman1991} 
in Figure~\ref{fig:boost_15m_extracted_hccs}, revealing further structure within the $kNN$ networks, each of which consisted of a single connected component. 
To examine the structure of the single component more closely, we applied Louvain analysis~\citep{blondel2008} and coloured the largest detected clusters. The clustering reveals distinct communities within both the lone $kNN$ HCC found in each of the datasets. It is possible the DS2 ones are more easily discernible either due to the smaller number of accounts ($675$ compared with $1,041$) or because the accounts were, in fact, organised teams of malicious actors acting over a longer time frame.

% \item How do the discovered communities differ, depending on the method used?

\subsubsection{HCC Differentiation (RQ2)}%Differences among HCCs, based on discovery methods}%How do the discovered communities differ, depending on the method used?}

% Describe content of what's found - GT, DS1, DS2

% method vs
% HCC stats, hashtags, domains, retweet proportions, replies, mentions, URLs?, temporally

%\textit{Distinguishing HCCs.}
\textit{How similar are the discovered HCCs to each other and to the rest of the corpus?}
% notably different to each other, and whether they are different to other groups of accounts in the corpus. 
The HCC detection methods used relied on network information; in contrast we examine content, metadata and temporal information to validate the results. We contrast DS1 and DS2 results with GT %\citep[\emph{cf.},][]{KellerICWSM2017} 
and a RANDOM dataset, %\citep[\emph{cf.},][]{CaoCLGC2015urlsh}, 
constructed %by randomly assigning non-HCC accounts from DS1 to groups matching the distribution of its HCCs
to match the HCC distributions in DS1 
(FSA\_V, $\gamma$=$15$). As DS2 consisted entirely of bad actors, and GT consisted entirely of political accounts, it was felt non-HCC accounts from DS1 would be more representative of non-coordinating `normal' accounts.
%Though the HCC detection methods rely on network connectivity and simple attribute information, comparison can also consider the content and metadata of the posts used to build the networks. Here we examine the content, metadata, and temporal aspects of the discovered HCCs' posts.

% \marginpar{Check the phrasing of this carefully to ensure it's clear.}
%If HCCs are coordinating their behaviour strongly, especially through co-retweeting, the content of their posts ought to be highly internally similar. The more coordination, the stronger this signal should be. This is, in fact, what we find when we plot the similarity matrix of discovered HCC members, shown in Figure~\ref{fig:hcc_sim_mtxs}. To create this visualisation for each corpus, we first create a document term matrix using all the tweets posted by the HCC members, in which each `document' corresponds to an account and `terms' are $5$ character n-grams. The term vectors in the resulting semantic space are the $5$ character n-grams in the combined account's tweets. N-grams of this size were chosen for their performance in language independent text similarity studies~\citep{Damashek1995} and DS2 contains non-English text. Using tweets as documents resulted in too sparse a matrix. The term vectors are then compared with cosine similarity, and used to colour the cells in the matrix visualisation, which are grouped by HCC and ordered by HCC size.

\begin{figure*}[t!]
    \centering
    \subfloat[GT.]{
        \includegraphics[width=0.41\textwidth]{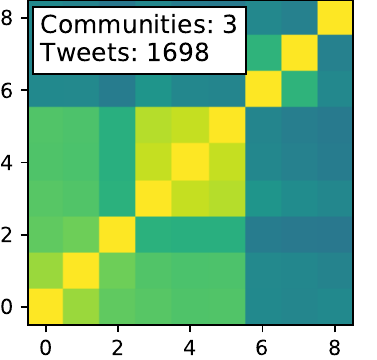}%
        \label{fig:sapol_sim_mtx_15m}
    } %\hfill
    \subfloat[DS1.]{
        \includegraphics[width=0.435\textwidth]{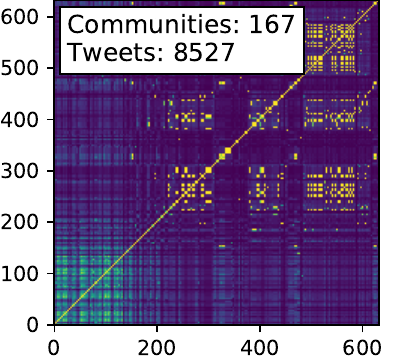}%
        \label{fig:saelec_sim_mtx_15m}
    } \\ %\hfill
    \vspace{-0.5em}
    \subfloat[DS2.]{
        \includegraphics[width=0.42\textwidth]{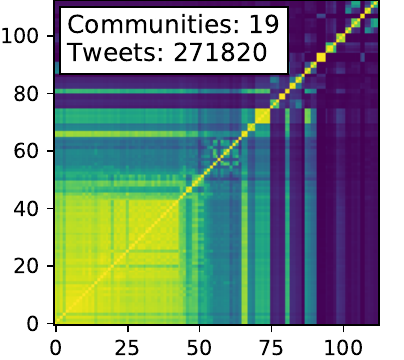}%
        \label{fig:ira_sim_mtx_15m}
    } %\hfill
    \subfloat[RANDOM.]{
        \includegraphics[width=0.465\textwidth]{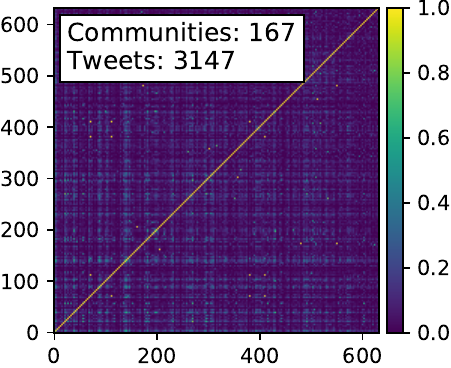}%
        \label{fig:random_sim_mtx_15m}
    }
    \caption{Similarity matrices of content posted by HCC accounts ($\gamma$=15, FSA\_V). Each axis has an entry for each account, grouped by HCC. Each cell represents the similarity between the two corresponding accounts' content, calculated using cosine similarity %as cosine similarity of their content 
    (yellow = high similarity). Each account's content is represented as a vector of $5$ character n-grams of their combined tweets.}
    \label{fig:hcc_sim_mtxs}
\end{figure*}

\paragraph{\textbf{Internal consistency.}} 
% If HCCs are boosting a message, it is reasonable to assume the content of HCCs members will be more similar
% % when they are compared to each other than when compared to non-members.
% internally that when compared externally, to the content of non-members. 
% % internally than to external messaging. 
% Treating each HCC member's tweets as a single document, we created a doc-term matrix using $5$~character n-grams for terms, and then compared the members' document vectors using cosine similarity. This approach was chosen for its performance with non-English corpora~\citep{Damashek1995}, and because using individual tweets as documents produced too sparse a matrix. 
Visualising the similarities between accounts using the method in Section~\ref{sec:method_int_messaging} %, grouping them by HCC 
(Figure~\ref{fig:hcc_sim_mtxs}), the HCCs are discernible as being internally similar. 
%This method ignores the number of tweets HCCs post, so we can draw no conclusions about connections between HCC size and the internal similarity of their content, though more active HCCs (i.e., with more tweets) are more likely to be similar, through co-occurrence of n-grams.
%{\color{blue} This method also may not hold for HCCs detecting using different coordination strategies: there is no reason to assume that co-hashtag or co-mention HCC members will post highly similar content, whereas retweets consist of identical content.} % As there is no consideration of the number of tweets an HCC posts using this method, we cannot say if large or small HCCs are more or less likely to have high internal similarity. 
The RANDOM groupings demonstrated little to no similarity, internal or external, as expected, while the DS2 HCCs demonstrated high internal similarity, as expected of organised accounts over an extended period. 
The internal consistency of the DS1 HCCs is not as clear as for DS2, possibly due to the greater number of HCCs. Where HCCs are highly similar to others (indicated by yellow cells off the diagonal), it is highly likely these are due to small HCCs, e.g., with two or three members, retweeting the same small number of tweets (fewer than ten) as each other. The use of filtering in conjunction with FSA\_V may help remove potentially spurious HCCs, as could a final merge phase, joining HCC candidates whose evidence for coordination matches closely (e.g., two small HCCs retweeting $90\%$ of the same tweets, kept separate by FSA\_V but clearly similar).

%Though only the count of HCCs is offered in each subfigure, the HCCs can be visually/subjectively discerned as the brighter squares. The diagonal is, of course, brightest, as it is the comparison of each term vector against itself. The ground truth's $4$ HCCs can be seen, even based on $2,077$ tweets. Though there are very many ($160$) HCCs in DS1, and all are small (the largest has $5$ members, and $138$ have fewer than that), some can be easily distinguished and there are patches of high similarity amongst the smallest HCCs, likely because they consist almost entirely of retweets ($113$ posted only retweets). Reintroducing FSA's `stitching' process may help join HCCs with highly similar content.

%To ensure that these groups are ``meaningful'', we adopt Cao \emph{et al.}'s approach of creating randomly assigned account groups as a baseline~\citep{CaoCLGC2015urlsh}. The RANDOM groups match the sizes of DS1's HCCs ($\gamma$=$15$, FSA\_V). As expected, their members' content is not similar (Figure~\ref{fig:random_sim_mtx_15m}), even though they tweeted nearly $2,500$ times.  

\begin{figure*}[t!]
    \centering
    \subfloat[GT, DS1 and RANDOM.]{
        \includegraphics[width=0.45\textwidth]{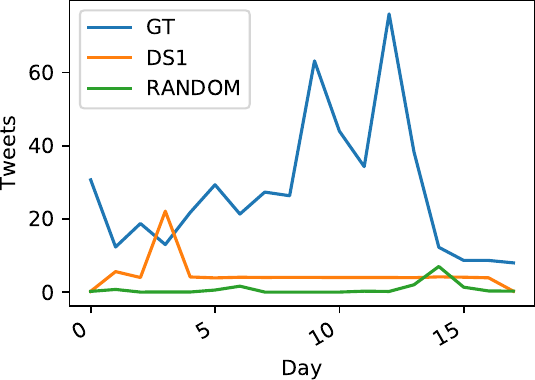}%
        \label{fig:other_hcc_timelines}
    } %\hfill
    \subfloat[DS2.]{
        \includegraphics[width=0.47\textwidth]{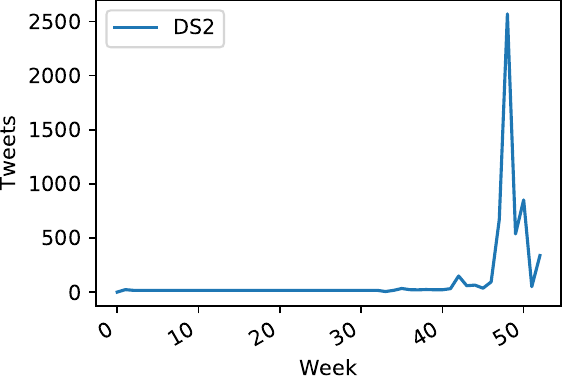}%
        \label{fig:ira_hcc_timeline}
    }
    \caption{Averaged temporal graphs of HCC activities %, each set of activities combined using dynamic time warping barycenter averaging 
    ($\gamma$=$15$, FSA\_V).}
    \label{fig:hcc_timelines}
\end{figure*}

\paragraph{\textbf{Temporal patterns.}} %Campaign types can exhibit different temporal patterns~\citep{LeeCCS2013campext}. We used the same temporal averaging technique as \citet{LeeCCS2013campext} {\color{blue}(dynamic time warping barycenter averaging)} 
We applied the temporal averaging technique described in Section~\ref{sec:method_temp_avg} 
to compare 
the daily activities of the HCCs found in GT, DS1 and RANDOM (all of which occur over the same time period) in Figure~\ref{fig:other_hcc_timelines} and weekly activities in DS2 in Figure~\ref{fig:ira_hcc_timeline}. The GT accounts were clearly most active at two points prior to the election (around day $15$), during the last leaders' debate and just prior to the mandatory electoral advertising blackout. DS1 and RANDOM HCCs were only consistently active at different times: around the day $3$ leaders' debate and on election day, respectively.
% DS1 and RANDOM HCCs were only consistently active on election day. 
Inter-HCC variation may have dragged the mean activity value down, as many small HCCs were inactive each day. Reintroducing FSA's stitching element to FSA\_V may avoid this. 
In DS2, HCC activity increased in the second half of $2016$, culminating in a peak around the election, %peaks around the candidate debate, the election and after, 
inflated by two very active HCCs, both of which used many predominantly benign hashtags over the year. %The most active used many Russian hashtags, while the second used English ones.

\begin{figure*}[ht!]
    \centering
    \subfloat[GT.]{
        \includegraphics[height=0.3\textheight]{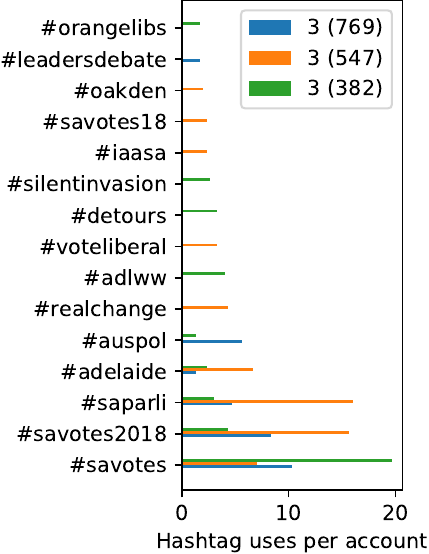}%
        \label{fig:sapol_top_hts_15m}
    } \quad %\hfill
    \subfloat[DS1.]{
        \includegraphics[height=0.3\textheight]{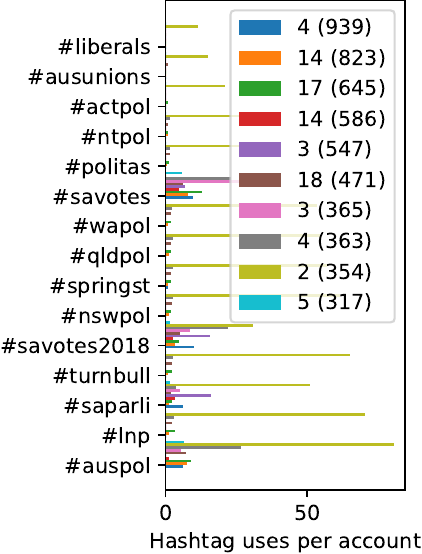}%
        \label{fig:saelec_top_hts_15m}
    } \\ %hfill
    % \vspace{-0.5em}
    \subfloat[DS2.]{
        \includegraphics[height=0.3\textheight]{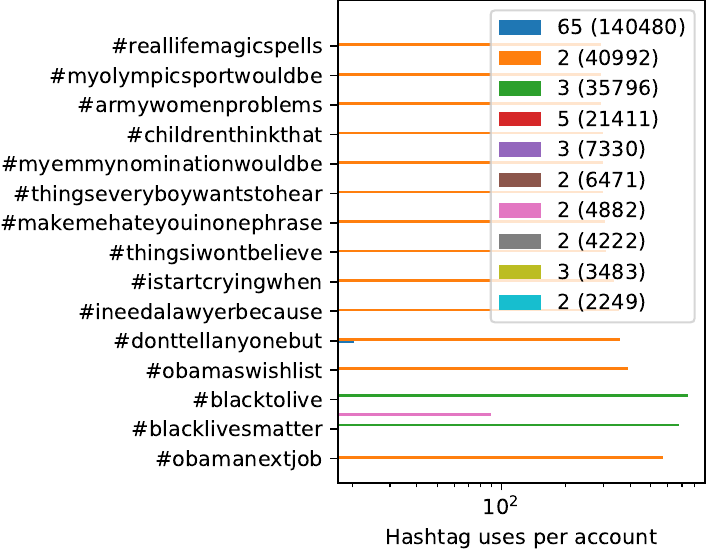}%
        \label{fig:ira_top_hts_15m}
    }
    \caption{Most used hashtags (per account) of the most active HCCs ($\gamma$=15, FSA\_V). The labels indicate HCC identifier and the number of tweets they posted. Not all HCCs used a hashtag often enough to be visible.}
    \label{fig:top_hashtags}
\end{figure*}

\paragraph{\textbf{Hashtag use.}} The most frequent hashtags in the most active HCCs revealed the most in GT (Figure~\ref{fig:sapol_top_hts_15m}). 
% are revealing in all datasets. 
It is possible to assign some HCCs to political parties via the partisan hashtags (e.g., \hashtag{voteliberals} and \hashtag{orangelibs}), %{\small \texttt{\#voteliberals}}), 
although the hashtags of contemporaneous cultural events are also prominent: e.g., \hashtag{silentinvasion}, \hashtag{detours} and \hashtag{adlww} all relate to a contemporaneous international writers' festival. 
DS1 hashtags are all politically relevant, but are dominated by a single small HCC (rendered in pale green) which used many hashtags very often (Figure~\ref{fig:saelec_top_hts_15m}). These accounts clearly attempted to disseminate their tweets through using $1,621$ hashtags in $354$ tweets.  Furthermore, the hashtags they use relate to political discussions in many regions around the country (all listed hashtags that end in `pol' relate to Australian states' or the national political discussion communities). Their prominence in hashtag use effectively hampers our ability to analyse the hashtag use of other HCCs, however, but seeing the results in context is important, as it helps to confirm that the pale green HCC is likely engaging in inauthentic behaviour. We can still see that a large portion of hashtag use amongst the other listed HCCs relates to \hashtag{savotes}, \hashtag{savotes2018}, and \hashtag{saparli}, focussing on the South Australian election. If the hashtags had been irrelevant to the election, that could have provided evidence of accounts attempting to divert the discussion to other topics (because those tweets would still have needed to include the collection filter terms -- i.e., ones relating to the election -- to have been captured in the first place).
Similarly, DS2 hashtags were dominated by a single HCC (using $41,317$ relatively general hashtags in $40,992$ tweets) and one issue-motivated HCC (Figure~\ref{fig:ira_top_hts_15m}).
% . Although more hashtags in DS1 are politically relevant, there was clear influence from elsewhere in Australia, where a by-election was held on the same day (Figure~\ref{fig:saelec_top_hts_15m}). Also, \emph{Boost}ing HCCs discovered with FSA\_V often correspond to a party. The messaging strategies of particular HCCs are clearer in DS2 (Figure~\ref{fig:ira_top_hts_15m}), with {\small \texttt{\#blacklivesmatter}}-related hashtags dominating one HCC (red) and {\small \texttt{\#maga}} and {\small \texttt{\#tcot}} (pro-Republican) hashtags dominating another (purple). 
Given DS2 covers an entire year, it is unsurprising that the largest HCCs use such a variety of hashtags that their hashtags do not appear on the chart (very little evidence of most of the HCCs listed in the legend appear visible in the barchart, even though the $x$ axis is a log scale), but it is revealing that at least a small number of HCCs devoted much of their content to using hashtags, while the other most active HCCs did not, indicating that different HCCs detected by searching for one coordination strategy (co-retweet) are engaging (perhaps even more strongly) in other strategies. Perhaps these hashtag disseminator HCCs acted as general distraction or supporter groups, contributing messages sporadically but not consistently.

% % Unsurprisingly, given the data extends across a whole year, the third most active HCC uses a variety of hashtags, while the most active uses only one hashtag enough to be rendered on the plot, and second none at all. Their hashtag use is sufficiently diverse that they may have acted as general distraction or supporting groups, contributing to messaging where needed, but ensuring their activity is not so focused as to drawn the attention of authorities. 
% They also consisted of many more members, so better community detection may have split them further.

\begin{figure}[t!]
    \centering
    \includegraphics[width=0.99\columnwidth]{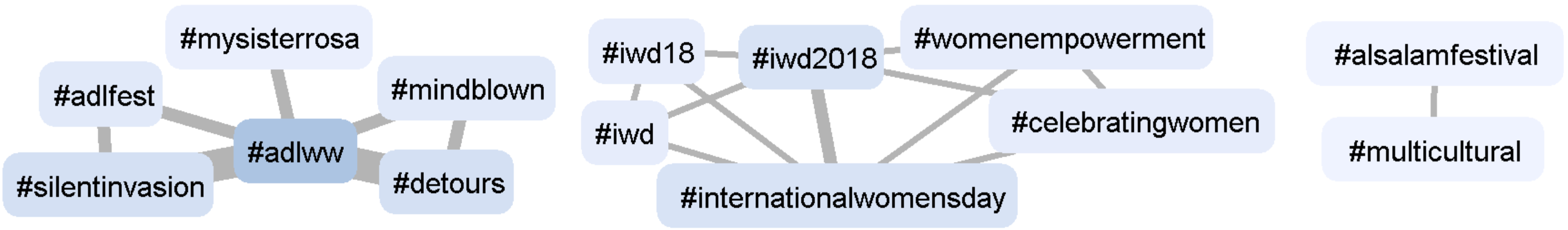}
    \caption{Clusters of hashtags relating to non-election events, including a writers festival, International Women's Day, and a multicultural festival, connected only when they appeared in the same tweet (GT). Wider edges represent a higher tweet count. Node colour implies the frequency of hashtag occurrences (darker means more).}
    \label{fig:sapol_cultural_co-hashtags}
\end{figure}

Analysing hashtag co-occurrences %, by creating a network of hashtags, which are linked when mentioned in the same tweets \citep[sometimes referred to as `semantic networks',][]{Radicioni2020semnet}}, 
can help further explore the HCC discussions to determine if HCCs are truly single groups or merged ones. Applied to GT HCC activities (Figure~\ref{fig:sapol_top_hts_15m}), it was possible to delineate subsets of hashtags in use: e.g., one HCC promoted a political narrative in some tweets with \hashtag{orangelibs} (which is an attack on another political party) and discussed cultural events such as the writers' festival in others with \hashtag{adlww} (Figure~\ref{fig:sapol_cultural_co-hashtags}), but was definitely one group.
%For example, we can delineate sets of hashtags (such as the Adelaide Writer's Week and Greens-related hashtags in Figure~\ref{fig:sapol_top_hts_15m}) and gain insight into the different discussion themes an HCC's members engage in.

\begin{figure}[t!]
    \centering
    \includegraphics[width=0.99\columnwidth]{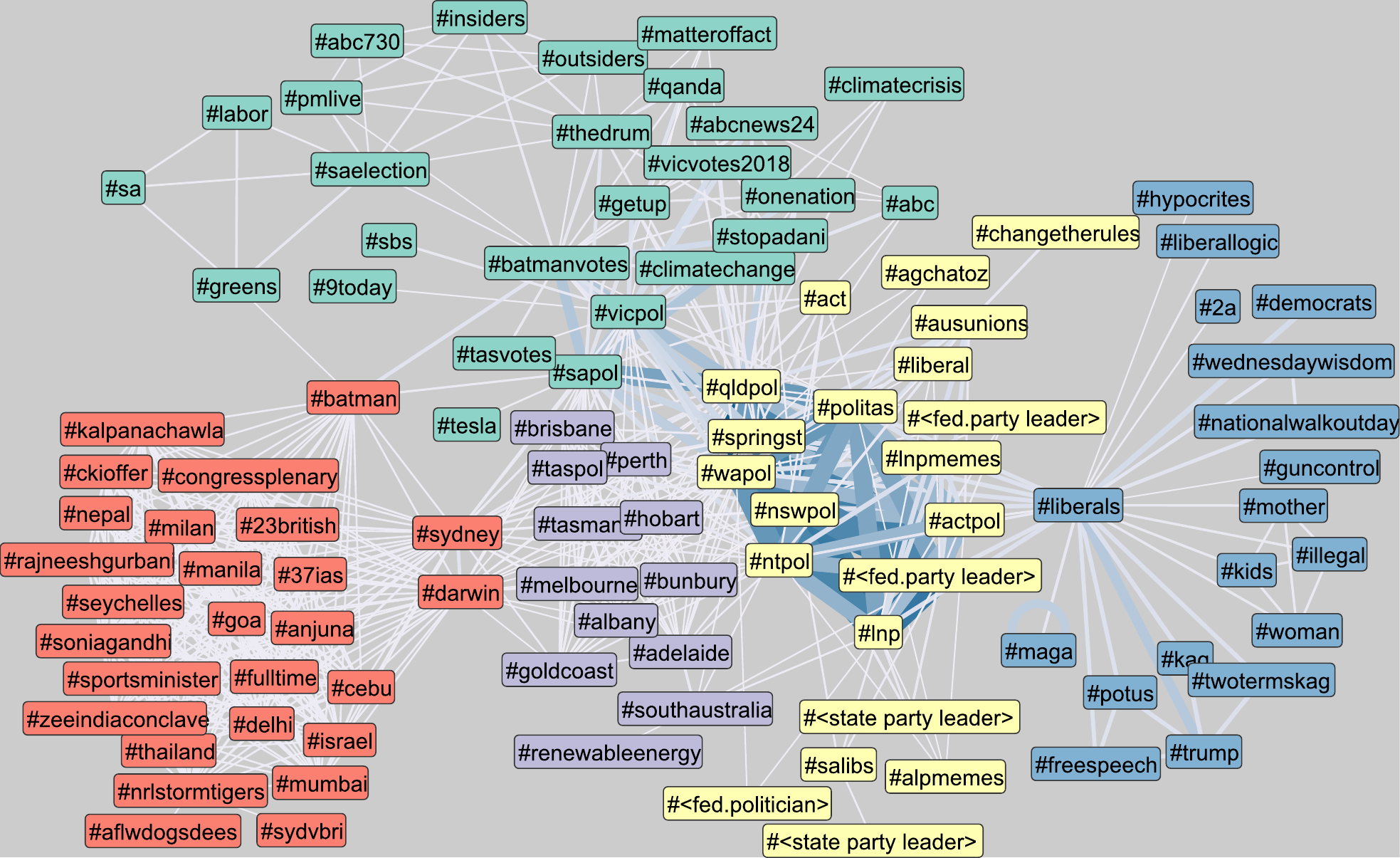}
    \caption{Clusters of hashtags used in DS1, connected only when they appeared in the same tweet. The minimum edge weight is $100$ and the most highly co-occurring hashtags (\hashtag{savotes}, \hashtag{savotes2018}, \hashtag{saparli} and \hashtag{auspol}) have been excluded. Nodes are coloured according to Louvain clustering \citep{blondel2008}, and some hashtags have been anonymised. Wider and darker edges represent a higher tweet count, and a darker background has been provided to improve contrast. Visualised with \emph{visone} (https://visone.info/).}
    \label{fig:saelec_co-hashtags_min100_excl_anon}
\end{figure}

\sloppy Given the great number of hashtags used in even moderate sized datasets such as DS1, using hashtag co-occurrence analysis to examine the broader election discussion in DS1 requires filtering to reveal the underlying core structure of the hashtag co-occurrence network. We limited the minimum frequency of co-occurrences to $100$ and also removed the most frequently occurring hashtags (\hashtag{savotes}, \hashtag{savotes2018}, \hashtag{saparli} and \hashtag{auspol}) to produce Figure~\ref{fig:saelec_co-hashtags_min100_excl_anon}. Application of Louvain cluster detection \citep{blondel2008} exposes five clear clusters, though domain knowledge tells us that there is interesting conflation of topics within some of the clusters. The green cluster contains a subclusters of clusters relating to current affairs television programmes (\hashtag{pmlive}, \hashtag{abc730}, \hashtag{insiders}, \hashtag{outsiders}, \hashtag{qanda} and \hashtag{thedrum}), political party and advocacy groups and relevant issues (\hashtag{onenation}, \hashtag{getup}, \hashtag{climatechange}, \hashtag{climatecrisis}, and \hashtag{stopadani}). It also includes political hashtags (e.g., hashtags ending with \hashtagsanshash{pol} and \hashtag{votes}) that might fit better in the yellow cluster, which is dominated by them and forms the core of the co-hashtag network by including the heaviest edges. The purple cluster consists primarily of location names, apart from \hashtag{renewableenergy} which hangs off \hashtag{southaustralia}, the focus of the election collection. 

The other two clusters make apparent the fact that Twitter is an international network and clashes of hashtags can draw in content irrelevant to local issues. The hashtag \hashtag{liberals} in the blue cluster can refer either to the Liberal party in South Australia (the major party that ultimately won the election) but also is used as a focus in American politics, especially rightwing politics, as reflected by the links to \hashtag{maga}, \hashtag{guncontrol} and \hashtag{2a} (i.e., the 2nd Amendment of the United States' Constitution, which refers to the right to bare arms), as well as \hashtag{nationalwalkoutday}. During the collection period, high school students in the United States staged a national day of protest against gun violence following a mass school shooting\footnote{\url{https://www.nytimes.com/2018/03/14/us/school-walkout.html}}. The red cluster also highlights content from outside the area of interest, with many terms relating to locations in other countries, possibly bound by sports, given the presence of \hashtag{fulltime}, \hashtag{nrlstormtigers}, \hashtag{aflwdogsdees}, and \hashtag{sydvbri}, the last three of which refer to Australian sporting matches between specific teams.

\begin{figure}[t!]
    \centering
    \includegraphics[width=0.99\columnwidth]{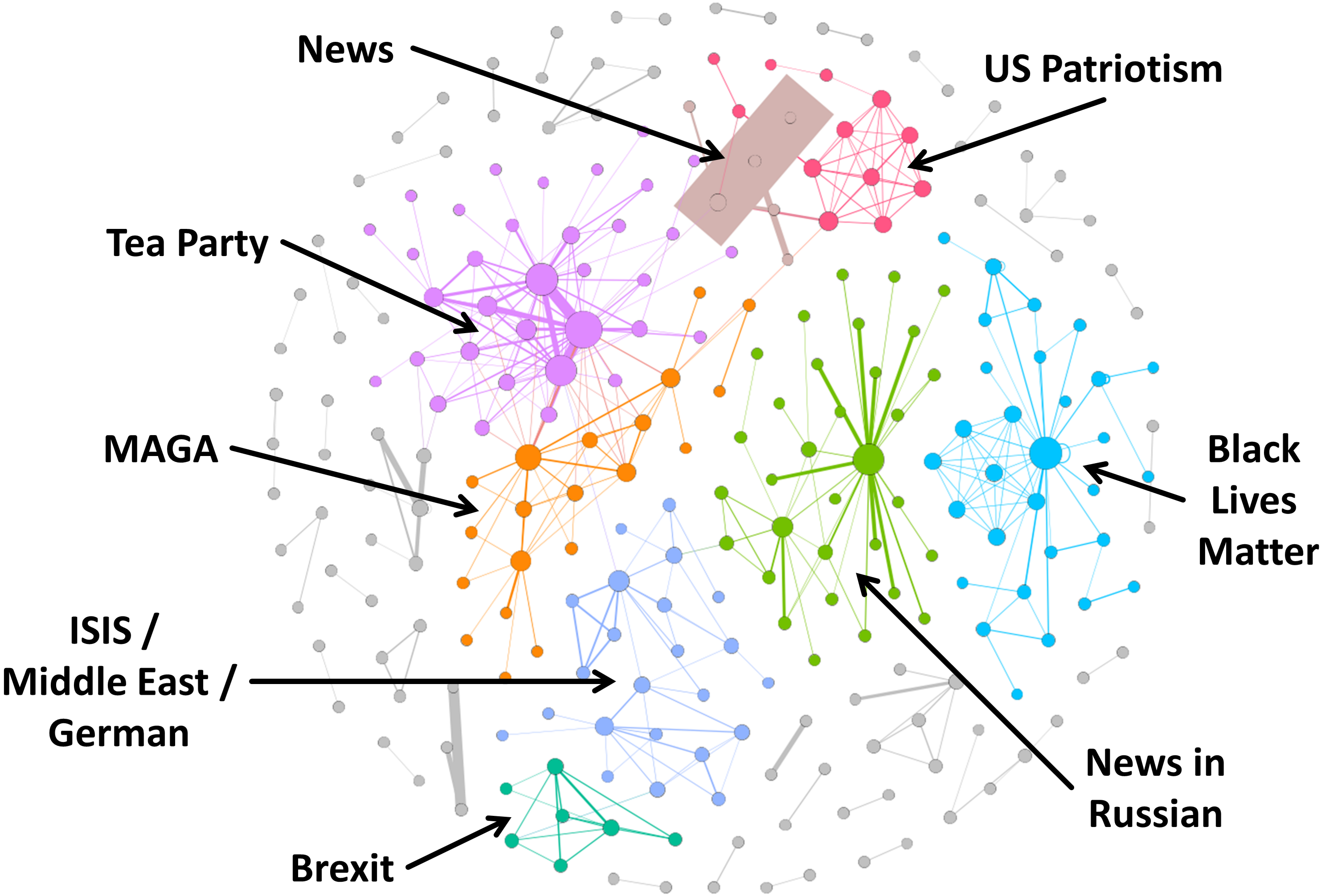}
    \caption{Clusters of hashtags used in DS2, connected only when they appeared in the same tweet. The minimum edge weight is $100$. Nodes are coloured according to Louvain clustering \citep{blondel2008}, the most prominent of which have been annotated with their topic of discussion. Wider and darker edges represent a higher tweet count. Visualised with Gephi (https://gephi.org/).}
    \label{fig:ira_co-hashtags_min100}
\end{figure}

DS2 covers a longer period and seemed to consist of different teams of accounts driving different topics. As a consequence, its co-hashtag network reveals clearly delineated (but often connected) discussion topics, as shown in Figure~\ref{fig:ira_co-hashtags_min100}. It is immediately notable that although the accounts in the dataset were flagged as trolls implicated in attempting to influence the US election, a lot of content is not in English and, in fact, appears to target other countries. Three examples are apparent: 
\begin{itemize}
    \item The green cluster in the centre consists primarily of Russian news-related hashtags, perhaps aimed at a Russian audience to direct their attention to US election-related content.
    \item The pale blue central cluster has many hashtags related to the Middle East, including the ISIS terrorist group, but also German politicians and German names for nearby countries, such as Turkey. Germany's response to refugees from Syria escaping ISIS was politically contentious and may have been seen as an opportunity to foster divisions in the European Union and within Germany.
    \item The green cluster on the lower left is aimed at discussions of the United Kingdom's (UK) exit from the European Union (EU), otherwise referred to as Brexit. The UK held a referendum in 2016 on whether it should leave the EU and the campaigning caused significant division within the UK and European communities.
\end{itemize}
Other significant communities in the co-hashtag network are the pink Tea Party / Conservatives Online (\hashtag{tcot} and \hashtag{ccot}) cluster, tightly connected to the emerging \hashtag{MAGA} cluster supporting Donald Trump, the red cluster focused on American patriotism and the highly active brown cluster including the terms \hashtag{news}, \hashtag{local}, \hashtag{business} and \hashtag{world}. The activity of HCCs shown in Figure~\ref{fig:ira_top_hts_15m} above presents a different and complementary view into hashtag use in the dataset, as very little of it apparent in the co-hashtag network.

% Have we found genuine HCCs? Look at GT and content
\paragraph{\textbf{Examining the Ground Truth.}} 
The importance of having ground truth in context is demonstrated by \citet{KellerICWSM2017,Keller2019}. 
By analysing the actions of known bad actors in a broad dataset, they could identify not just different subteams within the actors and their strategies, but their effect on the broader discussion. Many datasets comprising only bad actors (e.g., DS2) miss this context.

\begin{figure}[t!]
    \centering
    \subfloat[$\gamma$=15.]{
        \includegraphics[width=0.17\columnwidth]{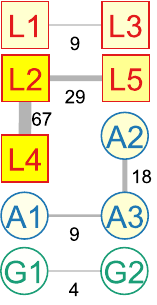}%
        \label{fig:sapol_hccs_15m}
    } \hfill
    \subfloat[$\gamma$=60.]{
        \includegraphics[width=0.25\columnwidth]{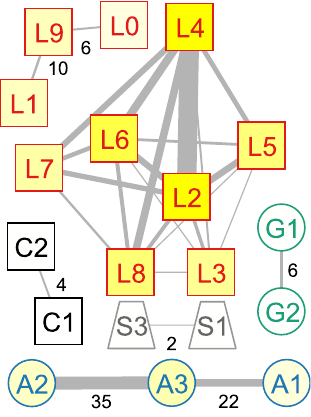}%
        \label{fig:sapol_hccs_60m}
    } \hfill
    \subfloat[$\gamma$=360.]{
        \includegraphics[width=0.19\columnwidth]{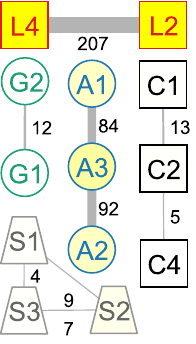}%
        \label{fig:sapol_hccs_360m}
    } \hfill
    \subfloat[$\gamma$=1440.]{
        \includegraphics[width=0.22\columnwidth]{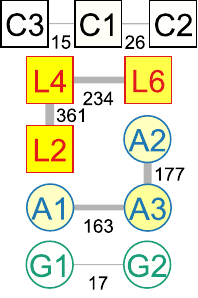}%
        \label{fig:sapol_hccs_1440m}
    }
    \caption{Ground truth HCCs identified with FSA\_V. Vertex shape = ideology (centre, left, right), colour = activity (brighter = higher), label and border colour = political party (L = red, A = blue, G = green, C = black), label = party and account identifier (e.g., `G1' is Greens account \#1), link width = co-retweet count (some omitted for clarity).}
    \label{fig:sapol_hccs}
\end{figure}

Considering GT alone, the HCCs identified consist only of members within the same political party, across all values of $\gamma$. 
Some accounts appeared in each window size.  
HCCs of six major parties were identified. 
% Examination of these HCCs' content confirmed they were genuine {\color{blue} communities using the co-retweet strategy (not necessarily deliberately). 
Figure~\ref{fig:sapol_hccs} shows the HCCs for each $\gamma$ value. %Each node uniquely identifies itself, indicating its political party (e.g., G1 is Greens account~\#1), but node shape indicates political alignment. High activity (node brightness) correlates with link weight, which is the number of co-retweets of the link's adjacent nodes. 
Some accounts and parties appeared at each window size, (e.g., parties L, A, G, and nodes L2, A1, G2), while some only appear in a few (e.g., parties C and S). 
This shows that different parties exhibited different approaches to retweeting and different members were involved over different time frames. 
Although party S members co-retweeted enough to appear in two time windows, they were not consistently active enough to re-appear in the largest time window, where their activity was overtaken by other accounts. It is particularly noticeable that the L party had two core cooperating accounts, L2 and L4, who were active enough to appear in each time window, and then a large team active in the hour-long window, implying that a deliberate strategy of team-based co-retweeting was employed (rather than a coincidental one). Rather than the posting times being highly coordinated (so that retweets could appear nearly simultaneously), it appears as if the L accounts were simply relatively highly attentive to their colleagues' tweets and retweets and retweeted them when they saw them appear (which often occurred within an hour), as could be expected of a modern social media-savvy party.

Examining the content of these HCCs confirmed that they were genuine communities using the co-retweet strategy (not necessarily deliberately). 
The top retweeted tweets of each HCC ($\gamma$=15) are:
\sloppy \begin{itemize}
    \item ``RT \mention{alpsa}: A message from former \mention{AustralianLabor} Prime Minister, \redact.     https://t.co/\aurl\footnote{URLs starting with `https://t.co/' refer back to the original retweeted tweet's URL, and are obscured here for readability and anonymity.}''
    \item ``RT \mention{\redact}: Liberals promise \$40m to tackle elective surgery waiting times in South Australian hospitals. \hashtag{SAVotes2018}… https://t.co/\aurl''
    \item ``RT \mention{SALibMedia}: Under Labor there aren’t enough job opportunities for young South Australians. Here’s what they are saying about \mention{\redact} and \mention{alpsa} \hashtag{saparli} https://t.co/\aurl''
    \item ``RT \mention{\redact}: The results of this state election are clear - celebrity candidates and pop up parties come and go, but the Greens\ldots https://t.co/\aurl''
\end{itemize}
Using the tweets each HCC posted, it is possible to attribute each to a political affiliation, if not directly to a party, without resorting to inspecting the identities of the members. %The same can be done with domain knowledge using the top hashtags (Figure~\ref{fig:sapol_top_hts_15m}).

\subsubsection{Focus of Connectivity (RQ3)} \label{sec:connectivity_focus}

\begin{figure*}[t!]
    \centering
    \subfloat[Retweets.]{
        \includegraphics[width=0.45\columnwidth]{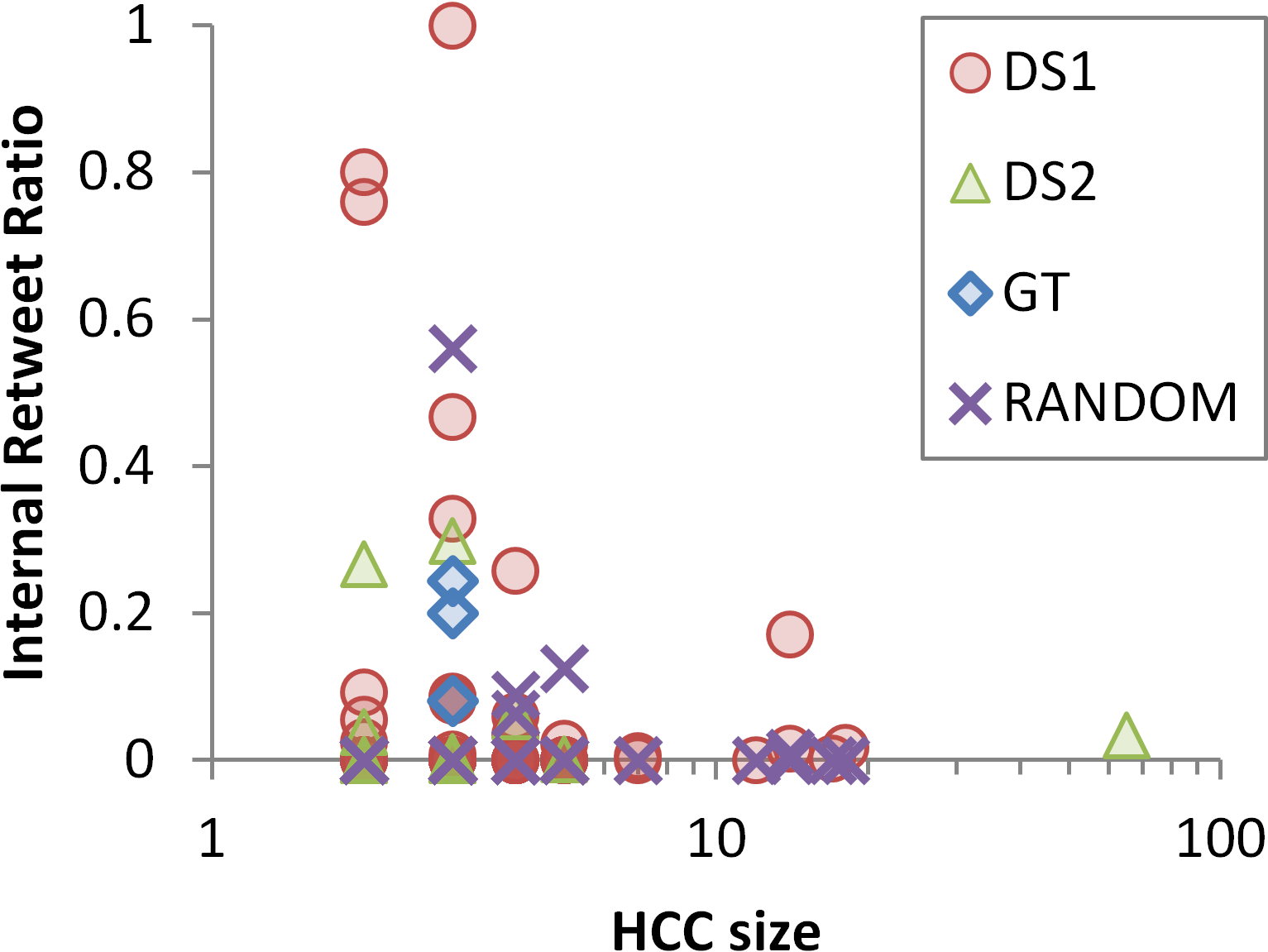}%
        \label{fig:int_rt_ratio_15m_fsa_v}
    } \hfill
    \subfloat[Mentions.]{
        \includegraphics[width=0.45\columnwidth]{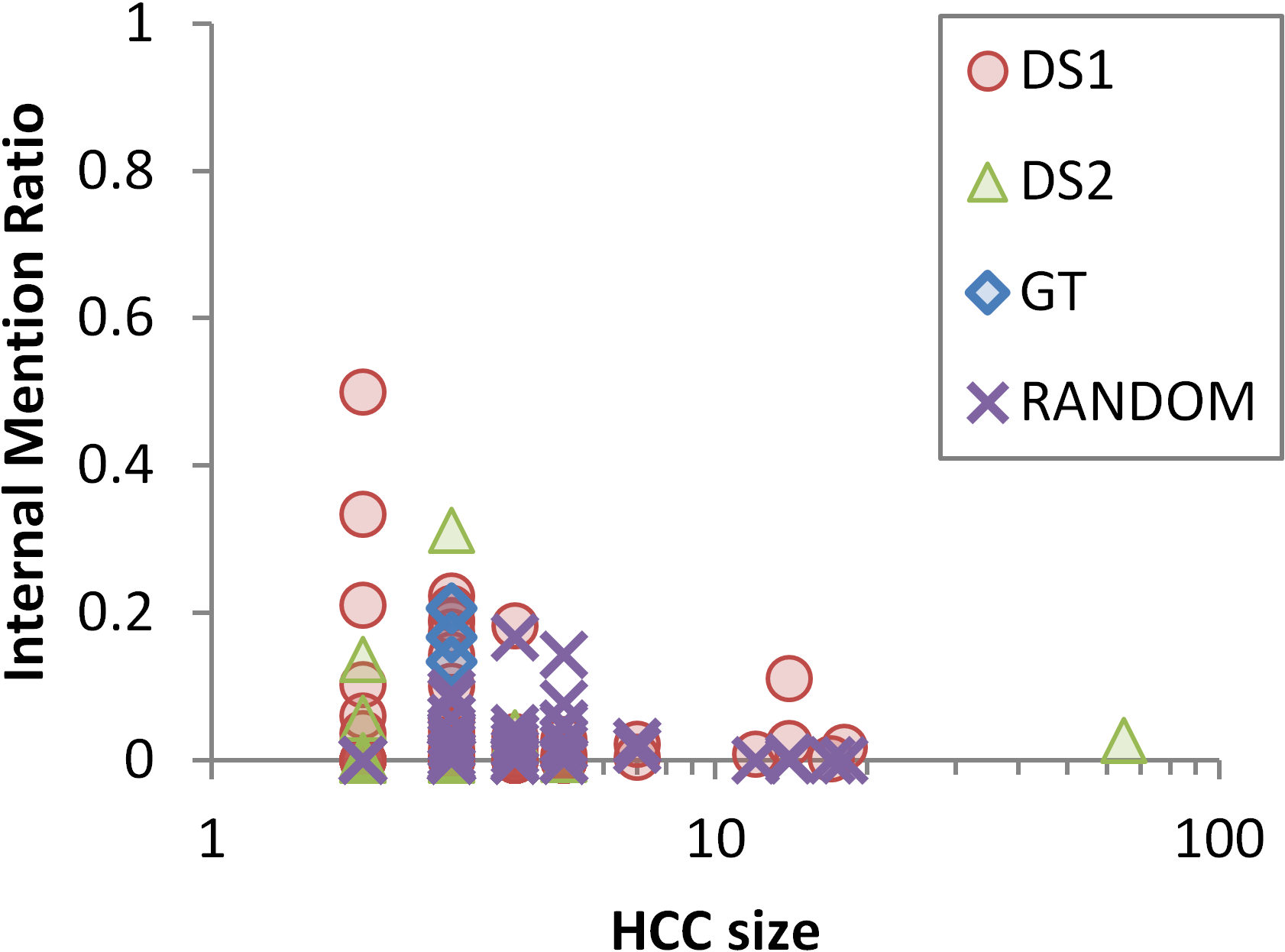}%
        \label{fig:int_m_ratio_15m_fsa_v}
    }
    \caption{The proportions of each HCCs retweets and mentions referring to accounts within the HCC ($\gamma$=15, FSA\_V).}
    \label{fig:hcc_int_ratios}
\end{figure*}

The IRRs and IMRs for the HCCs in the DS1, DS2, GT and RANDOM datasets is shown in Figure~\ref{fig:hcc_int_ratios}. The larger the HCC size, the greater the likelihood of retweeting or mentioning internally, so it is notable that DS2's largest HCC has IRR and IMR's of around $0$, though even the smaller HCCs have low ratios. Ratios for the smallest HCCs seem largest, possibly due to low numbers of posts, many of which may be retweets or include a mention, inflating the ratios. The hypothesis that political accounts would retweet and mention themselves frequently is not confirmed by these results, possibly because they are retweeting and mentioning official or party accounts outside the HCCs.

% {\color{blue}!! I want to see distributions of these - maybe as a multi-histogram?}
% didn't really work
% \begin{figure*}[t!]
%     \centering
%     \subfloat[Internal Retweet Ratios.]{
%         % \includegraphics[width=0.45\columnwidth]{resources/int_rt_ratio-15m-fsa_v_0.3.png}%
%         \includegraphics[width=0.99\columnwidth]{resources/irr_histograms.pdf}%
%         \label{fig:irr_histograms}
%     } \\ %hfill
%     \subfloat[Internal Mention Ratios.]{
%         % \includegraphics[width=0.45\columnwidth]{resources/int_mention_ratio-15m-fsa_v_0.3.png}%
%         \includegraphics[width=0.99\columnwidth]{resources/imr_histograms.pdf}%
%         \label{fig:imr_histograms}
%     }
%     \caption{Histograms of the HCCs internal retweet and mention ratios (i.e., referring to accounts within the HCC) ($\gamma$=15, FSA\_V).}
%     \label{fig:irr_imr_histograms}
% \end{figure*}

% \item How much variation is there in the content posted by the communities?
% entropy, hashtags, mentions

\subsubsection{Content Variation (RQ4)}% in HCC content}% How much variation is there in the content posted by the communities?}

\begin{figure*}[t!]
    \centering
    \subfloat[DS1.]{
        \includegraphics[width=0.31\columnwidth]{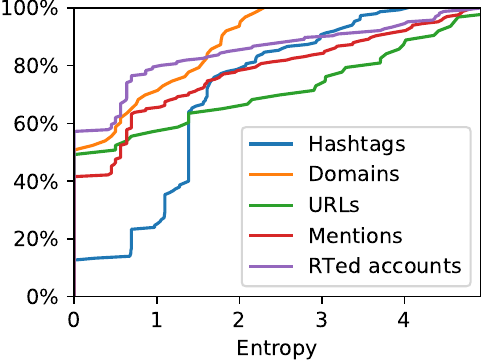}%
        \label{fig:saelec_15m_features_cf}
    } %\hfill
    \subfloat[DS2.]{
        \includegraphics[width=0.31\columnwidth]{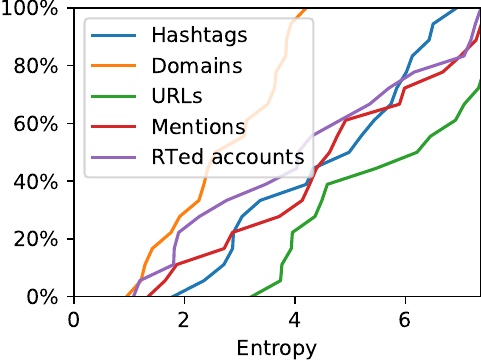}%
        \label{fig:ira_15m_features_cf}
    } %\hfill
    \subfloat[RANDOM.]{
        \includegraphics[width=0.31\columnwidth]{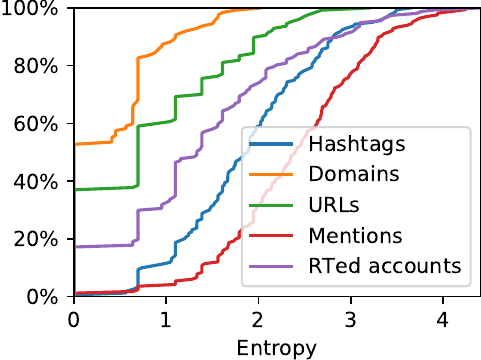}%
        \label{fig:random_15m_features_cf}
    }
    \caption{Cumulative frequency of HCCs' entropy scores for five tweet features, comparing DS1 and DS2 with RANDOM ($\gamma$=$15$, FSA\_V). Feature variation increases moving right on the x axis.}
    \label{fig:features_cfs}
\end{figure*}

\begin{figure*}[t!]
    \centering
    \includegraphics[width=0.99\columnwidth]{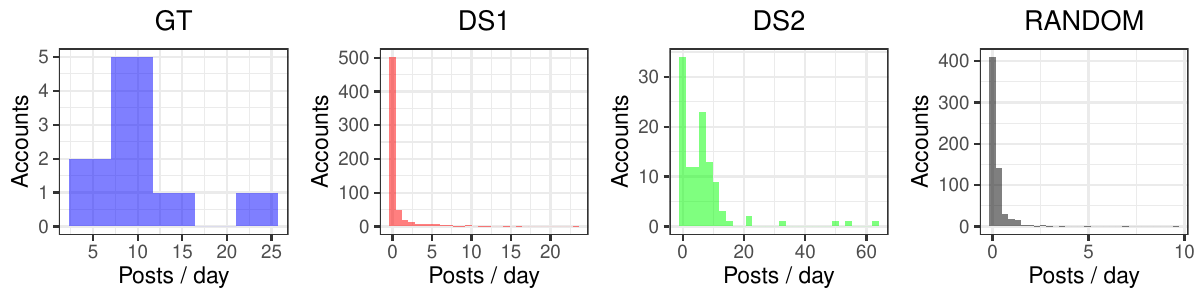}%
    \caption{Histograms of the daily posting rates of accounts in the GT, DS1, DS2, and RANDOM HCCs ($\gamma$=$15$, FSA\_V). Because the datasets cover different periods of time, the posting rate enables a fairer comparison. The distributions of DS1 and RANDOM posting rates are similar and notably different to those of DS2, while GT includes a higher proportion of more active than the other datasets.}
    \label{fig:hccs_daily_posting_rates}
\end{figure*}

% \textit{Variation within HCCs.}
% Highly coordinated reposting involves reusing the same content frequently, resulting in low feature variation (e.g., hashtags, URLs, mentioned accounts), which can be measured as entropy~\citep{CaoCLGC2015urlsh}. A frequency distribution of each HCC's use of each feature type was used to calculate each entropy score. Low feature variation corresponds to low entropy values. As per~\citet{CaoCLGC2015urlsh}, we compared the entropy of features used by DS1 and DS2 HCCs to RANDOM ones (Figure~\ref{fig:features_cfs}). Entries for HCCs which did not use a particular feature are omitted, as their scores would inflate the number of groups with $0$ entropy. 
We compared the entropy of features used by DS1 and DS2 HCCs to RANDOM ones (Figure~\ref{fig:features_cfs}). 
Many of DS1's small HCCs used only one of a particular feature, resulting in an entropy score of $0$ (Figure~\ref{fig:saelec_15m_features_cf}). In contrast, DS2's fewer HCCs have higher entropy values (Figure~\ref{fig:ira_15m_features_cf}), likely because they were active for longer (over $365$, not $18$, days) and had more opportunity to use more feature values. 
The majority of HCCs used few hashtags and URL domains, which is to be expected as the dominating domain is \emph{twitter.com}, which is embedded in all retweets as links back to the original tweet. Compared to the RANDOM HCCs (Figure~\ref{fig:random_15m_features_cf}), DS1 HCCs had lower variation in all features, while the longer activity period of DS2 resulted in distinctly different entropy distributions. Because DS1 HCC activity appears to have been more deliberate, and perhaps coordinated, it may be that the HCCs were more focused on their topic of conversation (especially when contrasted with RANDOM HCCs). Compared with RANDOM HCCs, DS1 HCCs retweeted fewer accounts, used fewer URLs (though they were from a similar distribution of domains), and many fewer mentions and hashtags. Many non-HCC accounts posted only a single retweet as their contribution to the discussion, and so it may be that a relatively high proportion the RANDOM HCC members only posted a single tweet, causing the distributions observed. As noted previously, the RANDOM HCC members posted $3,147$ tweets compared with DS1 HCCs' $8,527$ tweets, despite having the same number of members, so DS1 HCC members posted more than $2.7$ times as often. DS1 accounts clearly posted more tweets per individual than the RANDOM accounts, yet their distribution appears similar, and notably different to those of both DS2 and GT (Figure~\ref{fig:hccs_daily_posting_rates}).

% \item How consistent is the behaviour? To what extent does a sliding window mechanism emphasise HCCs? %Do the groups vary their membership 
% squares of alpha vs gamma showing similarities (cf overlap)
% \iffalse
\subsubsection{Consistent Coordination (RQ5)}%How consistent is the behaviour? To what extent does a sliding window mechanism emphasise HCCs?}

% The method presented, as evaluated so far, will highlight HCCs that coordinate their activity at a high level over an entire collection period. Further steps can be taken to determine which HCCs are coordinating their behaviour repeatedly and consistently across adjacent time windows. In this case, for each time window, we consider not just the nodes and edges from the current LCN, but additionally from previous windows, applying a degrading factor the contribution of their edge weights.
% To build an LCN from a sliding frame of $T$ time windows, the new LCN includes the union of the nodes and edges of the individual LCNs from the current and previous windows, but to calculate the edge weights, we apply a decay factor, $\alpha$, to the weights of edges appearing in windows before the current one. In this way, we apply a multiplier of $\alpha^x$ to the edge weights, where $x$ is the number of windows into the past: the current window is $0$ windows into the past, so its edges are multiplied by $\alpha^0 = 1$; the immediate previous window is $1$ window back, so its edge multiplier is $\alpha^1$; the one before that uses $\alpha^2$, and so on until the farthest window back uses $\alpha^{T-1}$. Generalising from Step~\ref{alg:step4_lcn}, the weight $w^{c,t}_{\{u,v\}}$ for an edge between $u$ and $v$ for criterion $c$ at window $t$ and a sliding window $T$ windows wide is given by

% \begin{equation} \label{eq:decayed_lcn_weight}
%   w^{c,t}_{\{u,v\}} = \sum_{x=0}^{T-1} w^{c,(t-x)}_{\{u,v\}} \cdot \alpha^{x}.
% \end{equation}

\begin{figure}[t!]
    \centering
    % \subfloat[GT.]{
    %     \includegraphics[width=\columnwidth]{resources/sapol-alpha-jaccard.pdf}%
    %     \label{fig:gt_alpha_overlap}
    % } \\
    \subfloat[DS1.]{
        \includegraphics[width=0.9\columnwidth]{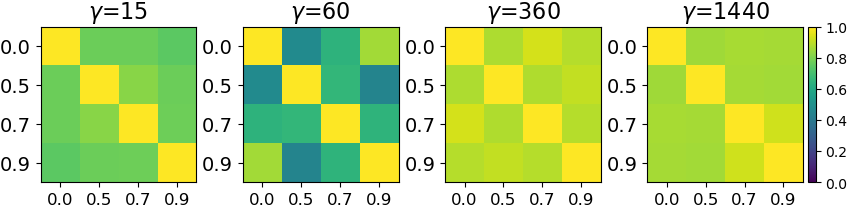}%
        \label{fig:ds1_alpha_overlap}
    } \\
    \subfloat[DS2.]{
        \includegraphics[width=0.9\columnwidth]{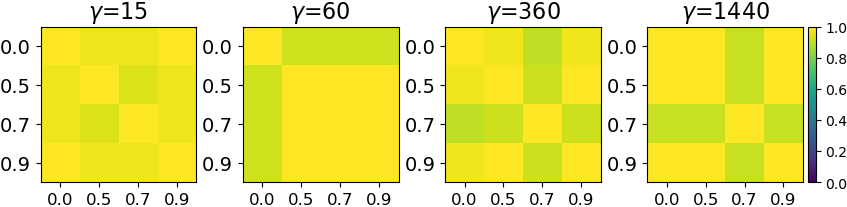}%
        \label{fig:ds2_alpha_overlap}
    }
    \caption{Jaccard similarity of HCC membership when varying $\alpha$ (0.0 is used for the Baseline condition).}
    \label{fig:alpha_overlaps}
\end{figure}

\begin{table}[th]
    \centering
    \caption{Statistics of discovered HCCs while varying $\alpha$ (Boost, FSA\_V). T=$5$ except in the Baseline condition. N = node count, E = edge count, C = HCC count.}
    \label{tab:graph_stats_across_alpha}
    \resizebox{\columnwidth}{!}{%
    \begin{tabular}{@{}cl|rrr|rrr|rrr|rrr@{}}
    % \begin{tabular}{@{}clrrrrrrrrrrrr@{}}
        \toprule
        \multicolumn{2}{l}{} & \multicolumn{3}{c}{$\gamma$=15} & \multicolumn{3}{c}{$\gamma$=60} & \multicolumn{3}{c}{$\gamma$=360} & \multicolumn{3}{c}{$\gamma$=1440} \\ 
        \cmidrule(r){3-5} \cmidrule(rl){6-8} \cmidrule(rl){9-11} \cmidrule(l){12-14}
        \multicolumn{2}{l}{}                
                       & N      & E     & C     & N      & E      & C    & N     & E       & C     & N      & E       & C     \\
        \midrule
        \multirow{4}{*}{DS1} 
        & Baseline     & 633    & 753   & 167   & 619    & 1,293  & 151  & 503   & 1,119   & 127   & 815    & 2,019   & 141   \\
        & $\alpha$=0.5 & 604    & 711   & 168   & 1,178  & 2,121  & 149  & 519   & 1,183   & 129   & 800    & 2,037   & 137   \\
        & $\alpha$=0.7 & 578    & 697   & 160   & 847    & 1,569  & 149  & 518   & 1,155   & 130   & 792    & 1,997   & 136   \\
        & $\alpha$=0.9 & 596    & 706   & 165   & 585    & 1,223  & 145  & 530   & 1,188   & 134   & 796    & 1,995   & 141   \\
        \midrule
        \multirow{4}{*}{DS2} 
        & Baseline     & 113    & 758   & 19    & 77     & 394    & 18   & 98    & 775     & 15    & 69     & 380     & 15    \\
        & $\alpha$=0.5 & 116    & 760   & 20    & 79     & 396    & 18   & 100   & 776     & 16    & 69     & 380     & 15    \\
        & $\alpha$=0.7 & 110    & 756   & 18    & 79     & 395    & 18   & 102   & 777     & 17    & 69     & 381     & 15    \\
        & $\alpha$=0.9 & 113    & 758   & 19    & 79     & 396    & 18   & 100   & 776     & 16    & 69     & 381     & 15    \\
        \bottomrule
    \end{tabular}%
    }
\end{table}

The sliding frame technique from Section~\ref{sec:sliding_window} was applied to DS1 and DS2 to reveal HCCs engaging in coordination consistently in adjacent time windows. The baseline used $T$=$1$ (i.e., a sliding frame a single time window wide) and $\alpha$=$0.0$. For the three other conditions, $T$ was set to $5$ as $\gamma$ increases approximately $5$ times each time, and $\alpha$=$\{0.5,0.7,0.9\}$. In this way, the choice of $\alpha$=$0.9$ would most strongly consider the contribution of LCNs from preceding time windows. Once applied for each time window, the aggregated LCNs were mined for HCCs and then the membership of these were compared in the same manner as in Section~\ref{sec:hcc_detection} using Jaccard similarity (Equation~\ref{eq:jaccard}), which, as noted earlier, is stricter about set matching than the Overlap method (Equation~\ref{eq:overlap}). Even so, as can be seen in Figure~\ref{fig:alpha_overlaps}, changes introduced by using the decaying sliding frame with different $\alpha$ values were insignificant in all cases, except for DS1 and $\gamma$=$60$. The implication, which is borne out when the exact network sizes (in nodes) are compared in Table~\ref{tab:graph_stats_across_alpha}, is that the previous windows did not add significant numbers of nodes, but instead increased the weight of existing edges, so the HCCs that were detected consisted of the same members working together over time, rather than splitting into subsets. To hide a team's coordination, one might expect that its members would associate separately in different time window, but that does not appear to have happened significantly in these datasets, except in the shorter time windows in DS1.

The greatest variation in node and edge count occurs in the shorter windows in DS1 ($\gamma$=$\{15,60\}$), probably because of the greater number of accounts active in DS1 (compared to DS2): accounts have more alters to form HCCs with in DS1, which has $20.5$k accounts, whereas choice in DS2 is limited to $1.3$k accounts. The near doubling of accounts in DS1's HCCs when $\gamma$=$60$ implies accounts co-retweeted often just within a single hour, and then not again (at least not for $T$=$5$ hours). This effect is swamped by the much more active consistent co-retweeting of a smaller set of users when $\alpha$ is increased to $0.7$ and above. Given the membership varies so little in the other conditions, an analysis of how these HCCs form and change over time is required. It is clear, however, that this approach would be best suited to filter-based collections, as they are likely to capture more accounts.

\subsubsection{Validation via HCC Classification (RQ6)}
% Points to make
% - use political HCCs as positive examples to build a classifier to validate
% - define features to use
% - explain use of one class classifier
% - offer results

% Points to make
% - use political HCCs as positive examples
% - 3 classifiers, including a bagging positive unlabeled implementation (why? explain)
% - comment on DS1 HCCs still being detectable, like the DS2 ones
Our final validation method relies on the HCCs in GT as positive examples of coordinating sets of accounts, given it is reasonable to assume that they ought to be coordinating their activities during an election campaign. 
% As our intent is to classify sets of users, rather than their content \citep[\emph{cf.} campaign detection and classification,][]{LeeCCS2013campext,Chu2012,Wu2018}, but members of such sets may have distinguishing common features, we choose HCC member accounts as our instances for classification. In this way, their features can be drawn not only from the accounts but also from their groupings. In contrast, \citet{Vargas2020} relied only on grouping features for their classifiers of coordinating communities.

% We used the GT HCCs to train 
% %To validate that the HCCs discovered were genuine, we used the ground truth HCCs in DS2 as positive examples and trained 
% three one-class account classifiers to categorise HCC accounts as similar to the political accounts or not. 
% These classifiers are like binary classifiers, but their training emphasises recognition of the positive class, and they are sometimes referred to as Positive/Unlabelled (PU) classifiers for this reason. %Such classifiers are required to
% They can, for example, suggest a new book from a wide range (such as a library) based on a person's reading history. In such a circumstance, the classifier designer has access to positive examples (books the reader likes) but all other instances (books here) are either positive or negative. When the classifiers that we have constructed recognise the other HCC accounts as positive instances, it provides confidence that the HCC members are coordinating their behaviour in the same manner as the political accounts. 

\paragraph{\textbf{Feature selection.}} As mentioned in Section~ \ref{sec:method_classification}, features are drawn from individual accounts and their groupings, specifically based on their individual and collective behaviour and homophily. For this reason, we select account-level features as well as group-level features to make up each account's feature vector, meaning that some of the values for HCC co-members will be identical. The following are the selected account-level features (all are regarding posts made by the account within the collection period):
\begin{itemize}
    % Account   & Post count, repost count, reply count, posting rate per minute, unique mentions, \\
    %           & mention count, unique hashtags, hashtag uses, unique URLs, URL uses, default profile \\
    %           & image (boolean), length of profile description (characters), and profile URL length \\
    \item \textit{Post count:} Number of posts they made.
    \item \textit{Repost count:} Number of reposts (retweets) included in the account's posts.
    \item \textit{Reply count:} Number of replies included in the account's posts.
    \item \textit{Posting rate:} Number of posts in the corpus divided by the length of the collection period in minutes.
    \item \textit{Unique mentions:} The number of unique accounts the account in question mentioned. \textbf{N.B.} A post may include multiple mentions, so the number of mentioned accounts and mention uses may exceed the number of posts.
    \item \textit{Mention count:} The number of mentions of accounts made by this account.
    \item \textit{Unique hashtags:} The number of unique hashtags (in lower case) used by the account. \textbf{N.B.} A post can include multiple hashtags, so the number of unique hashtags and their uses may exceed the number of posts.
    \item \textit{Hashtag uses:} The number of times hashtags were used by this account.
    \item \textit{Unique URLs:} The number of unique URLs used by this account. \textbf{N.B.} A post can include multiple URLs, so the number of unique URLs and their uses may exceed the number of posts.
    \item \textit{URL uses:} The number of times URLs were used by this account.
    \item \textit{Default profile image:} A Boolean representing whether or not the account is using the default profile image provided by the social media platform (i.e., whether the account has customised their profile image).
    \item \textit{Length of profile description:} The length, in characters, of the account's profile description in their first post in the corpus.
    \item \textit{Profile URL length:} The length, in characters, of the account's profile URL in their first post in the corpus.
\end{itemize}

The following features are drawn from the HCC's activity network (see Section~ \ref{sec:method_classification}) and included in the feature vector of each member of the HCC:
% We select the following group-level features based on the activity graph of the HCC of which the account is a member:
\begin{itemize}
    \item \textit{Aggregate post count:} The total number of posts by the members of the HCC in the corpus.
    \item \textit{Member count:} The number of accounts in the HCC.
    \item \textit{Interaction count:} The number of edges in the HCC's activity network.
    \item \textit{User count:} The number of account nodes in the HCC's activity network (includes accounts outside the HCC with whom the members interacted).
    \item \textit{Author count:} The number of the accounts in the HCC. %'s activity network who posted content (i.e., rather than being included simply because they were replied to or were retweeted from outside the collection period).
    \item \textit{Unique hashtag count:} The number of unique hashtags used in the HCC's members' posts.
    \item \textit{Hashtag uses:} The number of hashtag use interactions in the HCC's activity network.
    \item \textit{Unique URL count:} The number of unique URLs used in the HCC's members' posts.
    \item \textit{hashtag uses:} The number of URL use interactions in the HCC's activity network.
    \item \textit{Repost count:} The number of HCC's members' posts that were reposts (retweets).
    \item \textit{Quote count:} The number of HCC's members' posts that were quotes (i.e., reposts with comments).
    \item \textit{Mention count:} The number of mention interactions in the HCC's interaction network.
    \item \textit{Reply count:} The number of reply interactions in the HCC's interaction network.
    \item \textit{`In conversation' counts:} The number of times HCC members participated in `conversations' (trees of reply posts rooted in a single post) in the corpus.
    \item \textit{Proportion of HCC members reposted:} The frequency of reposts of HCC members divided by the overall number of reposts within the corpus (\emph{cf.} IRR, see Section~\ref{sec:connectivity_focus}). 
    \item \textit{Proportion of HCC members mentioned:} The frequency of mentions of HCC members divided by the overall number of mentions within the corpus (\emph{cf.} IMR, see Section~\ref{sec:connectivity_focus}).
    \item \textit{Proportion of HCC members replied to:} The frequency of replies to HCC members divided by the overall number of replies within the corpus.
\end{itemize}

\paragraph{\textbf{Classification results.}}
After being trained on the GT HCCs, the classifiers were then applied to the HCCs in DS1 and DS2. %We use COORDINATING and NON-COORDINATING to represent the positive and unlabeled classes. Again, upsampling was used to balance the positive and unlabeled instances. 
Unfortunately, the SVM classifier entirely failed to identify the positive class (i.e., coordinating accounts) in DS2, labelling every instance as non-coordinating. The SVM results for DS2 are therefore not considered beyond this point.

\begin{table}%[t]
    \centering%\small
    \caption{Accuracy (Acc.), positive class (COORDINATING) F\textsubscript{1}-scores ($F_{1P}$) and unlabeled class (NON-COORDINATING) F\textsubscript{1}-scores ($F_{1U}$) from the HCC classifiers. \textbf{N.B.} SVM scores for DS2 were uncomputable, as the classifier labeled all instances as NON-COORDINATING, and are marked with an asterisk (*) and only included for completeness.}
    \label{tab:classifier_results}
    \resizebox{\columnwidth}{!}{%
    \begin{tabular}{@{}llrrrrrrrrrrrr@{}}
        \toprule 
            &            & \multicolumn{3}{c}{$\gamma$=$15$} & \multicolumn{3}{c}{$\gamma$=$60$} & \multicolumn{3}{c}{$\gamma$=$360$} & \multicolumn{3}{c}{$\gamma$=$1440$} \\
            % \cmidrule{3-14}
                         \cmidrule(lr){3-5} \cmidrule(lr){6-8} \cmidrule(lr){9-11} \cmidrule(l){12-14}
            & Classifier & Acc. & $F_{1P}$ & $F_{1U}$ & Acc. & $F_{1P}$ & $F_{1U}$ & Acc. & $F_{1P}$ & $F_{1U}$ & Acc. & $F_{1P}$ & $F_{1U}$  \\
            % \cmidrule{2-14}
            \cmidrule(lr){2-2} \cmidrule(lr){3-5} \cmidrule(lr){6-8} \cmidrule(lr){9-11} \cmidrule(l){12-14}
        \multirow{3}{*}{DS1} % puts DS1 across 3 rows on the left
            & SVM & \bb{0.91} & \bb{0.92} & \bb{0.91} & \bb{0.76} & \bb{0.73} & \bb{0.79} & 0.58      & 0.37      & 0.68      & 0.87      & 0.86      & 0.88      \\
            & RF  & 0.65      & 0.70      & 0.59      & 0.59      & 0.60      & 0.58      & 0.59      & \bb{0.38} & 0.70      & \bb{0.92} & \bb{0.92} & \bb{0.93} \\
            & BPU & 0.69      & 0.72      & 0.66      & 0.65      & 0.63      & 0.66      & \bb{0.60} & 0.37      & \bb{0.71} & 0.88      & 0.86      & 0.89      \\
            % \cmidrule{2-14}
            \cmidrule(lr){2-2} \cmidrule(lr){3-5} \cmidrule(lr){6-8} \cmidrule(lr){9-11} \cmidrule(l){12-14}
        \multirow{3}{*}{DS2} % puts DS2 across 3 rows on the left
            & SVM &     0.50* &     0.00* &     0.67  &     0.50* &     0.00* &     0.67  &     0.50* &     0.00* &     0.67  &     0.50* &     0.00* &     0.67  \\
            & RF  &     0.89  &     0.88  &     0.90  & \bb{0.94} & \bb{0.94} & \bb{0.94} & \bb{0.97} & \bb{0.97} & \bb{0.97} & \bb{0.91} & \bb{0.90} & \bb{0.91} \\
            & BPU & \bb{0.93} & \bb{0.93} & \bb{0.94} &     0.92  &     0.91  &     0.92  &     0.95  &     0.95  &     0.96  &     0.54  &     0.18  &     0.68  \\
        
            % & SVM & \bb{0.85} & \bb{0.87} & \bb{0.82} & \bb{0.83} & \bb{0.85} & \bb{0.79} & 0.80      & 0.84      & 0.76      & 0.84      & 0.86      & 0.80      \\ 
            % & RF  & 0.78      & 0.82      & 0.73      & 0.71      & 0.78      & 0.60      & 0.76      & 0.81      & 0.69      & 0.77      & 0.81      & 0.70      \\
            % & BPU & 0.79      & 0.83      & 0.74      & 0.79      & 0.82      & 0.73      & \bb{0.84} & \bb{0.86} & \bb{0.80} & \bb{0.86} & \bb{0.88} & \bb{0.83} \\

        %     \cmidrule(lr){2-2} \cmidrule(lr){3-5} \cmidrule(lr){6-8} \cmidrule(lr){9-11} \cmidrule(l){12-14}
        % \multirow{3}{*}{DS3} % puts DS3 across 3 rows on the left
        %     & SVM & 0.16      & 0.17      & 0.16      & 0.31      & 0.24      & 0.36      & ---       & ---       & ---       & ---       & ---       & ---       \\
        %     & RF  & \bb{0.42} & \bb{0.59} & 0.00      & 0.34      & 0.32      & 0.36      & ---       & ---       & ---       & ---       & ---       & ---       \\
        %     & BPU & \bb{0.42} & 0.37      & \bb{0.46} & \bb{0.45} & \bb{0.35} & \bb{0.53} & ---       & ---       & ---       & ---       & ---       & ---       \\

        \bottomrule
    \end{tabular}
    } % end resizebox/textwidth
\end{table}

% The performance measured considered include the classifier's accuracy, $F_1$ scores for each class (COORDINATING and NON-COORDINATING), and the Precision and Recall measures that the $F_1$ scores are based upon. High precision implies the classifier is good at recognising samples correctly, and high Recall implies that a classifier does not miss instances of the class they are trained on in any testing data. For example, a good apple classifier will successfully recognise an apple when it is presented with one, and when presented with a bowl of fruit, it will successfully find all the apples in it. The $F_1$ score combines these two measures:

% \begin{equation}
%     F_1 = 2 \cdot \frac{Precision \cdot Recall}{Precision + Recall}
% \end{equation}

% and provides insight into to the balance between the classifiers precision and recall. The accuracy of a classifier is the proportion of instances in a test data set that the classifier labeled correctly. In this way, the accuracy is the most coarse of these measures, because it offers little understanding of whether the classifier is missing instances it should find (false negatives) or labeling non-matching instances incorrectly (false positives). The $F_1$ score begins to address this failing, but direct examination of the Precision and Recall provides the most insight into each classifier's performance.

The accuracy of the best classifier for each dataset and time window ranged from $0.60$ to $0.97$ (shown in Table~\ref{tab:classifier_results}), with better performance identifying HCC members in DS2 than in DS1. $F_1$ scores (outside $\gamma$=$360$) for the COORDINATING ($F_{1P}$) and NON-COORDINATING ($F_{1U}$) instances ranged from $0.73$ to $0.97$ and $0.79$ to $0.97$, respectively. Each classifier performed best for DS1 in different time windows, and RF and BPU performed similarly well for DS2 in all time windows, except for BPU struggling in the $\gamma$=1440 time window. All classifiers also performed the least well in the six hour time window for DS1, possibly because the GT HCCs' activity coordination was most prominent over the short time frames of an hour or less, and otherwise at the day level. Even so, $F_{1U}$ scores consistently hover around $0.7$, which is significantly better than random, though the $F_{1P}$ scores around $0.37$ indicate difficulty identifying all COORDINATING HCC members, a detail which is discussed in more detail below. The accuracy and $F_1$ results show that the the classifiers could all be successfully trained to recognise GT HCCs in most time windows and that the GT HCCs represented most of the HCCs in DS1 and DS2, despite the different levels of activity (DS2 HCC members interacted more than DS1 or GT HCC members in their corpus, primarily because the collection period was so much greater). 
%HCC results for $\alpha$$\neq$$0$ only varied edge weights for $\gamma$=$1440$, and the classifier results were almost identical and are not reported.

%The results in Table \ref{tab:classifier_results} show that accuracy and $F_1$ scores above $0.7$ and as high as $0.93$ were achieved for both datasets and all values of $\gamma$ except for $\gamma$=$360$ in DS2. SVM performed best for the shorter windows, while BPU and RF worked better for the longer windows, though not by a great margin. As most of the HCCs in DS2 remained identical across the decay conditions, only the $\gamma \neq 1440$ variants were tested, though the results matched those for $\gamma$=$1440$ and $T$=$1$ very closely and are not reported in detail here.

\begin{table}%[t]
    \centering
    \caption{Precision and Recall for the positive (COORDINATING) class. \textbf{N.B.} SVM scores for DS2 were uncomputable, as the classifier labeled all instances as NON-COORDINATING, and are marked with an asterisk (*) and only included for completeness.}
    \label{tab:classifier_pr_values}
    \resizebox{\columnwidth}{!}{%
    \begin{tabular}{@{}llrrrrrrrr@{}}
        \toprule
        & & \multicolumn{2}{c}{$\gamma$=$15$} & \multicolumn{2}{c}{$\gamma$=$60$} & \multicolumn{2}{c}{$\gamma$=$360$} & \multicolumn{2}{c}{$\gamma$=$1440$} \\
        & Classifier & Precision & Recall & Precision & Recall & Precision & Recall & Precision & Recall \\
        \cmidrule(lr){2-2} \cmidrule(lr){3-4} \cmidrule(lr){5-6} \cmidrule(lr){7-8} \cmidrule(l){9-10}
        \multirow{3}{*}{DS1} % puts DS1 across 3 rows on the left
        & SVM & \bb{0.86} & \bb{0.99} & \bb{0.85} & \bb{0.63} & 0.73      & \bb{0.25} & 0.91      & 0.82      \\
        & RF  & 0.61      & 0.80      & 0.59      & 0.62      & 0.80      & \bb{0.25} & \bb{1.00} & \bb{0.85} \\
        & BPU & 0.66      & 0.78      & 0.66      & 0.60      & \bb{0.86} & 0.23      & \bb{1.00} & 0.76      \\
        \cmidrule(lr){2-2} \cmidrule(lr){3-4} \cmidrule(lr){5-6} \cmidrule(lr){7-8} \cmidrule(l){9-10}
        \multirow{3}{*}{DS2} % puts DS2 across 3 rows on the left
        & SVM &     0.00* &     0.00  &     0.00* &     0.00  &     0.00* &     0.00  &     0.00* &     0.00  \\
        & RF  & \bb{1.00} &     0.79  & \bb{1.00} & \bb{0.88} &     0.96  & \bb{0.98} & \bb{1.00} & \bb{0.81} \\
        & BPU &     0.97  & \bb{0.89} &     0.94  & \bb{0.88} & \bb{1.00} &     0.91  &     0.78  &     0.10  \\

        % & SVM & \bb{0.77} & \bb{1.00} & \bb{0.74} & \bb{1.00} & 0.72      & \bb{1.00} & 0.75      & \bb{1.00} \\
        % & RF  & 0.70      & \bb{1.00} & 0.64      & \bb{1.00} & 0.68      & \bb{1.00} & 0.68      & \bb{1.00} \\
        % & BPU & 0.71      & \bb{1.00} & 0.70      & \bb{1.00} & \bb{0.75} & \bb{1.00} & \bb{0.78} & \bb{1.00} \\

        % \cmidrule(lr){2-2} \cmidrule(lr){3-4} \cmidrule(lr){5-6} \cmidrule(lr){7-8} \cmidrule(l){9-10}
        % \multirow{3}{*}{DS3} % puts DS3 across 3 rows on the left
        % & SVC & 0.16      & 0.17      & 0.27      & 0.22      & ---       & ---       & ---       & ---       \\
        % & RF  & \bb{0.46} & \bb{0.84} & 0.33      & \bb{0.31} & ---       & ---       & ---       & ---       \\
        % & BPU & 0.41      & 0.34      & \bb{0.43} & 0.29      & ---       & ---       & ---       & ---       \\

        \bottomrule
    \end{tabular}%
    } % end resizebox
\end{table}

Table~\ref{tab:classifier_pr_values} shows Precision and Recall across all classifiers and datasets for the COORDINATING class. (Given our emphasis on recognising COORDINATING instances, we do not present the corresponding results for the NON-COORDINATING class here.)
For all time windows, Precision is high for the classifiers against DS1 (ranging from $0.85$ to $1.0$) and is $1.0$ for all DS2, meaning that the HCCs are clearly discernible from the NON-COORDINATING instances. Recall varies significantly for DS1 (between $0.25$ and $0.99$), but less so for DS2 (between $0.81$ and $0.98$), meaning that some DS1 HCCs were misclassified but most were correctly identified in DS2. The Recall scores for $\gamma$=$360$ explain why the $F_{1P}$ scores were so low in Table~\ref{tab:classifier_results}, because the corresponding Precision scores are still relatively high. As mentioned above, there is something particular about the six hour time window ($\gamma$=$360$), as the GT HCC members (via their account features and group behavioural features) were less easily distinguishable from the randomised NON-COORDINATING accounts, resulting in poorer performing classifiers. The reason for this is possibly the choice of window boundaries. The time window boundaries rested at 0000, 0600, 1200, and 1800 hours, while boundaries defined more by work activity (e.g., 0200, 0800, 1400, 2000 hours) may better match human activity patterns. For other, less geographically bound datasets (i.e., ones where the activity comes from around the world, rather than from a single or small group of adjacent timezones), other ground truth may be required.

SVM was the best performing classifier for COORDINATING accounts in DS1 in the shorter time windows ($\gamma$=$\{15,60\}$), but RF performed best for the day long time window. Although Precision is high for $\gamma$=$360$, it is hard to argue the BPU classifier performed significantly better than the others. For DS2, both RF and BPU performed well, and similarly, except for BPU's Recall in the $\gamma$=$1440$ time window. For that reason, we can argue that RF performed best overall, but no particular consistently can be observed.

%Consequently, the classifiers provide confidence that the non-proponent HCCs appear similar to the proponent HCCs and thus may be behaving in similar ways, though the question of intent remains.
Consequently, by accepting the assumption that the ground truth HCCs exhibited at least one type of coordination, these classifiers provide confidence that the other HCCs appear similar to the GT ones and thus may have behaved in similar ways. The question of intent remains, however. Examining the content subjected to coordination will likely provide clues, but deeper examination of behaviour to identify, e.g., Princpal-Actor patterns \citep{Giglietto2020}, may also be enlightening. 
More examples of similar coordination activities as well as other coordination types would bolster the positive training and testing sets, as well as expand knowledge regarding coordination strategies in use online.

\subsubsection{Multiple Criteria: \emph{Bully}ing}

Some strategies can involve a combination of actions. Behaviours that contribute to \emph{Bully}ing by dogpiling, for example, include joining conversations started by the target's posts and mentioning the target repeatedly, within a confined time frame. As DS1 included all replied to tweets, we investigated it inferring links via co-mentions and co-convs with a window size of $10$ minutes, and FSA\_V with $\theta$=$0.001$, having maximised the ratio of MEW to HCC size. Of $142$ HCCs discovered, the largest had five accounts and most only had two. Only $32$ had more than ten inferred connections, but five have more than $1,000$. These heavily connected accounts, after deep analysis, were simply very active Twitter users who engaged others in conversation via mentions, which outweighed the more strict co-conv criterion of participants \emph{reply}ing into the same conversation reply tree. % (formed by starting at a single root tweet). % Very little evidence of co-conv was observed.
%The resulting $142$ HCCs included one with $5$ accounts and the remainder had only $2$ or $3$, and only $32$ had $10$ or more inferred connections. Notably, $5$ HCCs (all size $2$) had between $1,171$ and $2,553$ inferred connections, and $21$ had more than $120$ connections. Deeper inspection revealed co-mentions dominated co-conv links, and that the tweets of identified pairs indicated they were active, genuine Twitter users. In fact, the entire LCN only contained $123$ accounts found to join the same conversation within the $10$ minute window, only  $4$ of which did not also co-mention another account.

\begin{figure}[t!]
    \centering
    \includegraphics[width=0.99\columnwidth]{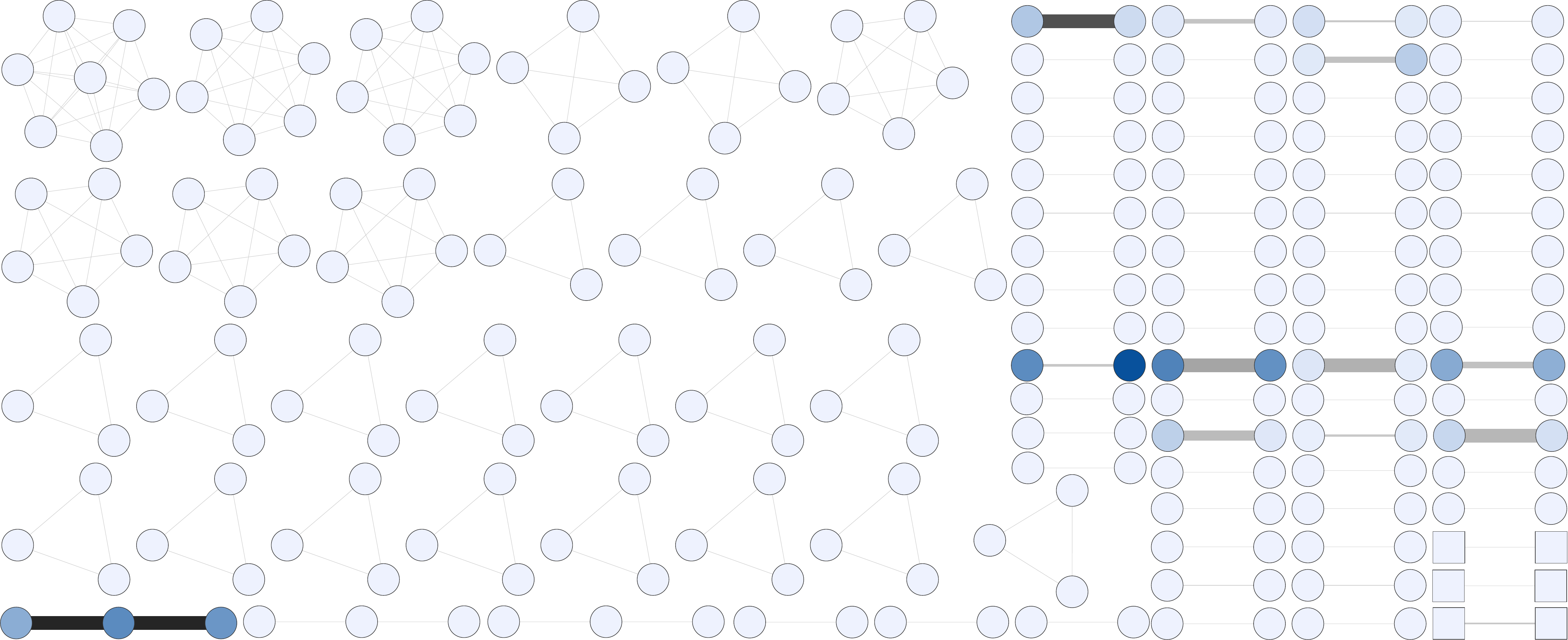}%
    \caption{While searching for \emph{Bully}ing behaviour in DS1, these are HCCs of accounts found engaging in co-mentions (circles) and co-mentions plus co-convs, i.e., engaged in both (square vertices in bottom right) ($\gamma$=360, FSA\_V, $\theta$=$0.01$). Edge thickness and darkness = inferred connections (darker = more). Vertex colour = tweets posted by that account (darker = more). Created with \emph{visone} (https://visone.info).}
    \label{fig:ds1_bully_360m}
\end{figure}

A larger window size was considered ($\gamma$=$360$) in case co-conv interactions were more prevalent. FSA\_V ($\theta$=$0.01$) exposed little further evidence of co-conv (Figure~\ref{fig:ds1_bully_360m}), finding $98$ small HCCs again dominated by co-mentions, not many of which had more than one inferred connection, implying most links were incidental; FSA\_V did not filter these out.
%We considered that bullies may also operate in longer time frames, consistently joining target's conversations throughout the day rather than in quick succession, and so we re-analysed DS1 using $\gamma$=$360$ and $\theta$=$0.01$, but again revealed few co-conv instances, all of which were significantly more connected by co-mentions (Figure~\ref{fig:ds1_bully_360m}). Of $98$ HCCs (the largest of which had $7$ accounts), only ones with $2$ or $3$ members had edges with a weight greater than $1$, implying most were only incidentally connected, but their (relatively) large sizes kept their mean edge weight great enough to be retained by FSA\_V.

This provides an argument for a more sophisticated approach to combining LCN edge weights for analysis, instead of Equation (\ref{eq:lcn_uv_combined_weight}), and that FSA\_V could be modified to better balance HCC size and edge weight. Furthermore, it is likely that bullying accounts will not just co-mention accounts frequently, but have low diversity in the accounts they co-mention, i.e., they repeatedly co-mention a small set of accounts, and spend a disproportional number of their tweets doing so. A further consideration is that participants in long discussions (reply trees) often include the author of the original tweet that sparked the discussion, and it would be misleading to include their account in results, implying that they \emph{bull}ied themselves. Finally, patterns of behaviour that would clearly qualify as conversations were observed in the datasets that did not fit the strict `conversation tree' model: accounts would mention several collocutors at the start of every tweet, but only potentially reply a tweet of one of them to continue the conversation. Importantly, sometimes the mentioned accounts included in tweets were prominent individuals whose names were included not because they were active participants in the conversation, but because the tweeter wanted to draw their attention to the conversation (regardless of the likelihood that the attempt would succeed; e.g., some tweets included references to prominent and busy politicians who would be unlikely to wade into arbitrary online discussions). These nuances are not explored here.

\subsubsection{HCC inter-relationships}

\begin{figure}[ht!]
    \centering
    \includegraphics[width=0.8\columnwidth]{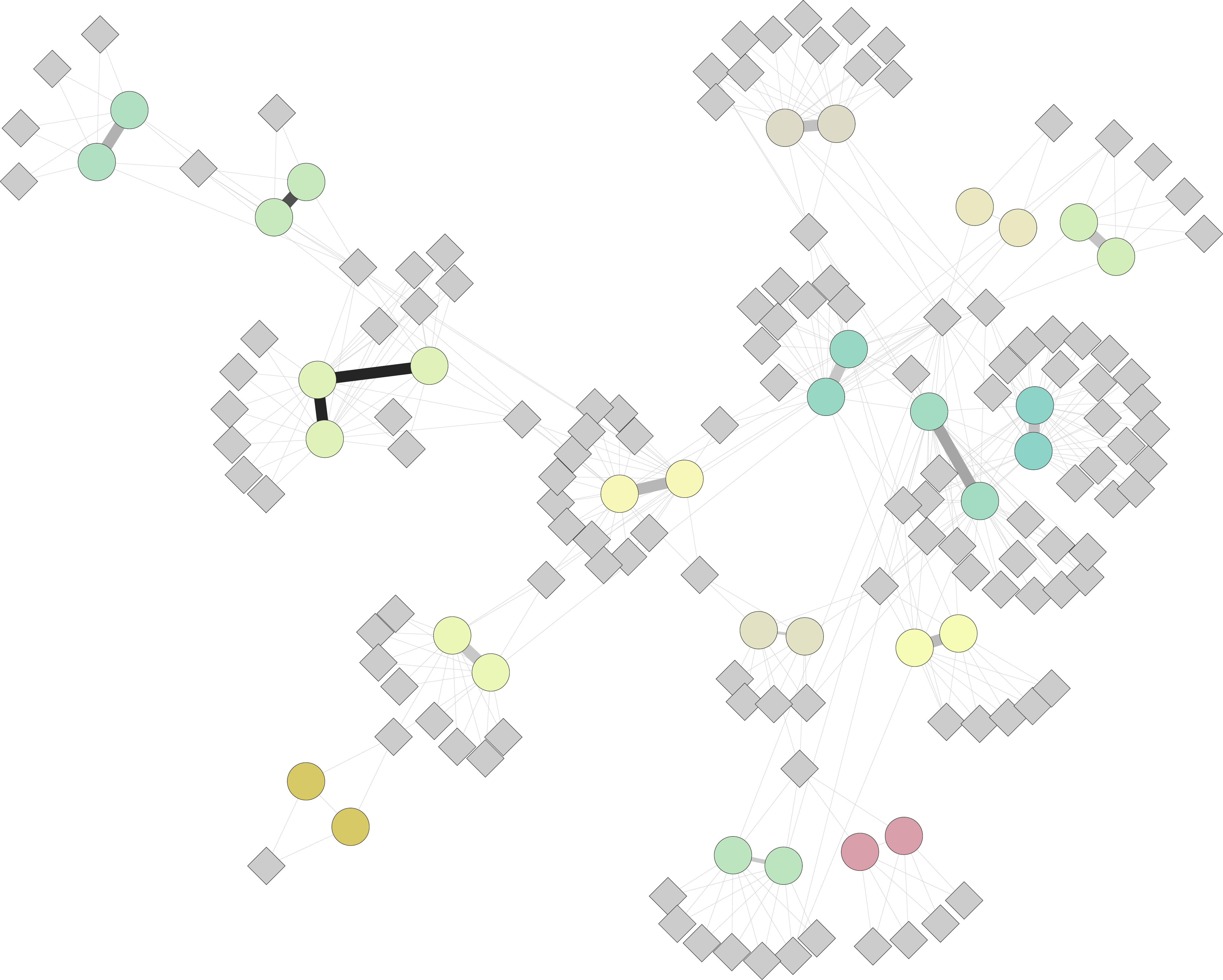}%
    \caption{A graph of DS1 HCC accounts (circle vertices) connected to the accounts they mention or conversations they join (diamonds). Accounts in the same HCC share a colour. Clear communities surrounding HCCs indicate who they converse with, and which conversants are co-mentioned by multiple HCC accounts. Edge width and darkness between HCC accounts indicates the weight of evidence joining them (darker implies more). Created with \emph{visone} (https://visone.info).}
    \label{fig:ds1_bully_360m_exp}
\end{figure}

%Introducing vertices to represent the \emph{reasons} HCC accounts are connected (e.g., who they co-mention, the conversations they join, {\color{blue} hashtags they use in common, or the tweets they co-retweet}) shows how the HCCs inter-relate {\color{blue} in the form of a two-level network (the nodes represent either \emph{accounts} or \emph{reasons})}. 
To study the relationships between HCCs, we create two-level networks starting with the HCC network and then adding nodes representing the elements of evidence linking them, known as \emph{reason} nodes (e.g., the tweets they co-retweet or the hashtags they use in common).
Figure~\ref{fig:ds1_bully_360m_exp} shows the largest component after such expansion was conducted on the HCCs in Figure~\ref{fig:ds1_bully_360m}. HCC accounts (circles) share colours and the distribution of the reasons for their connection (diamonds) show which other accounts are uniquely mentioned by an HCC and which are mentioned by more than one HCC. Heavy links between HCC accounts with few adjacent reason vertices imply these accounts are mentioning a small set of other accounts on many occasions.

\subsubsection{\emph{Boost}ing Accounts, not just Posts}

\begin{figure}[t!]
    \centering
    % \subfloat[GT.]{
    %     \includegraphics[width=0.37\columnwidth]{resources/sapol-top_hts-boost_users-hccs_10-hts_15-max_15-norm-w3h4.pdf}%
    %     \label{fig:gt_top_hts_boost_users}
    % } %\\
    % \subfloat[DS2.]{
    \includegraphics[width=0.8\columnwidth]{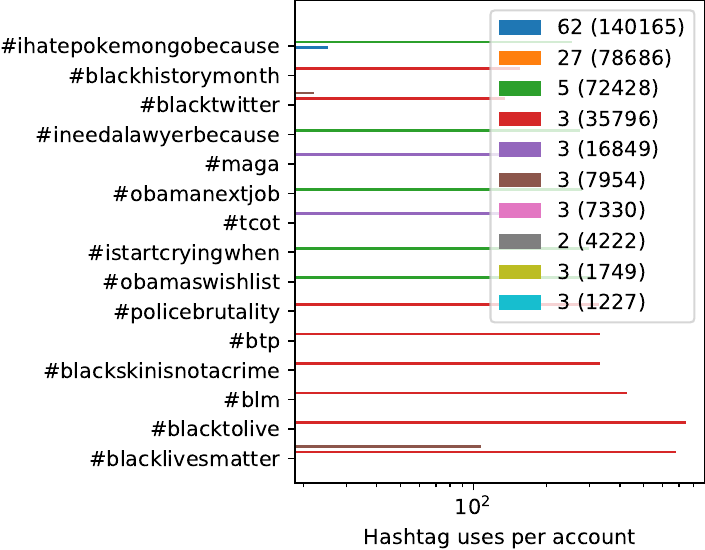}%
    %     \label{fig:ds2_top_hts_boost_users}
    % }
    \caption{Most used hashtags (per account) of the most active HCCs boosting accounts (FSA\_V, $\gamma$=$15$). The labels indicate member and tweet counts. Not all HCCs used a hashtag often enough to be visible.}
    \label{fig:top_hts_boost_users}
\end{figure}

\sloppy It is possible to \emph{Boost} an account rather than just a post. Returning to DS2, we sought HCCs from accounts retweeting the same account (FSA\_V, $\gamma$=$15$), and found that the hashtag use revealed further insights (Figure~\ref{fig:top_hts_boost_users}). No longer does one HCC dominate the hashtags. Instead clear themes are exhibited by different HCCs, but again, they are not the largest HCCs. The red HCC uses \hashtag{blacklivesmatter} and other Black rights-related hashtags (including \hashtag{blm}, \hashtag{blacktolive}, \hashtag{blackskinisnotacrime}, \hashtag{policebrutality} and \hashtag{btp}\footnote{BTP refers to the British Transport Police, the conduct of which was discussed in accounts of the arrest of a Black man at a London train station in mid-2016, e.g., \url{https://www.theguardian.com/uk-news/2016/jul/28/man-complains-after-police-place-spit-hood-over-head-during-arrest-london-bridge}.}), while the purple HCC uses pro-Republican ones (\hashtag{maga} and \hashtag{tcot}), and the green HCC is more general. Given the number of tweets these HCCs posted over 2016 (at least $16,849$), it is clear they concentrated their messaging on particular topics, some politically charged. It is arguable that their contributions helped inflame tensions and stoke divisions in socially sensitive topics, not just in the United States, but in the UK as well, and at the very least sought to draw the attention of others. 

The green HCC may be acting in a distractor role, as previously mentioned, given their contribution of $72,428$ tweets over the year.

% There were now four GT HCCs; one HCC \emph{Boost}ing tweets (Table~\ref{tab:hccs_info_1}) had split in two, and they used different hashtags (Figure~\ref{fig:gt_top_hts_boost_users}). The orange HCC concentrated on positive partisan hashtags ({\small\texttt{\#voteliberal}})

\subsubsection{Validation of Inauthentic Behaviour Detection}% of social bots and other inauthentic behaviour}

\begin{figure}[t!]
    \centering
    \includegraphics[width=0.3\columnwidth]{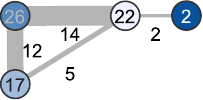}
    \caption[]{The most active DS1 co-retweet HCC ($\gamma$=10 seconds). Node label = number of posts, node colour = Botometer scores (higher = darker), link thickness and label = co-retweet occurrences.}
    \label{fig:ds1_10s_co-rt_hcc}
\end{figure}

The approach presented can be used to perform analytics similar to the Rapid Retweet Network used by \citet{PachecoFM2020whitehelmets}, who used it to expose tight clusters of bot-like accounts, which retweeted the same tweet within $10$ seconds of it appearing. We varied this for the DS1 dataset (due to its small nature) and searched for accounts which retweeted the same tweet within $10$ seconds, regardless of the age of the original tweet. We discovered a tight cluster of accounts, most with relatively high Botometer ratings\footnote{The English ``all features'' score was used as both our datasets are primarily either English speaking or are predominantly aimed at English speaking audiences.}~\citep{davis2016botornot}, shown in Figure~\ref{fig:ds1_10s_co-rt_hcc}.
The bot ratings were as follows: node 26: $0.787$; node 22: $0.381$; node 2: $0.949$; and node 17: $0.464$. All were high relative to the other accounts in the corpus, most of which had scores well below $0.2$; all four were had scores well above $0.2$, but the scores of two were also well above the `bot' threshold of $0.6$. 
On further inspection, they appeared to support vocational training and left-wing issues and posted retweets almost exclusively, but the content all related to the election. This finding enhances the bot ratings by making it clear which bots (or bot-like accounts) appear to work together. It also raises further questions regarding bot detection systems, however, as some of the accounts appeared to be genuine human, though abnormally active on Twitter. These accounts appeared to work together to actively disseminate messages aligned with their preferred narrative, though with a very low IRR (just shy of $10\%$) % 8 / 84 == 9.52%
despite most of their activity being retweets ($97.7\%$), % 84 / 86 == 97.67%
so to a certain degree it matters not whether they are automated or genuinely human-driven, but whether they are engaging in astroturfing or other inauthentic behaviour. In this circumstance, they may be genuine agenda-driven users, but they were definitely highly attentive to the same sources. Alternatively, when we consider their bot ratings more closely, it is possible that there is a mixture of account types, with node 26, in particular, acting as an automated `cheerleader' for nodes 22 and 17. Relative timings of their posts (to answer whether node 26 consistently was the second co-retweeter when paired with nodes 22 and 17) could reveal support for this hypothesis.

\subsubsection{Performance} \label{sec:performance}

In Table~\ref{tab:timings} we present the timings observed for the stages of processing for DS1 and DS2 conducted on a Dell Precision 5520 laptop equipped with an Intel Core i7-7820HQ CPU (2.9GHz), 32Gb RAM, and an NVMe PC300 480Gb SSD, running Windows 10. Parsing raw data is relatively cheap, with DS2's 1.5m tweets processed in just over a minute, and LCN construction dependent on the degree of activity within the time windows and the number of accounts. DS1's larger account pool increased the (node) size of the networks generated, and all associated post-processing. The size of DS1 LCNs were an order of magnitude greater than DS2's (in vertices and edges), resulting in correspondingly increasing execution times for aggregation and HCC extraction.

\begin{table}[t]
    \caption{Execution timings (in seconds).}
    \label{tab:timings}
    \resizebox{\columnwidth}{!}{%
        \begin{tabular}{@{}l|rrrr|rrrr@{}}
            \toprule
                                           & \multicolumn{4}{c}{DS1}     & \multicolumn{4}{c}{DS2}        \\ 
            \midrule
            Tweets                         & \multicolumn{4}{c}{115,913} & \multicolumn{4}{c}{1,571,245}  \\
            Parse raw (Step 1)             & \multicolumn{4}{c}{19.0 (from JSON)}    & \multicolumn{4}{c}{74.0 (from CSV)}       \\
            \midrule
            Window size $\gamma$ (minutes) & 15   & 60   & 360   & 1440  & 15    & 60    & 360   & 1440   \\
            \midrule
            Find evidence and build LCNs   & 15.0 & 28.0 & 123.0 & 427.0 & 121.0 & 106.0 & 246.0 & 567.0  \\
            Aggregate LCNs                 & 27.0 & 65.6 & 168.5 & 170.7 & 70.4  & 55.2  & 35.6  & 22.7   \\
            HCCs: FSA\_V                   & 28.3 & 58.2 & 126.1 & 209.3 & 6.3   & 4.2   & 5.8   & 5.0    \\
            HCCs: kNN                      & 9.0  & 22.7 & 97.5  & 206.4 & 4.3   & 4.3   & 4.7   & 4.6    \\
            HCCs: Threshold                & 5.2  & 11.9 & 34.6  & 64.0  & 2.2   & 2.3   & 2.7   & 2.7    \\ 
            \bottomrule
        \end{tabular}%
    }
\end{table}

\subsection{Applications}
% \begin{itemize}
%     \item ArsonEmergency co-url, co-domain
%     \item US Democratic and Republican Conventions, co-hashtag, co-url, co-domain
% \end{itemize}

Complementing the detailed validation presented above, in this section, we offer 
%In this subsection, we present 
two case studies in which our method has been used %to explore datasets 
in order to demonstrate its utility. %in different ways. 
Extending a study of polarised online communities in a discussion of bushfires in Australia \citep{WeberNFM2020bushfiresspringer}, the co-URL and co-URL domain analysis we conducted revealed how sources of were used by discussion participants, and how that use differed between the polarised communities. A second study of Twitter activity during the Democratic and Republican Conventions in the United States in August, 2020, %\citep{Weber2020asnacuspol}, 
making use of co-retweeting analysis in order to reveal influence attempts with social bots and co-hashtag analysis to discover discussion groups and their relations.

\subsubsection{\emph{\texttt{\#ArsonEmergency}} and Australia's ``Black Summer''}

During the Australian summer of 2019-2020, evidence of inauthentic Twitter behaviour emerged on the hashtag \hashtag{ArsonEmergency} \citep{WeberNFM2020bushfiresspringer} over an $18$ day period, including a week prior to and more than a week after a study on the matter \citep{GrahamKeller2020conv} was reported in the online tech and then mainstream media \citep{Stilgherrian2020zdnet}. Analysis of the tweets including the hashtag over the period revealed two clearly polarised retweeting communities, one supporting the narrative that arson was the cause of the bushfires and that eco-activism had prevented forest fuel load management (\emph{Supporters}) and one that countered the narrative, providing evidence that the fires were mostly started by natural or unintended causes (e.g., lightning and sparks from machinery) and the bushfires' ferocity was exacerbated by climate change (\emph{Opposers}). The remaining accounts in the dataset were referred to as \emph{Unaffiliated}. 

Differences between the communities' interaction and information sharing behaviour was apparent. \emph{Supporters} interacted with other accounts more by using replies, quotes, and mentions more than \emph{Opposers} or the \emph{Unaffilated}, as well as more hashtags and external URLs (i.e., referring to domains other than \texttt{twitter.com}), but they retweeted less. Notably, an analysis of the most shared URLs revealed that \emph{Opposers} shared articles debunking the narrative exclusively, while \emph{Supporters} shared a mixture of articles, mostly ones supporting or actively discussing their preferred narrative as well as some conspiratorial content. \emph{Unaffiliated} accounts shared narrative-focused URLs initially, but in the latter phase of the collection they shared debunking articles $9$ times more often.

\begin{figure}[t!]
    \centering
    \includegraphics[width=0.99\columnwidth]{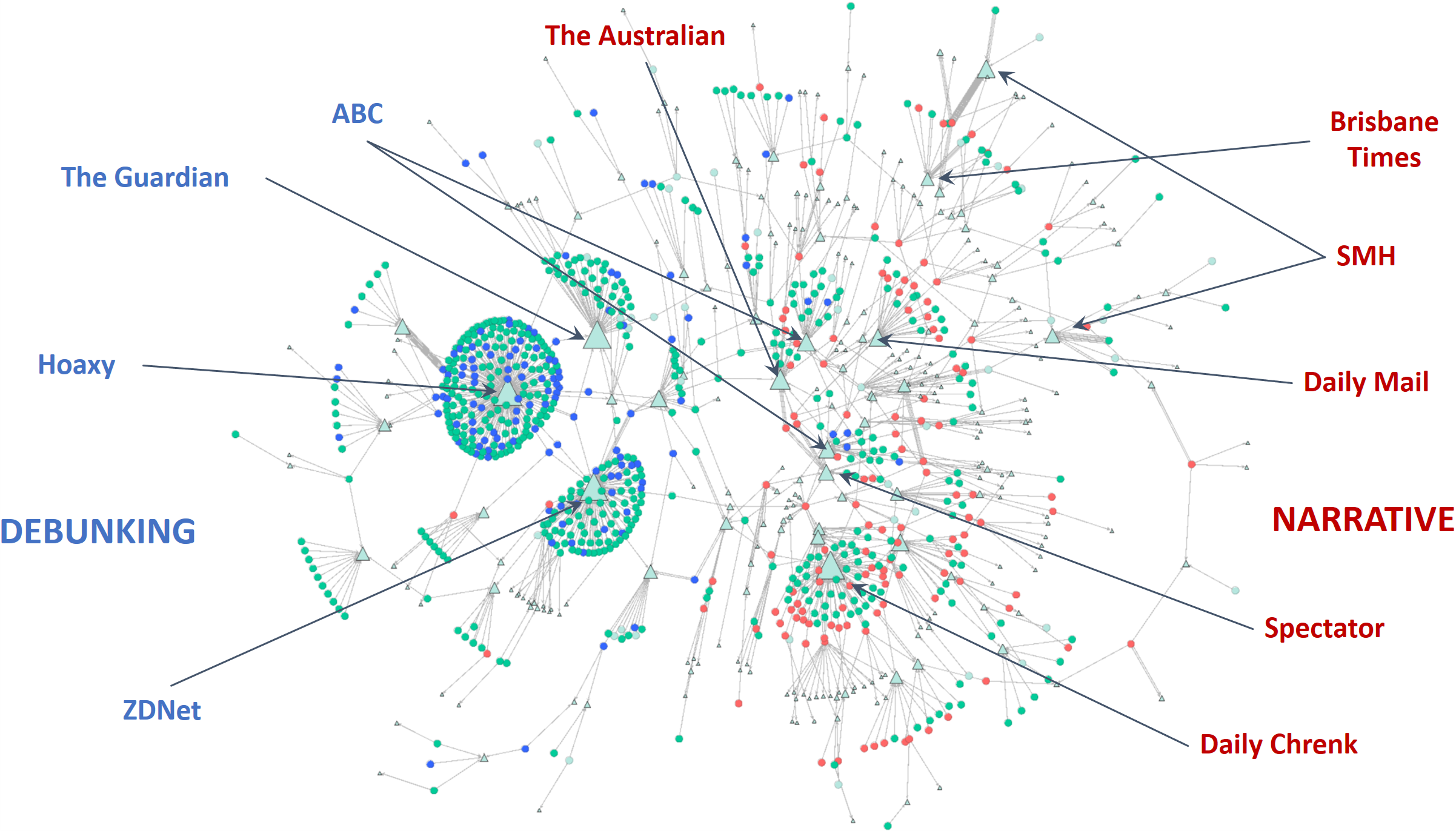}%
    \caption{Annotated account/URL bipartite network constructed from co-URL analysis of the ArsonEmergency dataset. %, adapted from \citep{Weber2020asnacarson}. 
    Circles represent HCC accounts (Threshold $t$=$0.1$, $\gamma$=$10$ seconds) and triangles represent URLs referring to articles. Accounts are linked to the URLs they shared, with multi-edges representing each use of a particular URL. URL nodes are sized by in-degree, and all coloured pale green. \emph{Supporter} nodes are coloured red, \emph{Opposer} nodes are blue, while \emph{Unaffiliated} ones are green. The most widely shared articles are annotated with the website on which they are hosted (\textbf{N.B.} ABC = Australian Broadcasting Corporation, SMH = Sydney Morning Herald). Blue annotated articles are categorised as DEBUNKING, while red ones are categorised as supporting or prominently discussing the `arson' NARRATIVE.}
    \label{fig:arson-co_urls}
\end{figure}

Analysis of the co-use of URLs revealed further behavioural differences between the communities (Figure~\ref{fig:arson-co_urls}). \emph{Opposers} were more focused that \emph{Supporters} in the URLs they shared, relying on a small set of debunking articles, also heavily shared by \emph{Unaffiliated} accounts. \emph{Supporters} were less tightly clustered around particular articles, and did share some debunking material as well as a variety of narrative-aligned articles.

\begin{figure}[t!]
    \centering
    \includegraphics[width=0.99\columnwidth]{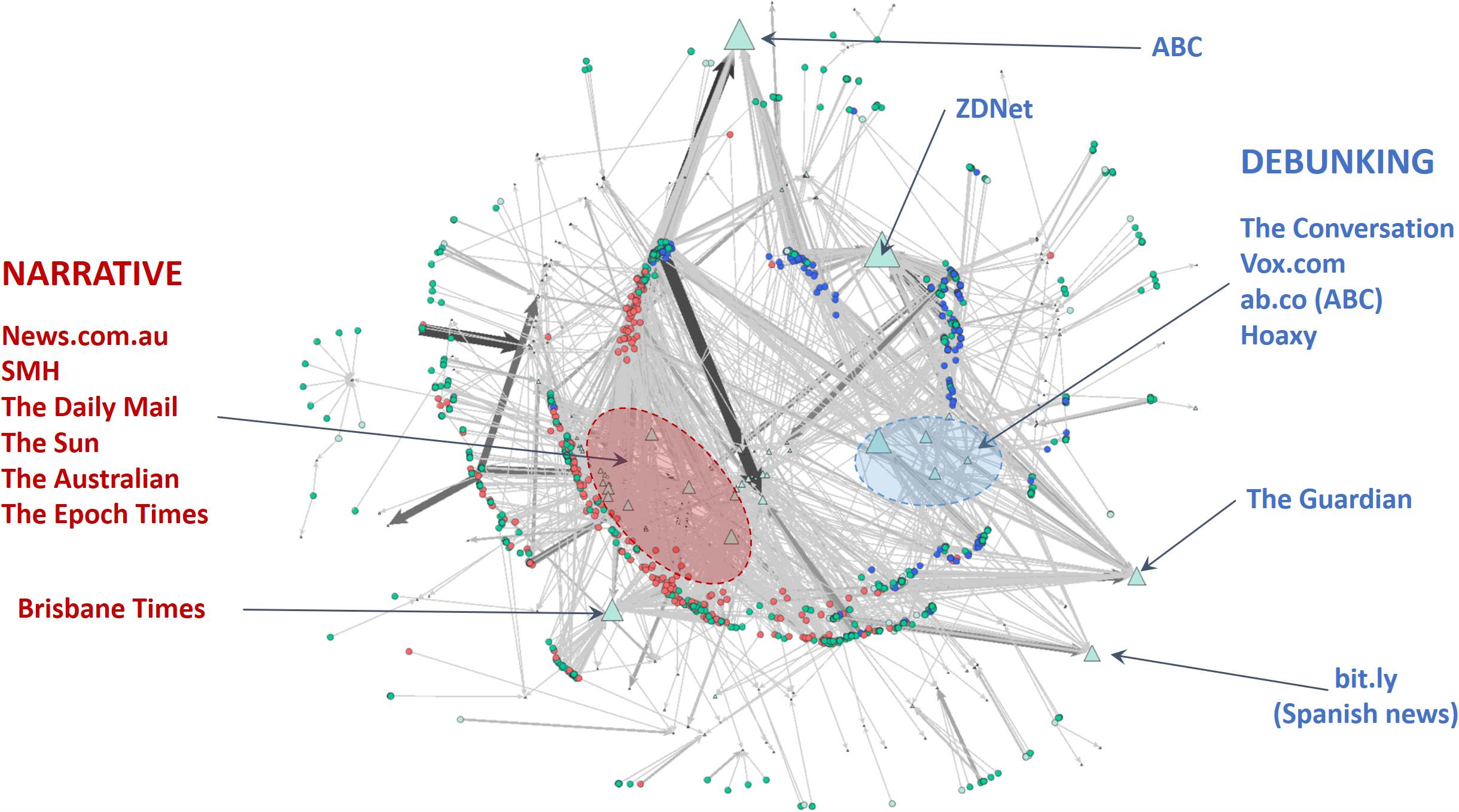}%
    \caption{Annotated account/URL domain bipartite network constructed from co-domain analysis of the ArsonEmergency dataset. %, adapted from \citep{Weber2020asnacarson}. 
    Circles are HCC accounts (Threshold $t$=$0.1$, $\gamma$=$10$ seconds) and triangles represent the domains of URLs used in tweets. Accounts are linked to the domains of URLs they shared, with thicker, darker edges representing frequent use of a particular domain. Domain nodes are sized by in-degree, and all coloured pale green. \emph{Supporter} nodes are coloured red, \emph{Opposer} nodes are blue, while \emph{Unaffiliated} ones are green. The most frequently referred to domains are annotated with the organisation to which they belong (\textbf{N.B.} ABC and ab.co = Australian Broadcasting Corporation, SMH = Sydney Morning Herald, News.com.au = News Corporation). Blue annotated domains are categorised as DEBUNKING, while red ones are categorised as supporting or prominently discussing the `arson' NARRATIVE. The red zone includes a number of DEBUNKING domains and is mostly referred to by \emph{Supporters} while the blue zone includes academic and centre and left wing domains categorised as DEBUNKING domains, which are referred to predominantly by \emph{Opposers}.}
    \label{fig:arson-co_domains}
\end{figure}

Co-domain analysis identified not just distinct URLs but distinct URL domains favoured by the different communities. Figure~\ref{fig:arson-co_domains} shows two clusters of domains: the red one contains domains from a number of conservative and right wing media organisations, while the blue one contains academic and centre and left wing media organisations. Although \emph{Supporters} mostly referred to the narrative-supporting domains while the \emph{Opposers} mostly referred to the debunking domains, it is notable that members of both communities referred heavily the ABC and the Guardian, which both published articles debunking the arson theory, often with reports from local fire fighting and law enforcement organisations. What is lost at this level of analysis is the way in which the articles were discussed when mentioned in tweets, including whether the tweets were agreeing or attacking the article content.

These co-analyses reveal how focused the polarised communities in their information sharing activities, contributing to the argument that the targeted efforts of the \emph{Opposer} community may have helped influence the broader \emph{Unaffiliated} community into sharing debunking articles.

\subsubsection{Twitter Discussion Groupings during the 2020 US Political Conventions}

A second case study making use of these techniques relates to the search for social bots attempting to influence the online discussion surrounding the Democratic and Republican National Conventions in August 2020, at which the parties formally nominate their candidates for the Presidential Election, later in the year. For a $96$ hour period over each $4$-day convention, tweets were filtered using RAPID, starting with \hashtag{demconvention} and \hashtag{rnc2020} for the Democratic National Convention (DNC) and the Republican National Convention (RNC), respectively. For the three hours prior to the formal collection, RAPID's topic tracking feature was enabled, which added hashtags used frequently in the discussions observed, resulting in larger sets of filter terms for each convention:

\begin{itemize}
    \item DNC: \textbf{\hashtag{demconvention}}, \hashtag{bidenharris}, \hashtag{bidenharris2020}, \hashtag{khive}, \newline \hashtag{signsacrossamerica}, \hashtag{unitedforbiden}, and \hashtag{wewantjoe};
    \item RNC: \textbf{\hashtag{rnc2020}}, \hashtag{rncconvention}, and \hashtag{nevertrump}.
\end{itemize}

Despite the disparity in hashtags used (i.e., found to occur sufficiently frequently alongside the seed terms), each dataset ultimately comprised approximately $1.5$ million tweets by over $400$ thousand unique users at each convention. Bots are often used to boost tweets, reaching other accounts that follow them, or by flooding hashtag communities or gaming trending algorithms \citep{woolley2016autopower,HegelichJ2016ukranianbotnet,Keller2019,GrahamKeller2020conv,graham2020asnac}. Social bots are specifically designed to mimic genuine human users, hiding the fact they are automated \citep{rise2016,GrimmeAA2018perspectives}. They do this to avoid detection, and in doing so can contribute to astroturfing campaigns, artificially boosting narratives while making them appear as simply popular grass roots movements. 

\begin{figure*}[t!]
    \centering
    \subfloat[DNC.]{
        \includegraphics[height=0.13\textheight]{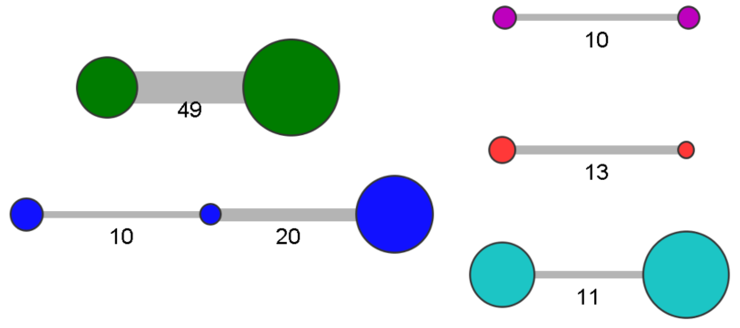}%
        \label{fig:uspol_dnc_co-rter_hccs}
    } \hfill
    \subfloat[RNC.]{
        \includegraphics[height=0.13\textheight]{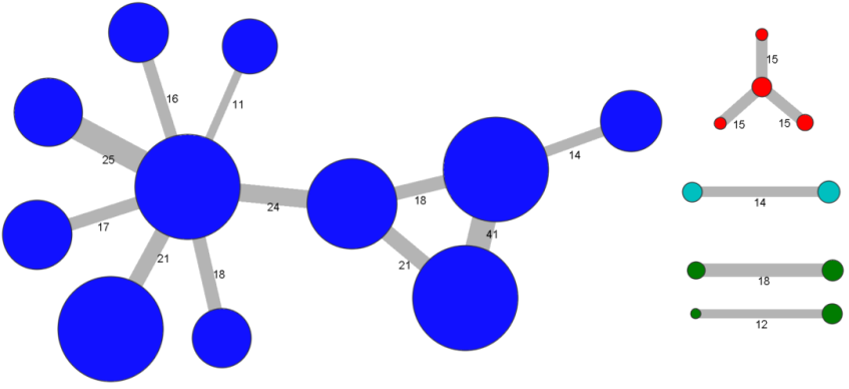}%
        \label{fig:uspol_rnc_co-rter_hccs}
    }
    \caption{Co-retweeting HCCs detected during the August 2020 DNC and RNC, active within the same $10$ second time window and thresholded to all edges with a normalised weight of $0.1$ and above. Nodes are HCC member accounts, sized by the number of tweets they contributed to the discussion, and joined by edges, which are sized and labelled according to the number of times they retweeted the same tweet. The nodes are coloured by Louvain cluster for convenience, but any matching colours between the DNC and RNC subfigures has no meaning.}
    \label{fig:uspol_co-rter_hccs}
\end{figure*}

By searching for accounts that co-retweeted within a ten second window of each other and then using a minimum normalised edge weight threshold of $0.1$, several HCCs were identified in each convention (see Figure~\ref{fig:uspol_co-rter_hccs}). Analysis of the HCC members using Botometer \citep{davis2016botornot} found the majority had a Complete Automation Probability (CAP) ratings above $0.6$ \citep[the relatively high threshold used by][as noted earlier]{DebateNightICWSM2018}. Further analysis of the HCCs' content provided some indication of their agendas, and examination of their account age and posting rates enabled categorisation into official accounts (verified by Twitter), unofficial reposters (topic-focused aggregators), and accounts that gave the appearance of typical human users. These `normal people', however, posted at very high average daily rates for years, often at far greater rates than previous automation detection methods have used \citep[e.g., 50 tweets a day, ][]{Neudert2018ger}. 

The largest HCC (the large blue HCC in Figure~\ref{fig:uspol_rnc_co-rter_hccs}) consisted of a cluster of potential social bot accounts supporting an official political campaign account, \mention{TrumpWarRoom}, responsible for $2,085$ tweets during the Republican Convention. For each pair of members in each HCC, we considered the proportion of time that one account retweeted a tweet before the other, to determine if both accounts were potentially working together (in which case, they would be equally likely to retweet a tweet first), or if one was a `cheerleader' for the other (in which case the cheered account would always retweet first, quickly followed by the other account). We found strong evidence that at least three of the accounts were cheerleaders for \mention{TrumpWarRoom}, retweeting the same tweet within ten seconds on $214$, $229$, and $89$ occasions over the four day collection period. These particular accounts had daily tweeting rates of $78.7$, $209.4$ and $147.4$ tweets per day for $0.9$, $8.5$ and $3.6$ years, respectively. Given the age of these accounts, it is clear that they have successfully avoided Twitter's bot scanning processes for some considerable time.

\begin{figure*}[t!]
    \centering
    \subfloat[DNC.]{
        \includegraphics[height=0.295\textheight]{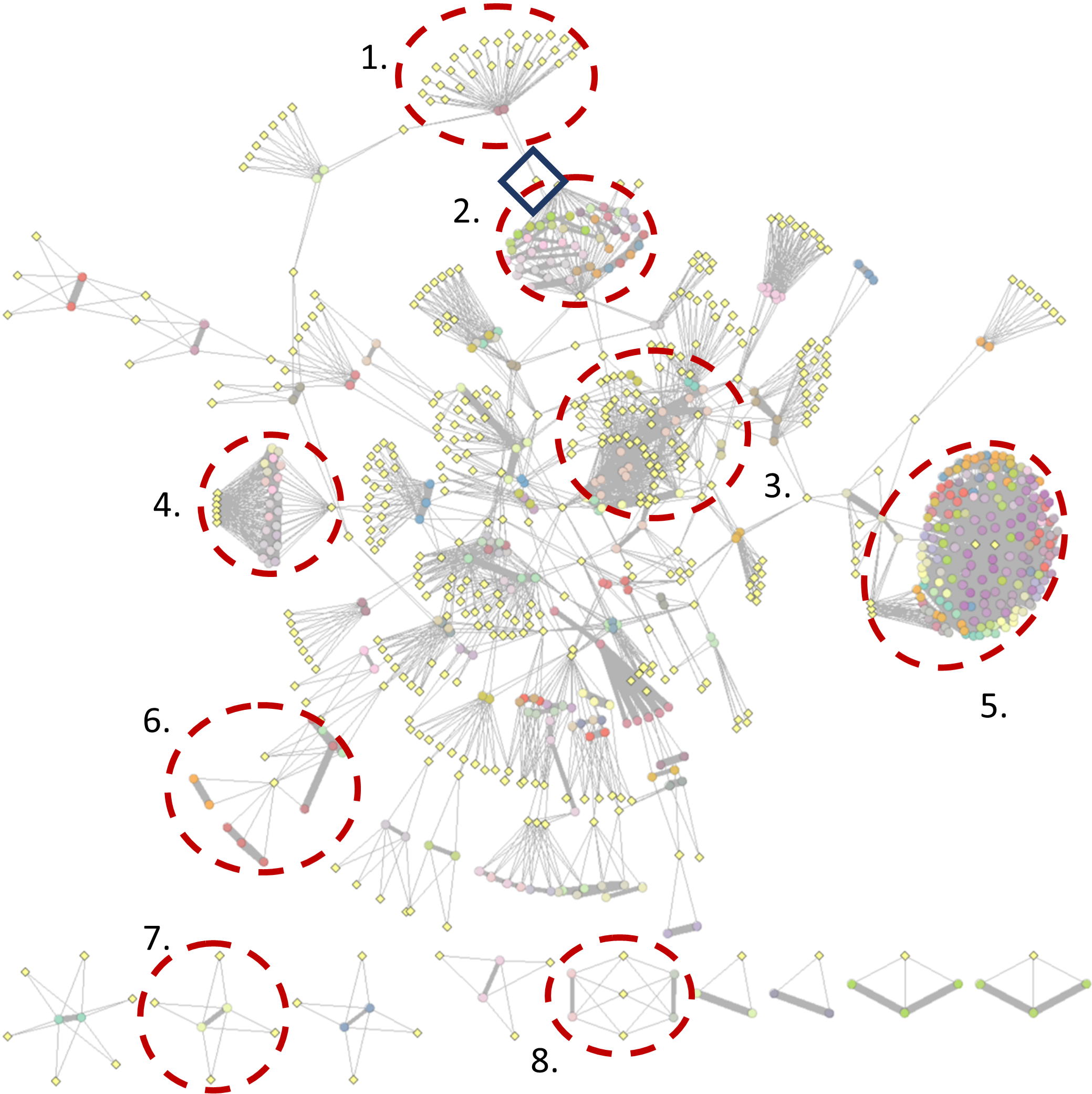}%
        \label{fig:uspol_dnc_co-hashtags}
    } \hfill
    \subfloat[RNC.]{
        \includegraphics[height=0.295\textheight]{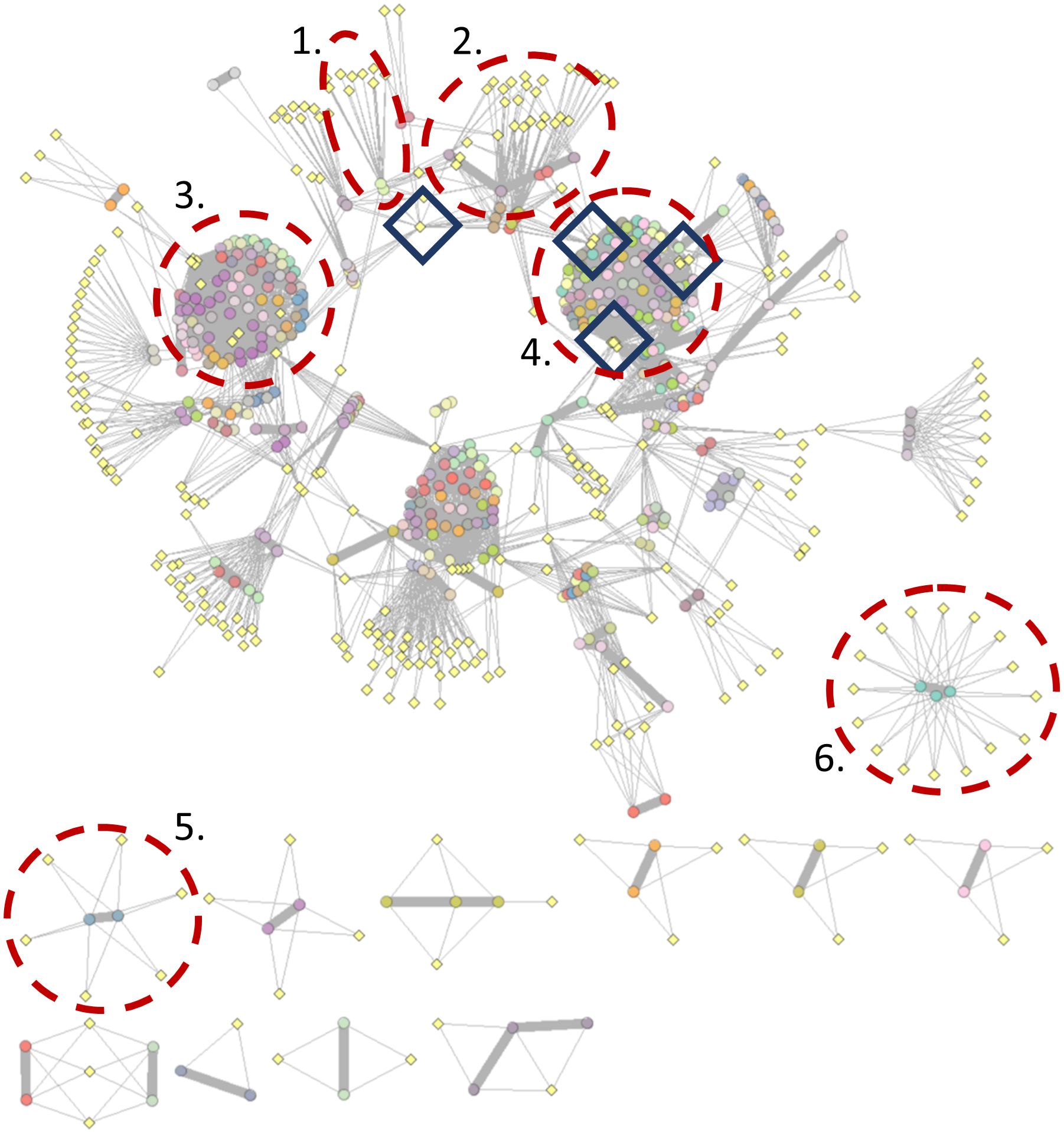}%
        \label{fig:uspol_rnc_co-hashtags}
    }
    \caption{Account/hashtag two-level networks of co-hashtag HCCs and the hashtags they used during the August 2020 DNC and RNC, active within the same $10$ second time window and extracted with FSA\_V ($\theta$=$0.3)$. Circular nodes are HCC member accounts, coloured by HCC, and hashtags are yellow diamond nodes. The links between accounts are sized by their co-hashtag frequency (i.e., how often they used the same hashtag in the same time window). \emph{visone}'s \emph{stress minimisation} layout was used for both networks. Notable clusters have been highlighted with red dashed ovals and numbered, while particular hashtag clusters have been highlighted with blue diamonds.}
    \label{fig:uspol_co-hashtags}
\end{figure*}

We also applied co-hashtag analysis (FSA\_V, $\theta$=$0.3$, $\gamma$=$10$ seconds) to the two datasets and plotted two-level networks of the resulting HCCs with the hashtags they used (Figure~\ref{fig:uspol_co-hashtags}). Regardless of the content, a number of structures are immediately apparent. These include:

\begin{itemize}
    \item clusters that are bound around a few yellow diamond hashtag nodes (e.g., DNC clusters 5, 6 and 8) or lie between hashtags (e.g., DNC clusters 2 and 4);
    \item fan shapes that consist of a small number of accounts using a wide variety of hashtags (e.g., DNC clusters 1 and 7);  
    \item island clusters that are bound by the hashtags they use but are isolated from the broader community which has ignored the hashtags they are using (e.g., DNC clusters 7 and 8).
\end{itemize}

The fact that the clusters are coloured according to their HCC in Figure~\ref{fig:uspol_co-hashtags} highlights how what FSA\_V is extracting what it regards as distinct clusters are, in fact, bound by the topics they are discussing (by the hashtags they are co-using). This indicates that there may be benefit in re-introducing the re-stitching step in FSA \citep{Sen2016fsa} that FSA\_V avoids, or also experimenting further with FSA itself. Using conductance cutting \citep{BrandesGW2015clustering} for cluster detection aligned better with the visible clusters, but these clustering may be misleading to varying degrees, as it may combine polarised HCCs, as can be seen on closer inspection below.

Several co-hashtag clusters in Figure~\ref{fig:uspol_dnc_co-hashtags} provide insight into the nature of parts of the online discussion. 
\begin{itemize}
    \item Cluster 5 is closely centred on two hashtags (\hashtag{goodyear} and \hashtag{ohio}) that relate to then US President Donald Trump's call for a boycott of Goodyear tires\footnote{The Goodyear factory in Ohio banned clothing with political messaging, including the Trump campaign's MAGA caps, during the election campaign. Source: \url{https://www.abc.net.au/news/2020-08-20/donald-trump-calls-for-goodyear-boycott-over-alleged-maga-ban/12577372}} though it is unclear whether the surrounding accounts are for or against the boycott. Several hashtags linked on the left edge of the cluster indicate that some are against, as they refer to support for the then Democratic candidate Vice President Joe Biden.
    \item The fan cluster at the top consists of two accounts that are attempting to disseminate their message across America, as each hashtag is a US state code (e.g., \hashtag{ga} for Georgia) or a minority group (e.g., \hashtag{latinos}). These hashtags are all apparently unique, apart from the one highlighted just below cluster 1 surrounded by a blue diamond, which is \hashtag{blm}, linking cluster 1 to cluster 2, and the one to the left, which is another state code (\hashtag{nc} for North Carolina).
    \item Cluster 2 binds a number of HCCs spanning two relatively disjoint hashtags, one being \hashtag{vote} (below the cluster) and the other being the name of a musician who had recently encouraged his fans to vote.
    \item Cluster 3 is more diffuse than the others and appears to relate to a discussion of data science and big data in the context of the election campaign.
    \item Cluster 4 appears to join a number of potentially opposed HCCs, as they refer to \hashtag{trump2020landslide} and \hashtag{snowflakes} as well as \hashtag{epstein}\footnote{Jeffrey Epstein was a billionaire arrested for sex crimes before dying in custody, however he was known to Donald Trump, and therefore this hashtag's use can be seen as an attack on his political campaign. \url{https://www.forbes.com/sites/lisettevoytko/2020/10/18/spider-book-excerpt-how-trumps-presidency-helped-expose-jeffrey-epstein/}} and \hashtag{trumpvirus} (a condemnation of the Trump administration's handling of the response to the COVID-19 pandemic), the final hashtag which links the cluster into the broader community. 
    \item The island clusters 7 and 8 are focused on groups of particular politicians, which were not picked up by the broader community: Republicans who had pledged to vote for the Democratic candidate and US Congress members known to campaign for social equality, respectively.
\end{itemize}

The links between the clusters are sometimes deceptive. Already, we observed that some single clusters include polarised HCCs, however it is also possible to see internally (politically) consistent clusters that are linked but also contrary in their views. Cluster 1 in Figure~\ref{fig:uspol_dnc_co-hashtags} is linked to the left by \hashtag{nc} to another left-leaning cluster (calling for gun control), which itself is linked to the left by \hashtag{america} to another small cluster, which is clearly right-leaning (one of its hashtags is \hashtag{voteredtosaveamerica}). These visualisations may highlight how HCCs can be merged, but care must be taken regarding the conclusions to draw from such an action.

Analysis of the co-hashtag HCCs and their hashtags in Figure~\ref{fig:uspol_rnc_co-hashtags} offers further examples of these observations and offers new insights. Clusters 1 and 2 are joined by the blue diamond-highlighted hashtag, \hashtag{blacklivesmatter}, but cluster 1 is a detractor group (using \hashtag{alllivesmatter}) while cluster 2 is a supporter group using several Black rights-related hashtags. Cluster 4 discusses riots following Black Lives Matter protests in Kenosha, Wisconsin, however while the two sets of hashtags highlighted at the top of the cluster relate mostly to current events (e.g., \hashtag{kenosha} and \hashtag{covid19} on the left, and \hashtag{kenoshariots} and \hashtag{thursdaythoughts}, plus \hashtag{walkaway}, which links to a small fan, as it is a pro-Republican statement to avoid conflict), the hashtags at the bottom of the cluster are more clearly right wing or conservative in nature, referring to a relevant media organisation, \hashtag{kag2020} (Keep America Great, a pro-Trump slogan) and \hashtag{ccot} (Christian Conservative on Twitter). Whereas cluster 4 in Figure~\ref{fig:uspol_dnc_co-hashtags} includes polarised HCCs, the placement of the hashtag nodes they are linked to offers no guidance on how they might be separated, cluster 4 in Figure~\ref{fig:uspol_rnc_co-hashtags} indicates that an alternative layout algorithm may aid analysis. Cluster 6 represents a concerted anti-Trump effort with many attacking hashtags, but the isolation of the HCC at the cluster's centre makes it clear that not many of the others tweeting during the RNC took its lead. Cluster~5 is an effort to draw attention to an instance of police brutality, which also did not gain traction with the broader co-hashtag community.

\section{Conceptual Comparison and Critique}

Methods to discover coordinated behaviour by inferring links between accounts based on related interactions is not unique. \citet{CaoCLGC2015urlsh} and \citet{Giglietto2019} identified groups of accounts based on the URLs they shared in common, while \citet{LeeCCS2013campext}, \citet{Keller2019} and \citet{graham2020asnac} relied on the similarity of the content posted by accounts to do the same. \citeauthor{Giglietto2019} explicitly add the temporal element by considering potential links only between accounts that share a URL within a constrained time frame. Their ``rational is that, while it may be common that several entities share the same URLs, it is unlikely, unless a consistent coordination exists, that this occurs within the time threshold and repeatedly.''\footnote{Quoted from the README of \citeauthor{Giglietto2019}'s open source code (as of 2021-01-19): \url{https://github.com/fabiogiglietto/CooRnet}}.
To the knowledge of the authors, only three other proposed approaches appear to generalise the idea to allow links between accounts to be inferred based on a variety of common behaviours: \citet{Pacheco2020arxiv}, \citet{graham2020asnac}, and \citet{Nizzoli2020}.

\citeauthor{Pacheco2020arxiv}'s method creates strong ties between accounts that share similar behavioural traits. Behavioural traits are extracted from social media data (e.g., hashtags or URLs) and, together with the accounts using them, a bigraph is created. An (weighted) account network is projected from this bipartite network, linking accounts that have edges to the same trait node. The more shared traits, the heavier the edge between accounts. Finally, the account network undergoes cluster analysis, specific to the nature of coordination sought. In their examples, Twitter accounts linked by sharing the same account handle are divided into clusters by virtue of the connected component in which they appear. A second example examining share market ``pump and dump'' scams links accounts based on the similarity of the text they post, using \emph{text frequency/inverse document frequency} (TF-IDF), and then clusters are discovered by simply filtering edges with a final weight less than $0.9$. A third example connects accounts that use multiple hashtags in the same order in their tweets. The approach was employed searching for co-retweeting communities spreading propaganda attacking the Syrian White Helmet movement by linking accounts that retweeted the same tweet within $10$ seconds of the original being posted \citep{PachecoFM2020whitehelmets}.

In contrast, \citeauthor{graham2020asnac}'s ``coordination network toolkit''\footnote{\url{https://pypi.org/project/coordination-network-toolkit/}} (CNT) is also written in Python, but relies on a populated database of information extract from tweets to search for coordinated retweeting (retweeting the same tweet), co-tweeting (tweeting identical text), co-similarity (tweeting similar text), co-linking (sharing the same URL), and co-replying (replying to the same tweet). The database implementation is used to improve the performance of searching for evidence of coordination between pairs of accounts (which, as in our approach, requires pairwise comparison of all accounts in the dataset) by using an inner join. This implementation would need to be modified to suit a streaming data source, but could conceptually be applied to data from a variety of OSNs.

The approach of \citet{Nizzoli2020} is very similar to ours, however it explicitly begins by selecting a set of users of interest, whereas we begin with a corpus of posts and our set of users is defined by those present in it. \citeauthor{Nizzoli2020} makes clear that users may be defined by the starting corpus of posts in the same way, or may be otherwise selected as superproducers or superspreaders or followers of a prominent account. They also introduce a filter step before the extraction of HCCs. \citet{Pacheco2020arxiv} filter their user similarity network with an arbitrary filter, which, as pointed out by \citeauthor{Nizzoli2020}, results in a binary classification of coordinating and non-coordinating users, but importantly disregards the effect of the network structure. Instead, \citeauthor{Nizzoli2020} rely on multiscale filtering approaches for complex networks, which retain network structures (not just individual edges) based on statistical significance. Furthermore, these can be scaled to retain more or less of the network, permitted examination of the `degree' of coordination, not just a binary decision on whether or not it is present. They propose an iterative algorithm at this point for detecting clusters of coordinating users, which makes use of an increasingly strict definition of user similarity (i.e., coordination) and each time relies on the communities found in the previous step as the starting point, guaranteeing they are kept in some form. This makes it possible to track communities at different levels of coordination, similar to how $k$ core decomposition provides insight into how deeply particular nodes and structures are embedded within a network. Finally, they apply a validation step, studying the resulting networks with network measures, and text analysis of the posts of the HCCs, but all as a function of the resolution at which the HCCs were detected. The FSA\_V algorithm is our alternative to their filtering and cluster detection steps. The ability for \citeauthor{Nizzoli2020} to examine different degrees of coordination is a distinguishing factor, however they also \citep[just like][]{Pacheco2020arxiv} must decide beforehand what similarity measure to connect users with -- this is equivalent to the behaviours that underpin the coordination strategies we discussed in Section~\ref{sec:coordination_strategies}, however they make the point that the similarity measure may involve any relevant information about the user profiles, not just their behaviour within the corpus. The temporal aspect of the coordination is not discussed, presumably as it is assumed to be a component of the user similarity measure. 

\citeauthor{Giglietto2019}'s CoorNet R package does not allow specification of a time window directly, but instead uses a proportion threshold to determine what to regard as an anomalously small but active time window, and thus requires access to an entire dataset. It is designed to study Coordinated Link Sharing Behaviour (CLSB)~\citep{Giglietto2020b} and thus only considers URLs in posts, however, it accepts URLs from a variety of OSNs and sources, including via CrowdTangle\footnote{\url{https://www.crowdtangle.com/}} and MediaCloud\footnote{\url{https://www.media.mit.edu/projects/media-cloud/overview/}}.

Our method is similar to all of these but is described in greater detail, relies upon a discrete window-based approach to apply temporal constraints, and we provide and evaluate a novel cluster extraction algorithm, and an open source implementation is available.
By apply time constraints in discrete windows, connections may be missed across windows, but this makes it easier to apply in near real-time streaming settings. If one were to infer connections between accounts as each new tweet is posted, it could create a potentially significant, ongoing processing cost depending on the number of unique accounts observed in the current time window. As new posts arrive, new nodes may need to be added to the account graph, while others may need to be removed, along with their adjacent edges (which, let us recall, represent evidence of coordination, not individual timestamped interactions as one might find in a social network based on direct retweets, mentions or replies). Furthermore, this constantly updated account graph must be complete, i.e., edges should always be added in case the evidence they represent may be consolidated by future posts. 

If the choice of time window is very short \citep[e.g., $10$ seconds, as per][]{PachecoFM2020whitehelmets}, and LCNs from adjacent windows are aggregated (as per our method), the absence of a truly sliding window like \citeauthor{graham2020asnac}'s may not significantly affect results, as ongoing high levels of coordination will appear over multiple windows. In contrast, if the time window is longer (e.g., $5$ or more minutes), then the hard boundary between windows may cause coordinated activities to be missed. The question is, then, what kind of coordination is being sought. Teams of bots tweeting or retweeting the same tweet within small time frames will be vulnerable to detection, however a deliberate covert human team with sockpuppet accounts may escape detection (at least initially) by varying the time frame over which retweets are posted (e.g., spread them unevenly over an hour or more), but if the same accounts cooperate for long, our method will find them once their activities are aggregated. One type of coordination that is very difficult to detect is single event boosts of a post: e.g., when, say, $1,000$ paid accounts retweet a single tweet. In a large discussion, $1,000$ tweets will not stand out especially, but, depending on how connected the paid accounts are to the broader discussion, they may spread the content a considerable distance through the network. Furthermore, gaming OSN trending algorithms may not be difficult\footnote{OSN gaming efforts of the form ``Let's get X trending'' are quite common in Australia, e.g.,  \url{https://twitter.com/Timothyjgraham/status/1351742513044807680}}, and even a thousand retweets may result in a valuable degree of influence in comparatively smaller communities (e.g., Australia).

As a final comment, all methods discussed in this section are suited to post-collection analysis, with \citeauthor{graham2020asnac}'s relying on the power of database systems to build the LCN but avoids clustering analysis for HCCs, \citeauthor{Giglietto2019}'s relies on R's expressivity and filtering based on anomaly detection, while our implementation uses Python and batch mode processing to enable flexibility in the choice of cluster analysis technique. \citeauthor{Pacheco2020arxiv}'s implementation is not available, but has been applied to very large datasets, and so may rely on a high performance programming language (e.g., Java) or distributed processing platform, such as Hadoop (\citeyear{Pacheco2020arxiv}). \citeauthor{Nizzoli2020} do not mention the availability of their implementation, only that their test dataset will be forthcoming (\citeyear{Nizzoli2020}).

Our paper is the only one of these to address the concept of searching for multiple coordination criteria, and how to treat the combination of their evidence, and the attendant complications explored in Step~\ref{alg:step4_lcn} of Section~\ref{sec:pipeline}. In fact, the other papers primarily treat the coordination criteria (i.e., user similarity measure) as entirely dependent on the current investigation and no generalisation of the concepts are discussed.

\section{Future Work}

The most important directions to take this work relate to the following aspects:
\begin{enumerate}
    \item \emph{\textbf{Temporal analysis.}} Temporal analysis of the evolution of HCCs and their influence on the broader discussion over time will provide insight into how the HCCs operate and achieve their goals, and perhaps will reveal distinct classes of HCCs and employed strategies. Further theoretical research relying on the temporal aspect is also required to determine the real meaning of edges in the LCN. Figure~\ref{fig:lcn_edge_semantics} provides an illustration of where there is ambiguity. If accounts A, B, and C all retweet tweet 1 within a single time window, or at least overlapping time windows, then we join A to B and B to C in the LCN, and there is a reasonable assumption that A and C may be related somehow. This is less clear when A and B retweet tweet 1 and B and C retweet tweet 2, both in different time windows; it is much less reasonable to assume a relationship between A and C in that case (especially if the time windows are far apart), but both situations result in the same LCN structure: A connects to B and B connects to C.
    
    \begin{figure*}[t!]
        \centering
        \subfloat[Situation 1.]{
            \includegraphics[height=0.24\textheight]{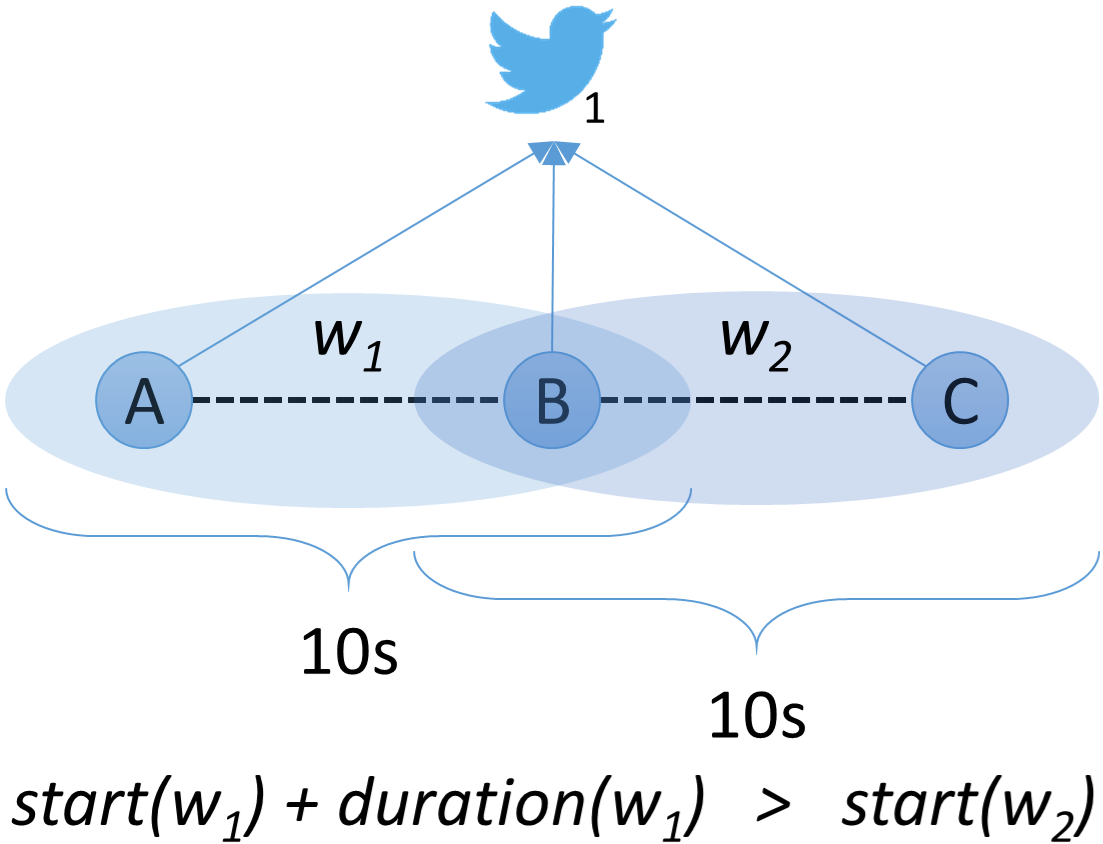}%
            \label{fig:lcn_edge_semantics_eg1}
        } \hfill
        \subfloat[Situation 2.]{
            \includegraphics[height=0.24\textheight]{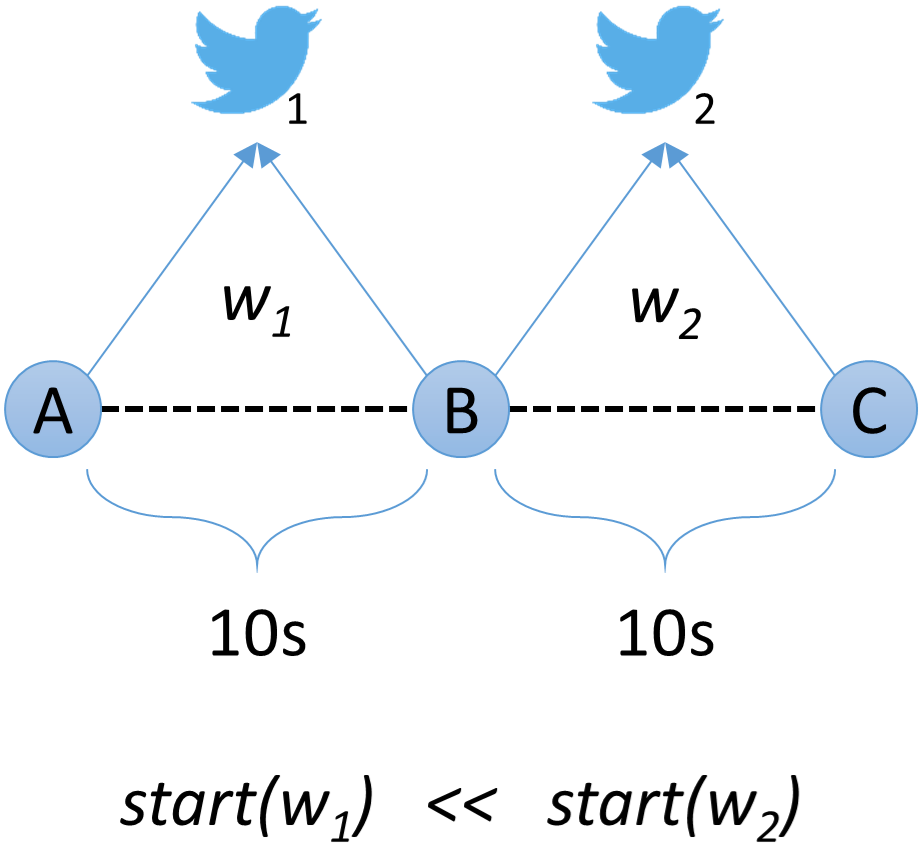}%
            \label{fig:lcn_edge_semantics_eg2}
        }
        \caption{The semantics of edges in LCNs requires clarification. If A is connected to B and B is connected to C, it may be due to the events in Situation 1 or Situation 2, but any inference of a relationship between A and C will be different depending on which it actually occurred. Each window, $w_x$, starts at timestamp $start(w_x)$ and has duration $duration(w_x)$. An arrow between, e.g., A and the Twitter bird 1 implies that account A retweeted tweet 1 within a specific window.}
        \label{fig:lcn_edge_semantics}
    \end{figure*}

    \item \emph{\textbf{Types of coordination.}} Not all coordination occurs in short periods of time nor in adjacent time windows. For example, an unscrupulous political campaign may purchase the services of $10,000$ bots to retweet a single campaign post once at a particular time to maximise its political effect. Alternatively, a small group of trolls may post the same tweets attacking a public figure each day at 5pm on weekends but not weekdays. Considering how to address broader definitions of coordination will be an ongoing challenge as OSNs change their features and people find new ways to use and abuse them.
    
    \item \emph{\textbf{Distinguishing authentic and inauthentic behaviour.}} The issue of astroturfing \cite[e.g.,][]{metaxas2012} brings this into sharp relief: some campaigns are genuine grass roots movements driven by a broad desire to see policies change (e.g., campaigning to address climate change), however some are artificially organised, aimed at gaming OSN trending algorithms to spread their narrative further and to give the appearance of genuine wide public support \cite[e.g., efforts to convince US Congress to release a politically controversial FBI memo,][]{mckew2018releasethememo}. As covert campaigns become more sophisticated and numerous, it will become more important for OSN, law enforcement and relevant agencies to focus their efforts on the relevant activities and therefore to be able to dismiss harmless fan campaigns from the next disinformation campaign. Others have noted that this problem remains unresolved at this time \citep{Vargas2020}.
    
    \item \emph{\textbf{Process improvement.}} Improving the implementation of the process, and how HCC extraction is performed, and how validation is conducted will be an ongoing activity. As demonstrated by \citet{Pacheco2020arxiv} and \citet{graham2020asnac}, different types of community extraction will suit different types of coordination strategy, just as will the choice of strategy to search for (e.g., pollution or boost). Introducing a genuine sliding window (\emph{cf.} our distinct, adjacent windows) will prevent missing further instances of coordination, but modification of the approach will be required to apply it in a near real-time setting \cite[e.g.,][]{CarneinAT2017textclust,AssenmacherATG2020}. Finally, to bring some statistical robustness to the question of validity, there are measures that can be used to determine if the accounts in HCCs engage in statistically significant greater or lesser levels of, say, retweeting than the general population. These measures will help determine in what ways HCC behaviour differs, but will leave unresolved the question of intent and authenticity of that behaviour.
\end{enumerate}

\section{Conclusion}

As online influence operations grow in sophistication, our automation and campaign detection methods must also expose the accounts covertly engaging in ``orchestrated activities''~\citep{GrimmeAA2018perspectives}. We have described several coordination strategies, their purpose and execution methods, and demonstrated a novel pipeline-based approach to finding sets of accounts engaging in such behaviours in two relevant datasets. Using discrete time windows, we temporally constrain potentially coordinated activities, successfully identifying groups operating over various time frames.
Guided by research questions posed in Section~\ref{sec:intro}, our results were validated by using a variety of techniques, including developing three one-class classifiers to compare the HCCs found in two relevant datasets, plus a randomised one, with HCCs from a ground truth subset. Two case studies were also presented, in which our technique was applied to reveal insights into the activity of polarised groups in one and the activity of social bots and bot-like accounts in the other. The algorithmic complexity of our approach was discussed, as well as comparison with several similar contemporary approaches. 
%In this paper, we have identified a number of coordination strategies used in online influence activities and the mechanisms by which they are executed, and then provided a pipeline-based processing methodology to identify sets of users engaging in such behaviours\footnote{See \url{https://github.com/weberdc/find_hccs} for code and data.}. Making use of a window-based approach, we can identify groups that operate over short or long time frames%%%{\color{red}, and find that a decayed sliding window technique highlights consistent behaviour and expands HCCs in filter-based stream collections}
%. The content of tweets posted by HCCs discovered in two datasets was revealed to be internally consistent, especially in a year-long dataset of known bad actors. Examination of hashtag use by HCCs also revealed their narrative foci. 
The temporal analysis of HCC evolution and their impact on the broader discussion, theoretical questions of the semantics of edges in LCNs, the ability to distinguish between authentic and inauthentic coordinated behaviour, improvement of HCC extraction and validation techniques and application in near real-time processing environments all provide opportunities for future research in this increasingly important field.

\section*{Declarations}

\subsection*{Funding} Not applicable.

\subsection*{Conflicts of interest/Competing interests} Not applicable.

\subsection*{Availability of data and material} The data (the identifiers of tweets only, as per Twitter's terms and conditions) used in this study are available at \url{https://github.com/weberdc/find_hccs}.

\subsection*{Code availability} The data manipulation and analysis software written for this study is available at \url{https://github.com/weberdc/find_hccs}.

\subsection*{Authors' contributions} Both authors contributed to the conception of the study, and Derek Weber performed the data collection, analysis and writing of the first draft. Both authors read and approved the final manuscript.

\subsection*{Ethics approval} All data were collected, stored, processed and analysed according to the ethics protocol H-2018-045, approved by the University of Adelaide's human research and ethics committee.

\subsection*{Consent to participate} (include appropriate statements)

\subsection*{Consent for publication} All authors consent to this work being published.

\bibliographystyle{spbasic}      % basic style, author-year citations
\bibliography{refs}   % name your BibTeX data base

% % Non-BibTeX users please use
% \begin{thebibliography}{}
% %
% % and use \bibitem to create references. Consult the Instructions
% % for authors for reference list style.
% % Format for Journal Reference
% Author, Article title, Journal, Volume, page numbers (year)
% Author, Book title, page numbers. Publisher, place (year)
% \end{thebibliography}

\end{document}